\documentstyle[12pt,epsf]{article}

\newcommand{\be}{\small\begin{equation}}
\newcommand{\ee}{\end{equation}\normalsize\vspace*{-0.1ex}}
\newcommand{\bea}{\small\begin{eqnarray}}

\newcommand{\eea}{\end{eqnarray}\normalsize\vspace*{-0.1ex}}
\newcommand{\bdm}{\small\begin{displaymath}}
\newcommand{\edm}{\end{displaymath}\normalsize\vspace*{-0.1ex}}
\newcommand{\beas}{\small\begin{eqnarray*}}

\newcommand{\eeas}{\end{eqnarray*}\normalsize\vspace*{-0.1ex}}

\newcommand{\n}{\noindent}
\newcommand{\nn}{\nonumber}
\newcommand{\eps}{\epsilon}
\newcommand{\intl}{\int\limits}

\newcommand{\dd}{{\mbox d}}
\newcommand{\ON}{{\cal{O}}\!\left(\frac{1}{n}\right)}
\newcommand{\al}{\alpha}
\newcommand{\dl}{\delta}
\newcommand{\om}{\omega}
\newcommand{\lm}{\lambda}
\newcommand{\ulm}{\underline{\lambda}}
\newcommand{\Gm}{\Gamma}
\newcommand{\gm}{\gamma}
\newcommand{\sg}{\sigma}
\newcommand{\HK}{\hat{K}}

\newcommand{\ka}{\kappa}

\newcommand{\up}{\underline{p}}
\newcommand{\uq}{\underline{q}}
\newcommand{\uu}{\underline{u}}
\newcommand{\uxi}{\underline{\xi}}
\newcommand{\um}{\underline{m}}
\newcommand{\ut}{\underline{t}}
\newcommand{\ual}{\underline{\al}}

\newcommand{\pa}{\partial}
\newcommand{\Qslash}{\not\! Q}

\newcommand{\paslash}{\not\! \partial}
\newcommand{\Dl}{\Delta}
\newcommand{\ODl}{\overline{\Delta}}
\begin{document}

%%%%% BEGIN  TITLEPAGE

\thispagestyle{empty}
\renewcommand{\thefootnote}{\fnsymbol{footnote}}

\setcounter{page}{0}
\begin{flushright}
UM-TH-95-22\\
hep-ph/9510437\\
October 1995  \end{flushright}

\begin{center}
\vspace*{0.8cm}
{\Large\bf
Ultraviolet Renormalons\\[0.2cm]
in Abelian Gauge Theories}\\
\vspace{1.8cm}
{\sc M.~Beneke\footnote{
Address after Oct, 1, 1995: SLAC,
P.O.~Box 4349, Stanford, CA~94309, U.S.A.
}} \\
\vspace*{0.3cm} {\it Randall Laboratory of Physics\\
University of Michigan\\ Ann Arbor, Michigan 48109, U.S.A.}\\[0.5cm]
and\\[0.5cm]
{\sc V.A.~Smirnov\footnote{E-mail: smirnov@theory.npi.msu.su}
} \\
\vspace*{0.3cm} {\it Nuclear Physics Institute\\
Moscow State University\\
119899 Moscow, Russia}\\[1.3cm]
{\bf Abstract}\\[0.3cm]
\end{center}

\n We analyze the large-order behaviour in perturbation theory
of classes of diagrams with an arbitrary
number of chains (i.e.
photon lines, dressed by vacuum
polarization insertions).
We derive explicit formulae for the leading and subleading divergence
as $n\to\infty$, and a complete result for the vacuum polarization
at the next-to-leading order in $1/N_f$. In general, diagrams with
more chains yield stronger divergence. We define an analogue of the
familiar diagrammatic $R$-operation, which extracts ultraviolet
renormalon counterterms as insertions of higher-dimension operators.
We then use renormalization group equations to sum the leading
$(\ln n/N_f)^k$-corrections to all orders in $1/N_f$ and find the
asymptotic behaviour in $n$ up to a constant that must be calculated
explicitly order by order in $1/N_f$.
\vspace*{0.3cm}

\begin{center}
{\it submitted to Nuclear Physics B}
\end{center}

\newpage
\renewcommand{\thefootnote}{\arabic{footnote}}
\setcounter{footnote}{0}

%%%%%%%%%%%%%%%%%%%% SECTION 1 %%%%%%%%%%%%%%%%%%%%%%%%%%%%%%%%%

\section{Introduction}

Perturbative calculations in quantum field theories involve integrations
over arbitrarily small distances and typically lead to divergent
results. A finite result is obtained, when a
theory is first regulated\footnote{Note that there are also
schemes without regularization, e.g.,
BPHZ, differential. Although they lead to physically meaningful results
there is no bare Lagrangian for them.}
and counterterms are added to the bare Lagrangian ${\cal L}_0$.
The Green functions computed from

\be\label{l1}
{\cal L}={\cal L}_0(\Lambda) + {\cal L}^{(4)}_{\rm ct}(\Lambda)
\ee

\n have perturbative expansions free from divergences order by order in
the renormalized coupling $\alpha(\mu)$ so that they are finite in the limit
when the cut-off $\Lambda$
is removed. A justification for this rather {\em ad hoc}-looking
procedure can be obtained from the philosophy of
`effective field theories':
Only the renormalized parameters are accessible to low-energy
experiments and the apparently divergent regions of integration are
in fact (almost) insignificant.

\mbox{From} the beginnings of renormalized quantum field theory it has been
recognized that the Green functions (in the limit $\Lambda\to \infty$)
obtained in this way can not be unambiguously defined (as certain
analytic functions in a neighbourhood of $\alpha=0$) through their
perturbative expansions alone, because they diverge for any $\alpha\neq0$
\cite{DYS52}. Although from a practical point of view one may consider
these expansions as asymptotic (to {\em nature}), the (non-perturbative)
existence of renormalized field theories remains a mathematically largely
unsolved problem, the divergence of perturbative expansions being one
face of this problem, the issue of triviality in non-asymptotically free
theories being another.

Without touching the profound problem of existence, the behaviour
of perturbative expansions as formal series is itself important.
In considering the perturbative expansion to all orders, one takes
in fact a glimpse beyond perturbation theory. Thus, although the
questions of triviality and Landau poles in general can not be answered
without knowledge of non-perturbative properties of the theory, some
aspects can be investigated strictly within perturbation theory. In
theories where the coupling can become large at low energies, the details
of the divergence of the perturbation series may provide some
hints to selecting the numerically most important corrections already
in moderately large orders.
The dominant source of divergence (at least with present knowledge)
was identified by Lautrup \cite{LAU77} and `t~Hooft \cite{THO77},
who investigated a particular class of diagrams with
an arbitrary
number of vacuum polarization insertions into a single gauge boson
line in a loop diagram. The sum of these diagrams has the generic behaviour

\be \label{renormalon}
\Gamma(\up,\mu,\alpha(\mu)) = \sum_{n} K \left(\frac{\beta_0}{a}
\right)^n n!\,
n^b \alpha(\mu)^{n+1} \left[1+\ON\right]\,.
\ee

\n Here $\up$ is a collection of external momenta and
$\beta_0$ is the first coefficient of the $\beta$-function.
With the definition given later,
the constant $a$ is integer (and
sometimes half-integer). The $n!$-behaviour arises as a consequence
of renormalization and for this reason has become known as renormalon
divergence. Note that the term `divergence' is applied to the divergence
of the perturbation {\em series} as well as to the divergences of
Feynman integrals, which are subtracted in the process of renormalization.

In strictly renormalizable theories the coupling depends
logarithmically on the renormalization scale $\mu$ and each vacuum
polarization loop
gives a $\ln k^2/\mu^2$. This logarithm is large whenever the loop
momentum $k$ is very different from $\mu$. In the present paper we
consider only large momentum regions $k\gg\mu$ and ultraviolet (UV)
renormalons, with $a>0$. The divergences associated with
regions of integration over large loop momenta,
%that give rise to explicit ultraviolet divergences,
like $\dd^4k/k^4$ for large $k$,
are removed
by the familiar renormalization procedure. The ultraviolet
renormalons occur when a large power of $\ln k^2/\mu^2$ is integrated
over the remaining ultraviolet regions of the subtracted integrands.
In particular, the strongest divergence ($a=1$) is exclusively due to the
remaining behaviour $\dd^4 k/k^6$ for large $k$. The fact that UV renormalons
are related to the large momentum {\em expansion} of Feynman
{\em integrands} is crucial for their understanding, since it allows to
describe UV renormalon divergence in large orders in terms of local
operators just as explicit divergences in finite orders in the usual
framework of renormalization. This observation is the basis for
Parisi's hypothesis \cite{PAR78} that UV renormalons can be removed
by adding higher dimensional operators to the Lagrangian. In particular,
the leading UV renormalon  ($a=1$) could be compensated by an additional
term

\be \label{dimsix}
{\cal L}_{\rm ct}^{(6)} = \frac{1}{\mu^2}\sum_i E_i(\alpha(\mu))
\,{\cal O}_i\,,
\ee

\n where the sum runs over all local operators ${\cal O}_i$ of dimension
{\em six}. To compensate all UV renormalons an infinite series of
higher-dimensional operators would have to be added to the Lagrangian.
In this paper
we deal explicitly only with the leading UV renormalon and restrict ourselves
to operators of dimension six.

The fact that the removal of ultraviolet renormalons, as
the removal of ultraviolet divergences, can be formulated at the
level of counterterms in the Lagrangian implies their universality:
Once the coefficients $E_i$ have been determined from a suitable set
of Green functions, the subtraction of the first
ultraviolet renormalon is
automatic for all Green functions. Another consequence is that the
$E_i$ satisfy renormalization group equations, because Green functions
with operator insertions satisfy them.
These considerations fix the constant $b$ in
eq.~(\ref{renormalon}) \cite{PAR78}. The solution of the renormalization
group equation depends on a boundary value which remains unconstrained
by general considerations and is related to the normalization $K$ of
the renormalon divergence in eq.~(\ref{renormalon}).

An alternative approach, based on the Lagrangian at a finite cut-off
rather than the renormalized Lagrangian, has been taken in \cite{BER84}.
The idea is that, since UV renormalons arise as a consequence of
the infinite cut-off limit, their information is encoded in the
large-cut-off expansion. Viewed in this way the compensation of
UV renormalons bears close resemblance to Symanzik improvement of
lattice actions \cite{SYM83}.

Although the absence of the first ultraviolet renormalon as a consequence
of eq.~(\ref{dimsix}) seems rather obvious from
the physical origin of UV renormalons, it has not yet been rigorously
established. Moreover, the diagrammatic interpretation of eq.~(\ref{dimsix})
is rather unclear: It does not give a clue, which diagrams contribute
to the coefficients $E_i$ or whether they can be calculated at all in
a systematic way. These are the questions which we address in this
paper.

The diagrammatic study of UV renormalons received new attention
only recently,
through the work of Zakharov \cite{ZAK92} and others [8--11]. The
main difficulty that the diagrammatic approach has to face is that, because
the object is to study perturbative expansions in large (that is,
to all) orders, there is in fact no natural expansion parameter that
would select a manageable subset of the infinity of all diagrams. In abelian
gauge theory it is useful to classify diagrams in terms of complete
gauge boson propagators. In first approximation,
where only the first coefficient in the $\beta$-function is kept,
the complete photon propagator reduces to a `chain', a string
of fermion loops. With few exceptions,
previous investigations of UV renormalons have focused on diagrams
with a single chain. In \cite{DIC95} it was shown that at this level
one could remove the first ultraviolet renormalon from Green functions
by counterterms of the form of eq.~(\ref{dimsix}). The full complexity
of calculating the normalization $K$ is already exhibited by
diagrams with one complete photon propagator:
To obtain the value of $K$, one can
not approximate the propagator by a string of fermion loops (chain).
The exact
photon propagator has to be kept \cite{GRU93,BEV93}. For practical
purposes this is equivalent to the statement that $K$ can not be
calculated exactly.

An important new insight comes from the work of Vainshtein and Zakharov
\cite{VAI94}, who investigated the dominant contributions to the
large-order behaviour of the photon vacuum polarization from diagrams
with two chains by making direct use of the fact that the UV renormalons
originate from the large-momentum regions in loops. The diagrams with
two chains display a qualitatively new behaviour, because the
four-fermion operators that appear in eq.~(\ref{dimsix}) do not
contribute to diagrams with a single chain. After insertion into the
photon vacuum polarization, they were found to yield the dominant
large-order behaviour.

In this paper we approach UV renormalons from an entirely diagrammatic
perspective within the expansion in the number of chains, or, to
be precise, in $1/N_f$, where $N_f$ is the number of fermions.
Guided by the interpretation of UV renormalon divergence as similar
to the usual ultraviolet divergences, we proceed in close analogy
with the usual renormalization program. We will see that the analysis
of UV renormalon divergence order by order in the expansion in chains
has much in common with the analysis of UV divergence order by order
in the coupling $\alpha$. The UV renormalon problem
then takes a form similar to usual UV divergences: While certain
properties like locality of counterterms and renormalization group
equations can be established to all orders (for an arbitrary number
of chains), the actual calculation of counterterms is limited to a few
first terms because of the increasing complexity of integrals. In the
present paper we proceed heuristically and do not give proofs which
could be worked out as generalizations of standard cumbersome proofs
of renormalization theory. Rather than giving proofs we
describe the properties of regularized Feynman integrals and subtraction
operators which would be essential ingredients to these proofs and
illustrate how the extraction of ultraviolet renormalon divergence
works in a number of examples.

Let us make a remark on the framework of the expansion in the
number of chains: It
might appear inconsequent to replace one expansion (the one
in the coupling $\alpha$) by another. Our attitude is that this
expansion can shed some light on the organization of UV renormalon
divergence in the same way as, for instance, the $1/N_c$-expansion
in QCD may reveal some information on the strong-coupling regime. In
addition, we will see that the dominant UV renormalon divergences
in each order of $1/N_f$ can be resummed to all orders in $1/N_f$
by solving the renormalization group equations. It is possible that
the conclusions drawn from the $1/N_f$-expansion are not valid
for any finite $N_f$ or valid only in a finite range of $N_f$. However,
in an abelian gauge theory we do not consider this a likely possibility
and rather expect a smooth continuation from the large-$N_f$ limit
to small $N_f$.

In Sect.~2 we begin with detailing the expansion in chains. We
introduce the Borel transform as generating function of perturbative
coefficients and show how the factorial divergence of perturbation
theory is encoded in the singularities of analytically regularized
Feynman integrals. We collect some of their properties and classify
the subgraphs that can contain the dominant UV renormalon.
The extraction of the renormalon divergence then reduces to the
extraction of pole parts of analytically regularized Feynman integrals
at certain positions in regularization parameter space.

In Sect.~3
we construct an operation that picks out the terms with the largest
number of singular factors. The leading large-$n$ behaviour at any
order in $1/N_f$ is then found by successive extraction of pole
parts of one-loop integrals. Beyond the leading large-$n$ behaviour
it is essential to apply the method of infrared rearrangement \cite{Vlad,CS}.
The operations introduced in this section are illustrated by the simplest
example of the fermion self-energy up to two chains. In Sect.~4 we
compute the fermion-photon vertex and the photon vacuum polarization
including all diagrams with two chains and those diagrams with three
chains that contain an additional fermion loop.

The universality of UV renormalons becomes most transparent by
elevating the diagrammatic subtractions to the level of
counterterms in the Lagrangian. We formulate the results of the
previous sections in this language in Sect.~5. The renormalization
group equations for Green functions with operator insertions are
then used to sum the leading UV renormalon singularities to
all orders in $1/N_f$. We summarize in Sect.~6 and discuss further
applications of the formalism.

The derivation of some results quoted in sections 2 and 3 is
collected in an appendix.

%%%%%%%%%%%%%%%%%%%% SECTION 2 %%%%%%%%%%%%%%%%%%%%%%%%%%%%%%%%%%%%%%%%

\section{Renormalons and analytic regularization}
\setcounter{equation}{0}

In this section we set up the reorganization of the perturbation
series in terms of chains and derive the Feynman rules for the
Borel transform of this expansion. We show how renormalons are related
to the singularities of analytically regularized Feynman integrals.
We consider only the abelian gauge theory (QED) with Lagrangian

\be\label{lagrangian}
{\cal L} = -\frac{1}{4} F_{\mu\nu} F^{\mu\nu} + \bar{\psi}
i \!\not\! \partial\psi + g \bar{\psi}\!\not\!\! A \psi + {\cal L}^
{(4)}_{\rm ct}\,.
\ee

\n Since we are interested in large-momentum regions, we can consider
the fermions as massless. We assume $N_f$ species of fermions, but
do not write the flavour index explicitly. Summation over repeated
flavour indices is understood. The $\beta$-function is given by
($\alpha=g^2/(4\pi)$)

\be\label{betafunction}
\beta(\alpha) = \mu^2 \frac{\partial \alpha}{\partial\mu^2} =
\beta_0\alpha^2+\beta_1\alpha^3+\ldots\,,
\qquad\quad
\beta_0 = \frac{N_f}{3\pi}\,,\qquad \beta_1=\frac{N_f}{4\pi^2}
\,.
\ee

\subsection{Chains}

Consider the perturbative expansion of a truncated Green function
$G(\uq;\alpha)$, where $\uq=(q_1,\ldots,q_M)$
denotes a collection of external momenta
and $\alpha=\alpha(\mu)$ the renormalized coupling. We can organize
this expansion in terms of diagrams with complete photon propagators

\be \label{completepropagator}
\frac{(-i)}{k^2} \left(g_{\mu\nu}-\frac{k_\mu k_\nu}{k^2}\right)\,
\frac{\alpha}{1+\Pi(k^2)} + (-i)\,\xi\,\frac{k_\mu k_\nu}{k^4}\,\alpha\,,
\ee

\n where $\Pi(k^2)$ is the photon vacuum polarization and $\xi$ the
gauge fixing parameter. Each such diagram corresponds to a class
of diagrams in the usual sense. Let $\bar{\Gamma}$ be such a class of
diagrams with $N$ complete photon propagators and let the lowest
power of $g$ that occurs in a diagram in $\bar{\Gamma}$ be
$g^e\alpha^N$. (In the abelian theory, $e$ depends only on the Green
function $G$ and equals the number of external photon lines.) Then
we write the contribution from $\bar{\Gamma}$ to the series expansion
of $G$ as

\be\label{G2}
G_{\bar{\Gamma}}(\uq;\alpha) = (-ig)^e\sum_{n=N-1}^\infty r_n
\alpha^{n+1}\,.
\ee

\n Note that if we consider a physical quantity, we may take the
complete photon propagator to be renormalized. Due to the Ward
identity, no further renormalization is required and the sum over
all diagrams with a given number of renormalized photon propagators
is finite.

Formally, the series can be written in the Borel representation

\be\label{borelrepresentation}
G_{\bar{\Gamma}}(\uq;\alpha) = (-ig)^e \frac{1}{\beta_0}\intl_0^\infty
d u\,e^{-u/(\beta_0\alpha)}\,B[G_{\Gm}](\uq;u)\,,
\ee

\n where

\be\label{seriesdef}
B[G_{\Gm}](\uq;u) =
\sum_{n=N-1}^\infty\frac{r_n}{n!}\,
\beta_0^{-n} u^n
\ee

\n is the Borel transform of the series. The factorial divergence
of the series then leads to singularities of the Borel transform
at finite values of $u$. For example,
the large-order behaviour of eq.~(\ref{renormalon}) results in
a singularity at $u=a$ and the constant $b$ determines the nature of
the singularity. The integral in eq.~(\ref{borelrepresentation})
does not exist due to these singularities. In the present context,
we use the Borel transform only as generating function for the
perturbative coefficients. We do not consider at all
the problem of summation of the perturbative series (e.g. by use
of the Borel transform).

Let $\Gamma$ be the skeleton diagram corresponding to $\bar{\Gamma}$.
Then $G_{\bar{\Gamma}}$ is represented as

\be\label{G}
G_{\bar{\Gamma}}(\uq;\alpha) = (-i g)^e \int\prod_l\frac{\dd^4 p_l}{(2\pi)^4}
\prod_{j=1}^N\frac{\dd^4 k_j}{(2\pi)^4}\,I_\Gamma(\uq;p_l,k_j) \,
\prod_{j=1}^N\frac{\alpha}{k_j^2 (1+\Pi(k_j^2))}\,.
\ee

\n The momenta $p_l$ are assigned to fermion lines and $k_j$
to photon lines. The function $I_{\Gm}$ is the Feynman integrand
(without the factors $1/k^2_j$) of the
skeleton diagram, including $\delta$-functions in momenta from
vertices. The complete
photon propagators are written explicitly, except for the Lorentz
structure $(-i) (g_{\mu\nu}-k_\mu k_\nu/k^2)$, which is included
in $I_\Gamma$. We can drop the piece proportional to $\xi$ in
eq.~(\ref{completepropagator}) by specifying Landau gauge
$\xi=0$. This will be assumed in the following unless stated
otherwise. The convolution theorem for Borel
transforms can be used to derive the identity

\be
B\!\left[\prod_{j=1}^N\frac{\alpha}{1+\Pi(k_j^2)}\right](u) =
\frac{1}{\beta_0^{N-1}}\intl_0^u\left[\prod_{j=1}^N d u_j\right]\,
\delta\!\left(u-\sum_{j=1}^N u_j\right)\,\prod_{j=1}^N
B\!\left[\frac{\alpha}{1+\Pi(k_j^2)}\right]\!(u_j)\,,
\ee

\n so that

\bea\label{someintermediateformula}
B[G_{\Gm}](\uq;u) &=& \frac{1}{\beta_0^{N-1}}\intl_0^u\left[
\prod_{j=1}^N d u_j\right]\,
\delta\!\left(u-\sum_{j=1}^N u_j\right)
\nonumber\\
&&\times \,\int\prod_l\frac{\dd^4 p_l}{(2\pi)^4}
\prod_{j=1}^N\frac{\dd^4 k_j}{(2\pi)^4}\,I_\Gamma(\uq;p_l,k_j) \,
\prod_{j=1}^N \frac{1}{k_j^2}\,
B\!\left[\frac{\alpha}{1+\Pi(k_j^2)}\right]\!(u_j)\,.
\eea

\n This expression is still too complicated, because it contains
the complete photon propagators. The complete propagators can
themselves be expanded, if we consider the limit, when the number
of fermion flavours $N_f$ is large. Then write

\be\label{fermionloop}
\Pi(k^2) = -\beta_0\alpha\left[\ln\left(-\frac{k^2}{\mu^2}\right)
+ C\right] + \Pi_1(k^2) + {\cal O}\!\left(\frac{1}{N_f}\right)\,.
\ee

\begin{figure}[t]
   \vspace{0cm}
   \centerline{\epsffile{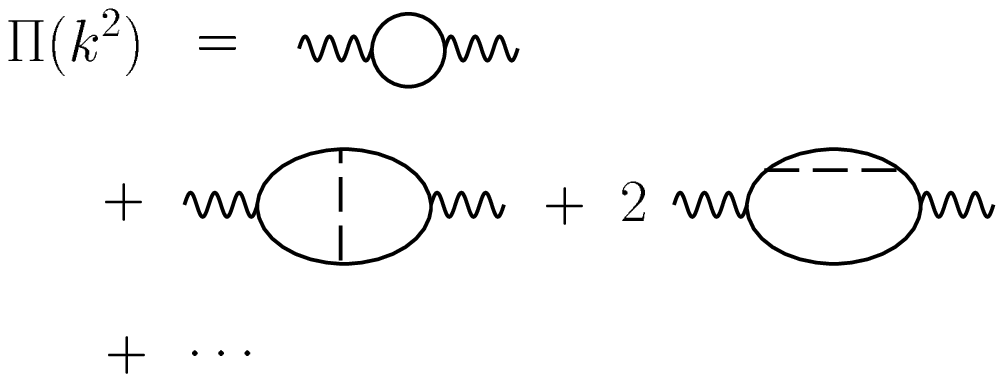}}
   \vspace*{0.5cm}
\caption{\label{chain} Expansion of the photon vacuum polarization in
chains ($1/N_f$). The dashed line denotes summation over a photon line
with an arbitrary number of fermion loops inserted.}
\end{figure}

\n This expansion in {\em chains} is depicted in Fig.~\ref{chain}.
The dashed line denotes a chain, i.e. a photon propagator with an arbitrary
number of simple fermion loops inserted. $C$ is a subtraction
constant for the fermion loop, $C=-5/3$ in the $\overline{\rm MS}$
scheme. Then we use again the convolution theorem to obtain

\be\label{borelcomplete}
B\!\left[\frac{\alpha}{1+\Pi(k^2)}\right]\!(u) = \left(-\frac{
k^2}{\mu^2}\,e^C\right)^u-\frac{1}{\beta_0^2}\intl_0^u
d v\,v\left(-\frac{k^2}{\mu^2}\,e^C\right)^v B\!\left[\frac{
\Pi_1}{\alpha}\right]\!(u-v) + \ldots
\ee

\n The expansion can be continued to include further terms. We will
not need them in this paper. The Borel transform of $\Pi_1$ has
been computed in \cite{BEN93,BRO93}. It can be written as

\bea\label{f}
B\!\left[\frac{\Pi_1}{\alpha}\right]\!(u) &=& \beta_1
\left[\left(-\frac{
k^2}{\mu^2}\,e^C\right)^u F(u) - G(u)\right]
\\
F(u) &=& -\frac{32}{3}\,\frac{1}{u (2+u)}
\sum_{k=2}^\infty \frac{(-1)^k k}{(k^2-(1+u)^2)^2}\,,
\nonumber\eea

\n where $G(u)$ is a scheme-dependent function, whose expression,
for example in the $\overline{\rm MS}$ scheme, can be found in
Sect.~5 of \cite{BBB1}. (Note that the parameter $u$ there, as in
most publications that deal mainly with infrared renormalons in
QCD, is defined with an opposite sign compared to the definition we
use in this paper.) We will not need an explicit form of $G(u)$.

Let us momentarily replace all complete photon propagators by
chains, so that we keep only the first term on the right hand
side of eq.~(\ref{borelcomplete}). Then

\be\label{analytic1}
B[G_{\Gm}](\uq;u) = \frac{1}{\beta_0^{N-1}}
\left(-\mu^2 e^{-C}\right)^{-u}\intl_0^u\left[
\prod_{j=1}^N d u_j\right]\,
\delta\!\left(u-\sum_{j=1}^N u_j\right)\, G_\Gamma(\uq;\uu)\,,
\ee

\n where $\uu = (u_1,\ldots,u_N)$ and

\be\label{analytic2}
G_\Gamma(\uq;\uu) = \int\prod_l\frac{\dd^4 p_l}{(2\pi)^4}
\prod_{j=1}^N\left[\frac{\dd^4 k_j}{(2\pi)^4}
\frac{1}{(k_j^2)^{1-u_j}}\right]\,I_\Gamma(\uq;p_l,k_j)
\ee

\n is the skeleton diagram with analytically regularized
photon propagators.\footnote{Note by comparison with eq.~(\ref{G})
that $G_\Gamma(\uq;\uu)$ contains no powers of the coupling $g$.}
Note that a different parameter is introduced
for each photon line and the fermion lines are not regularized.
For a physical quantity the sum over all skeleton diagrams is
finite. For individual diagrams $\Gamma$ all UV
divergences are explicitly regularized by the parameters $u_i$, since
the only loops without any regularization parameter in $\Gamma$
are fermion loops with four or more
photon lines attached, which are UV finite. Since fermion lines are
not analytically regularized, gauge invariance is preserved.

The factorial divergence of perturbation theory corresponds
to singularities of $B[G_{\Gm}](\uq;u)$. From
eq.~(\ref{analytic1}) together with eq.~(\ref{analytic2}) we
deduce that it is sufficient to know the singularities of the
analytically regularized skeleton diagram. We collect results
on its singularity structure in the next subsection.

Corrections to the approximation of replacing the complete photon
propagator by a chain can be incorporated through the
corrections to the first term in eq.~(\ref{borelcomplete}).
$B[G_{\Gm}](\uq;u)$ is then
expressed as a cer\-tain con\-vo\-lu\-tion integral of $G_\Gamma(\uq;u_i)$
and $B[\Pi_1/\alpha](u)$. Similar expressions follow,
if one includes yet
further terms in eq.~(\ref{borelcomplete}). Thus, provided the
vacuum polarization is known to the desired accuracy, one can
restrict attention to analytically regularized skeleton diagrams.

When the complete photon propagator is replaced by a chain, the calculation
of $B[G_{\Gm}](\uq;u)$ for a skeleton diagram with an arbitrary
number of chains follows from the usual Feynman rules with the following
modifications: The photon propagator is given by

\be\label{rulephoton}
\frac{(-i)}{k^2} \left(g_{\mu\nu}-\frac{k_\mu k_\nu}{k^2}\right)
\left(-\frac{\mu^2}{k^2} e^{-C}\right)^{-u_j}
\ee

\n (a $+i\eps$-prescription is understood)
and for internal vertices we have\footnote{The factor $\sqrt{4\pi}$
comes from $g=\sqrt{4\pi\alpha}$, since the
Borel transform is taken with respect to $\alpha$.}

\be (-i)\sqrt\frac{4\pi}{\beta_0}\,\gamma_\mu\,.
\ee

\n For vertices extending
to an external photon line we have $\gamma_\mu$ only, cf. eq.~(\ref{G2}).
Finally all $u_j$-parameters are integrated over with $\beta_0
\int_0^u \prod d u_j\,\delta(u-\sum u_j)$.

\subsection{Singularities of analytically regularized
Feynman \newline diagrams}

To characterize the singularities of $B[G_{\Gm}](\uq;u)$ we
use the results and techniques developed
for the analysis of divergences and singularities of
Feynman integrals \cite{Speer,Pohl,BM}. Here we summarize
the relevant statements regarding their analytical structure.
Details can be found in the appendix.

Let $F_{\Gm}(\uq,\um)$ be a Feynman
integral corresponding to a graph $\Gm$. It is a function of
external momenta $\uq=(q_1,\ldots,q_{M})$ and
masses $\um=(m_1,\ldots,m_L)$.
Below we really need only the pure massless case.
The Feynman integral is supposed to be constructed from
propagators of the form

\be
\frac{Z_l(p)}{(p^2+m_l^2)^{1+r_l}}.
\label{pr}
\ee

\n Here $Z_l$ is a polynomial of degree $a_l$ in the momentum of the $l$-th
line. For the moment we let $r_l$ be integer.
The UV divergences are characterized by the UV degrees of divergence
of subgraphs $\gm$ of $\Gm$,

\be
\om(\gm) = 4 h(\gm) - 2L(\gm) + a(\gm) - 2 r(\gm),
\label{om}
\ee

\n where $h(\gm)$ and $L(\gm)$ are respectively the numbers of loops and
lines of $\gm$, $a(\gm) = \sum_{l\in\gm} a_l$
$r(\gm) = \sum_{l\in\gm} r_l$. An analytically regularized Feynman
integral is defined by the replacement

\be
1/(p^2+m_l^2)^{1+r_l} \to 1/(p^2+m_l^2)^{1+r_l+\lm_l} .
\label{prl}
\ee

\n For some lines, the corresponding $\lm$-parameters can be equal
to zero.

Let us suppose that there are no
infrared (IR) divergences in the graph and that
the available $\lm$-parameters are sufficient to regularize UV divergences
in all UV divergent subgraphs (in which $\om(\gm)\geq 0$). The
second
assumption means that for any such $\gm$ at least one of the
corresponding regularization parameters is not identically zero.
Then the analytically regularized Feynman diagram $F_{\Gm}(\uq;\ulm)$
can be represented as

\be
\sum_{\cal F} \prod_{\gm\in {\cal F}:\,\om(\gm)\geq 0}
\frac{1}{\lm(\gm)} \; g_{\cal F} (\ulm)
\label{as} \ee

\n where $\ulm = (\lm_1,\ldots,\lm_L)$,
the functions $g_{\cal F} $ are analytical in a
vicinity of the point $\ulm=\underline{0}$,
and $\lm(\gm) = \sum_{\l\in\gm} \lm_l$.
Here the sum is over all maximal forests of $\Gm$. Remember that a forest
is a set of non-overlapping subgraphs. A forest ${\cal F}$ is maximal if
for any $\gm$ which does not belong to ${\cal F}$
the set ${\cal F} \cup \{\gm\}$ is no longer a forest. The steps that
lead to eq.~(\ref{as}) are detailed in the appendix.

\begin{figure}[t]
   \vspace{0cm}
   \centerline{\epsffile{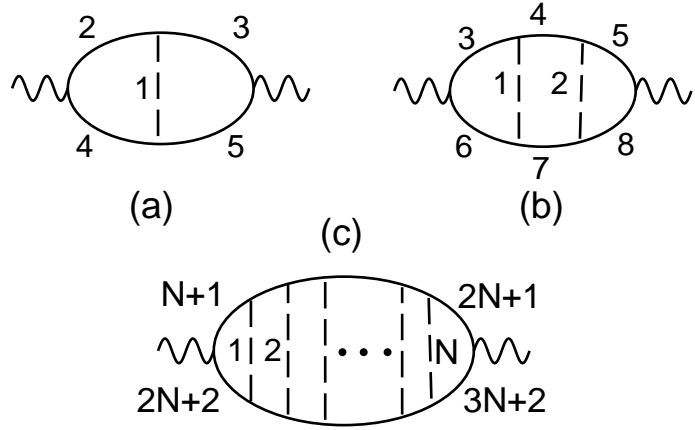}}
   \vspace*{0.7cm}
\caption{\label{vac2} Ladder contributions to the photon
vacuum polarization.}
\end{figure}

We now consider the behaviour of $G_{\Gm}(\uq;\uu)$ of
eq.~(\ref{analytic2}) in the vicinity of the point
$\uu_0=(u_{01},\ldots,u_{0N})$. The previous discussion is applicable
without modification, when all components of $\uu_0$ are integer. We
now allow them to be arbitrary complex numbers. Thus we set $r_l=-u_{0l}$ and
$\lambda_l=u_{0l}-u_l$. As specific examples, one may have in mind the
skeleton diagrams shown in Fig.~\ref{vac2}.
It is natural to define the UV degree of divergence
of a subgraph $\gm$ dependent on $\uu_0$ and given by

\be \label{uvdegree}
\om_{\uu_0}(\gm) = 4 h(\gm) - L_f(\gm) - 2L_{ph}(\gm) + 2 u_0(\gm)\,,
\ee

\n where $L_{f(ph)}(\gm)$ is the number of fermion (photon)
lines in $\gm$, $u_0(\gm)=\sum_{l\in\gm} {\rm Re\,}(u_{0l})$ and a
similar definition holds for $u(\gm)$.
Note that eq.~(\ref{as}) is derived from the factorization
of singularities in terms of sector variables, eq.~(\ref{Tfact}). The
same factorization shows that the analytically regularized
Feynman diagram $G_{\Gm}(\uq;\uu)$ is a meromorphic function of $\uu$
with poles described by the equations

\be
u(\gm) = u_0(\gm)-\lambda(\gm) =
-[\om_{\underline{0}}(\gm)/2] + k, \;\;  k=0,1,2,\ldots\,,
\label{poles}
\ee

\n where $[x]$ denotes the integer part of $x$. Since
$\lambda(\gm)$ is by definition small, we see that the subgraph $\gm$
contributes a singular factor $1/\lambda(\gm)$ to eq.~(\ref{as}) only
if $u_0(\gm)$ is integer. Thus $G_{\Gm}(\uq;\uu)$ can be represented
as

\be\label{singofg}
G_{\Gm}(\uq;\uu) = \sum_{\cal F} \prod_{
{\scriptsize \begin{array}{c} \gm\in {\cal F}:\om_{\uu_0}(\gm)\geq 0
\\[0.0cm] u_0(\gm)\,\,{\rm integer}
\end{array}}
}\frac{1}{u_0(\gm)-u(\gm)} \; g_{\cal F} (\uu)
\ee

\n where the functions $g_{\cal F} $ are analytical in a
vicinity of the point $\uu_0$. Note that because the fermion loop with
four photon lines attached is UV convergent in QED, the forests that
contain this subgraph as their smallest element should be excepted
from the sum above.

When the singularities of $G_{\Gm}(\uq;\uu)$ at $u(\gamma)=0$ are
removed by the usual process of renormalization, the condition that
$u(\gamma)$ is integer for at least one $\gamma\in {\cal F}$ in order
to obtain a singularity implies
that the Borel transform $B[G_{\Gm}](\uq;u)$ is analytic for
$0\leq u < 1$. Moreover, the singularity closest to the origin at $u=1$
arises only from the boundaries of the integration over the $u_j$ in
eq.~(\ref{analytic1}). The singularity at $u=1$ is called the leading
ultraviolet renormalon. The strongest singularity for a given set of
diagrams $\bar{\Gm}$ comes from those forests which contain the maximal
number of divergent subgraphs ($\omega_{\uu_0}(\gm)\ge 0$) and from those
points $\uu_0$ in the integration domain, where $u_0(\gamma)$ is
integer for all divergent subgraphs $\gm$ of the forest.

As long as we are interested only in the leading UV renormalon, we
are interested only in the domain
$0\leq u_l \leq 1$, $0\leq \sum u_l \leq 1$,
whence $u_0 (\gm) \leq 1$. Therefore
$\om_{\uu_0}(\gm) \geq 0$ implies $\om(\gm) \geq -2$ (here
$\om(\gm)$ denotes the usual degree of divergence, i.e.
$u_0(\gm)=0$ in eq.~(\ref{uvdegree})). In particular,
if a subgraph $\gm$ includes all the regularized photon lines,
we have exactly $\om_{\uu_0}(\gm) = \om(\gm) -2$.
Thus we have to
consider not only divergent (in the usual sense, i.e.
with $\om(\gm)\geq 0$)
subgraphs but also those convergent ones, that have
$\om(\gm)=-1$ or $\om(\gm)=-2$. An example of such a subgraph
is given by the `box',
consisting of lines $\{1,2,4,7\}$ in Fig.~\ref{vac2}b. Thus the
1PI Green functions with the following field content contain the leading
UV renormalon: $\bar{\psi}\psi$, $\bar{\psi}\psi\bar{\psi}\psi$,
$\bar{\psi} A^n\psi$ ($n=1,2,3$). All other Green functions are
superficially free from the leading UV renormalon and develop it
only through subgraphs that can be classified in terms of the Green
functions listed above.

Barring cancellations, eq.~(\ref{singofg}) allows us to derive in a
straightforward manner the $n$-dependence of the large-order behaviour of
a set of diagrams $\bar{\Gm}$. As an illustration, we consider the
contributions to the photon vacuum polarization depicted in Fig.~\ref{vac2}.
Referring to diagram (b), denote $\gamma_1=\{1,3,6\}$, $\gamma_2=\{1,2,
4,7\}$, $\gm_+$ the union of $\gm_1$ and $\gm_2$ and $\Gm$ the entire
graph. Thus $u(\gm_1)=u_1$ and $u(\gm_2)=u(\gm_+)=u(\Gm)=u_1+u_2=u$.
The last equality follows from the delta-function in eq.~(\ref{analytic1}).
Let ${\cal F}_1=\{\gm_1,\gm_+,\Gm\}$ and ${\cal F}_2=\{\gm_2,\gm_+,\Gm\}$.
When $u$ approaches unity, we find that in the vicinity of the point
$\uu_0=(1,0)$ the singular factors in eq.~(\ref{singofg}) for these
two forests are given by

\be
{\cal F}_1:\quad \frac{1}{(1-u)^2}\frac{1}{1-u_1}\qquad\quad
{\cal F}_2:\quad \frac{1}{(1-u)^3}\,.
\ee

\n The integration over $u_1$ and $u_2$ is trivial and leads to the
singularities of $B[G_{\Gm}](\uq;u)$

\bea
&&{\cal F}_1:\quad \frac{\ln(1-u)}{(1-u)^2}\quad\Rightarrow\quad
K_1\,\beta_0^n n!\, n\ln n\,\alpha^{n+1}\\
&&{\cal F}_2:\quad \frac{1}{(1-u)^3}\quad\Rightarrow\quad
K_2\,\beta_0^n n!\, n^2\,\alpha^{n+1}\,.
\eea

\n The constants $K_{1,2}$ will be determined in Sect.~4. To the right
of the arrows we have indicated the corresponding large-order contribution
to the perturbation expansion $(-ig)^2\sum r_n\alpha^{n+1}$ of the
vacuum polarization, which can be deduced from eq.~(\ref{seriesdef}).

We notice that the strongest divergence, $n!\, n^2$,
arises from the forest that contains
the box subgraph. The contribution from one chain, shown in Fig.~\ref{vac2}a,
produces only $1/(1-u)^2$ \cite{BEN93}. Therefore the diagrams with
two chains dominate by a factor of $n$ over the single chain ones
\cite{VAI94}. This enhancement has a simple interpretation in terms of
counting logarithms that are integrated over. In a given order
$g^2\alpha^{n+1}$ in perturbation theory the single chain in diagram (a)
gives $n$ logarithms from $n$ fermion loops. One additional logarithm is
generated from the UV behaviour of the reduced
diagram, when the subgraph $\{1,2,4\}$ is contracted.
The result is $n!\,n$. In case of (b), at the
same order in perturbation theory, we have only $n-1$ fermion loops, but
the reduced diagram after contraction of the box subgraph contains two
UV logarithms so that the total number of logarithms is the same as for
(a). However, there exist $n$ ways of distributing $n-1$ fermion loops
over two photon propagators and the stronger divergence, $n!\,n^2$
arises from this combinatorial enhancement. In the following we will
see that the singular factors in eq.~(\ref{singofg}) originate from
$\dd^4k/k^6$ terms in the expansion of Feynman integrands in external
(exceptional) momenta.
This will allow us to associate the box subgraph with a counterterm
proportional to a four-fermion operator in the sum of eq.~(\ref{dimsix})
and to give yet another interpretation of this enhancement.

What can be expected from diagrams with more than two chains? Consider
the ladder diagram with $N$ chains in Fig.~\ref{vac2}c. It is not
difficult to see that the strongest singularity at $u=1$ comes from a
forest built as follows: Start with the subgraph $\{1,2,N+2,2N+3\}$
and continue by including subsequent ladders to the right. The product
of $N+1$ singular factors for $\uu_0=(1,0,\ldots,0)$  is given by

\be
\frac{1}{1-u_1-u_2}\frac{1}{1-u_1-u_2-u_3}\,\ldots\,
\frac{1}{1-u_1-\ldots-u_{N-1}}\frac{1}{(1-u_1-\ldots-u_N)^3}\,,
\ee

\n resulting in the singularity $\ln^{N-2}(1-u)/(1-u)^3$. Thus each additional
chain yields an enhancement only by a factor of $\ln n$ for large $n$,
but no additional factors $n$ occur. If we anticipate the association
of the box subgraph with a four-fermion operator insertion, we can
interpret the dressing by ladders as a renormalization of this operator.
Thus we expect that the series of singularities generated by ladder
diagrams (and many others) can be summed by renormalization group
methods and exponentiates according to

\be
\sum_N \frac{A^{N-2}}{(N-2)!}\ln^{N-2}(1-u)/(1-u)^3 \longrightarrow
\frac{1}{(1-u)^{3+A}} .
\ee

\n This will be discussed further in Sect.~5.

Beyond the point $u=1$, the Borel transform $B[G_{\Gm}](\uq;u)$
defines a multi-valued function due to the cuts attached to the singular
points. The above analytic properties of $G_{\Gm}(\uq;\uu)$ imply that
after integration over $\uu$ the singular points in $u$ in the right
half-plane occur at integers with a cut attached to each such point. We
can therefore conclude that to any {\em finite} order in the expansion
in chains the only singular points in the right half of the Borel plane
are UV renormalons at integer $u$. Whether this reflects the correct
singularity structure of QED, depends on the behaviour of the $1/N_f$
expansion. For example, the number of skeleton
diagrams grows rapidly in higher
orders of $1/N_f$, so that the $1/N_f$-expansion of the normalization $K$
of UV renormalons could be combinatorically divergent. Factorial
divergence of the perturbative expansion due to the number of diagrams is
indeed expected for theories with bosonic self-interaction. For theories
with no bosonic self-interaction such as QED, the Pauli exclusion principle
enforces strong cancellations. For QED with finite UV cut-off, the
combinatorial divergence was found to be \cite{PAR77b,BOG78}

\be
r_n\sim \Gamma\!\left(\frac{n}{2}\right)\,\left(\frac{1}{a}\right)^n\,
n^b\,.
\ee

\n If this combinatorial behaviour persists when it interferes with
UV renormalization, the corresponding singularities occur at $|u|=\infty$
in the Borel plane. Thus it is reasonable to assume that in QED the
conclusions obtained from the chain expansion pertain to the full theory
in any finite domain in the Borel plane.

It is interesting to compare this with the non-abelian gauge theory. In this
case the existence of instantons and bosonic self-interaction leads to
singularities at finite values of $u$, connected with the value of the
action of an instanton-antiinstanton pair \cite{BOG77}.
Because the action is independent
of $N_f$, this singularity does not show up in any finite order in
the $1/N_f$ expansion. (The simplest way to see this is to rescale the
coupling as $a=\alpha N_f$ and to write $\exp(-S/\alpha)=\exp(-N_f S/a)$,
which has vanishing Taylor expansion in $1/N_f$.) Thus, as physically
expected (for other reasons as well), a strict $1/N_f$ expansion is
certainly in trouble for non-abelian theories.

%%%%%%%%%%%%%%%%%%%%%%%%% SECTION 3 %%%%%%%%%%%%%%%%%%%%%%%%%%%%%%%%%%%%

\section{Extraction of singularities}
\setcounter{equation}{0}

While the formulas given in the previous section allow us to derive
the $n$-dependence of the large-order behaviour for a given set of
diagrams, we are still lacking a method to compute the overall
normalization without having to compute the analytically regularized
Feynman integrals exactly. In this section we first derive a formula
that allows calculation of the {\em leading singularity}, defined
as the collection of terms with the maximal number of singular
factors in eq.~(\ref{singofg}), by consecutive extraction of
pole parts of one-loop integrals. The {\em next-to-leading} singularity
is defined by the collection of terms with one singular factor less
than the maximal number. We shall see that the computation of the
first correction to the normalization $K$ of the large-order
behaviour requires some specific contributions to the next-to-leading
singularity and corrections to the approximation of the complete
photon propagator by a chain. The techniques developed in this section
are applied to the fermion self-energy for illustration.

\subsection{The basic formula for the leading singularity}

As mentioned above the leading singularity (LS) of the given Feynman
integral is defined as
the sum of terms in eq.~(\ref{as}) with the maximal number of the factors
$\lm(\gm)$ in the denominator.
Eq.~(\ref{singofg})
shows that this number is equal to or less than
the maximal number of divergent
subgraphs that can belong to the same forest. Note that `divergent' for
given $\uu_0$ means $\om_{\uu_0}(\gm)\geq 0$. Suppose that these divergent
subgraphs form a nested sequence $\gamma_1\subset\ldots\subset
\gamma_{n_0}$ and
let $k_i$, $i=1,\ldots,n_0$, be the loop momentum of $\gamma_i/\gm_{i-1}$
($\gm_0$ is defined to be the empty set). The singular factors
in eq.~(\ref{singofg}) arise when the internal momenta of a given subgraph
are much larger than the external momenta of this subgraph. Therefore
we expect that the leading singularity arises from the strongly ordered
region

\be \label{ksector}
k_1 \gg k_2 \gg \ldots \gg k_{n_0}\,.
\ee

\n Because of this ordering every subgraph appears as insertion of a local
vertex with respect to the next loop integration. The leading singularity
can then be found from consecutive contraction of one-loop
subgraphs and insertion of the polynomial in external momenta associated
with the local vertex. This fact is succinctly expressed by
the following simple representation for the leading singularity:

\be
LS (F_\Gm) = \sum_{{\cal F}: |{\cal F}_{h,div} | = n_0}
\prod_{\gm\in {\cal F}_{h,div}}
\frac{1}{\lm(\gm)}
\left(
\mbox{res}_{\lm(\gm)} F_{\gm/\gm_-} (u(\gm))
\right) .
\label{lsing}
\ee

\n Recall that $\lambda(\gm)=u_0(\gm)-u(\gm)$ and that for the given point
$\uu_0$ the values $u_0(\gm)$ must be integer for all
$\gm\in {\cal F}_{h,div}$ to obtain
the maximal number of singular factors.
Here each maximal forest $\cal F$ is represented as
${\cal F}_h \cup {\cal F}_r \equiv {\cal F}_{h,div}
\cup {\cal F}_{h,conv} \cup {\cal F}_r $ where the subscripts `div' and
`conv' denote 1PI elements respectively with $\om_{\uu_0}(\gm) \geq 0$ and
$\om_{\uu_0}(\gm) < 0$ and ${\cal F}_r$ contains all non-1PI elements.
The number $n_0$ denotes
$\max_{\cal F} |{\cal F}_{h,div} | $,
i.e. the maximal possible number of divergent subgraphs that can belong
to the same maximal forest.
Furthermore $\gm_-$ is the set of maximal
elements  $\gm'\in {\cal F}$ with $\gm'\subset\gm$.
Each factor
$\mbox{res}_{\lm(\gm)} F_{\gm/\gm_-} (u(\gm))$ is a
polynomial with respect to external momenta and internal masses of
$\gm/\gm_-$.
It is implied that these factors are partially ordered
and before calculation of the residue
$\mbox{res}_{\lm(\gm)} F_{\gm/\gm_-} (u(\gm))$
all polynomials associated with the set $\gm_-$
are inserted into this `next' reduced diagram. Note also that for the
Feynman integral $F_{\gm/\gm_-} (u(\gm))$ the sum of regularization
parameters is $u(\gm)$, rather than $u(\gm/\gm_-)$. It does not
matter into which line of the reduced diagram $\gm/\gm_-$ the
regularization parameter $\lm\equiv u(\gm)$ is introduced,
since the corresponding pole part in this $\lm$ does not depend on
this choice. Eq.~(\ref{lsing})
formalizes the expectation that the leading singularities of the
Borel transform can be calculated by extracting pole parts of
{\em one-loop} subgraphs and reduces to the purely combinatorial
problem of writing down all maximal forests for a given diagram
$\Gm$.

A proof of eq.~(\ref{lsing}), based on
the $\al$-representation technique to resolve the singularities
of Feynman diagrams, can be found in the appendix.
Here we summarize only the main points of this approach.

The initial step is to represent the Feynman diagram
as an integral over $L$ positive parameters $\al_l$
corresponding to its lines.
Then one performs a decomposition of this $\al$-representation into
subdomains which are called {\em sectors}\footnote{
Eq.~(\ref{ksector}) provides an example of such a sector in the
momentum space language. While this language is useful for
heuristic arguments, the $\al$-parametric technique is
adequate for proofs because of the simplicity of singularities
in terms of sectors and sector variables.} and correspond directly to
one-particle-irreducible subgraphs of the given graph.
After introducing, in each sector, new variables associated with
the family of 1PI subgraphs of the given sector, the complicated
structure of the integrand is greatly simplified, and the analysis
of convergence and/or analytical structure with respect to the parameters
of analytic regularization reduces to power counting in one-dimensional
integrals over sector variables. At this point one observes that the singular
factors exactly correspond to divergent subgraphs of the given graph.

The leading singularity then appears from the sectors with maximal
number of divergent subgraphs. To calculate coefficients of the
products of these singular factors one uses the local nature
of UV divergences. Practically, this means that calculation of
residues with respect to the corresponding
linear combination of analytical parameters reduces to Taylor expansion
of the corresponding subdiagram in its external momenta and inserting
the resulting polynomial into the reduced diagram. Eventually one comes
to eq.~(\ref{lsing}).

An alternative proof could use\footnote{We are grateful
to K.G.~Chetyrkin pointing out the possibility of such a strategy.}
the method of glueing
\cite{glue} which rests on integration of a given diagram with an
additional (glueing) analytically regularized propagator. Within
this method, the information about the large momentum behaviour
is encoded in analytical properties of glued diagrams with
respect to parameters of analytical regularization. In our problem, the idea
would be to use an inverse translation of these properties to get the
analytical properties of a diagram in the $u$-parameters from the large
momentum expansion of specific subdiagrams.

\subsection{Self-energy: The leading singularity}

\begin{figure}[t]
   \vspace{0cm}
   \centerline{\epsffile{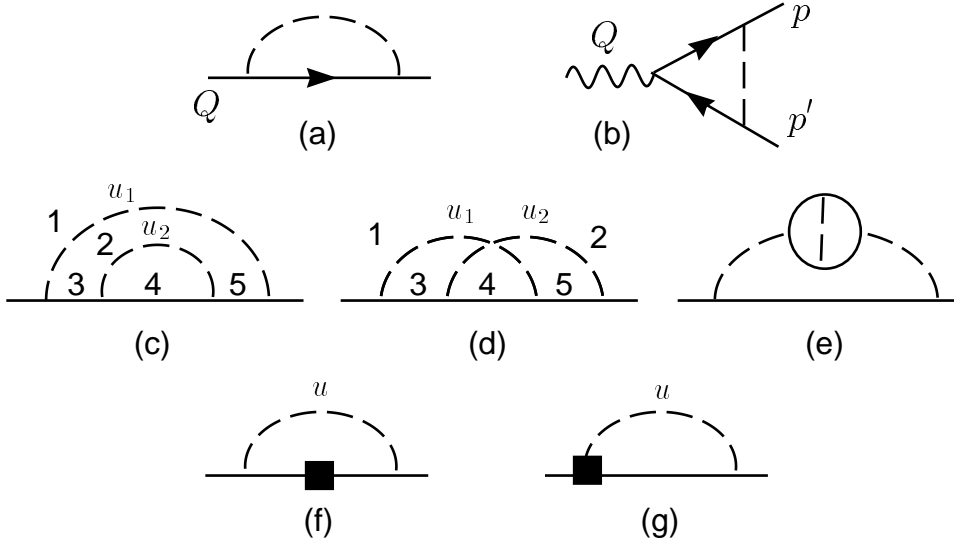}}
   \vspace*{0.4cm}
\caption{\label{self} Self-energy diagrams at next-to-leading order
in $1/N_f$ ((c)--(e)). (a) and (b) show the non-vanishing subgraphs
for diagrams (c), (d). (f) and (g) show the reduced diagrams with
insertion of a polynomial in external momenta of the contracted
subgraphs in (c) and (d).}
\end{figure}

In this subsection we exemplify the extraction of the leading
singular behaviour by self-energy diagrams with two chains. These
diagrams are shown in Fig.~\ref{self}c and d. Diagram e also
contributes at the same order in $1/N_f$. (For the counting in $N_f$
it is useful to think of $\alpha$ as being of order $1/N_f$.)
It corresponds to a correction to the photon propagator in
eq.~(\ref{borelcomplete})
and will be dealt with later. For completeness, we note that
the Borel transform of the leading order diagram
Fig.~\ref{self}a is given by

\bea\label{firstorderself}
B_{\bar{\Gamma}_a}[\Sigma](u) &=& (-i)^2 4\pi\int\frac{\dd^4 k}
{(2\pi)^4} \frac{\gamma_\mu\,i (\!\not\!Q-\!\not\!k)\gamma_\nu}
{(Q-k)^2}
\left(-\frac{\mu^2}{k^2} e^{-C}\right)^{-u}\frac{(-i)}{k^2}
\left(g_{\mu\nu}-\frac{k_\mu k_\nu}{k^2}\right)
\nonumber\\
&\stackrel{u\rightarrow 1}{=}&
\frac{i}{4\pi}\left(-\frac{Q^2}{\mu^2} e^C\right) \not\!Q
\,\frac{1}{2}\frac{1}{1-u}\,.
\eea

The self-energy requires renormalization beyond
renormalization of the coupling. In general this requirement is translated
into the necessity to subtract those singularities in the $u_j$ that
give rise to a singularity at $u=0$ of the Borel transform. (Such a
singularity prevents the calculation of coefficients $r_n$ of the
perturbative expansion as derivatives of the Borel transform at
$u=0$, see Sect.~2.1.) In the present case, the pole at $u=0$ is
absent, because the logarithmic UV divergence of the one-loop
self-energy cancels in Landau gauge. When fermion loops are inserted
in the photon line, logarithmic overall divergences are present. In terms
of the Borel transform their subtraction in a specified scheme
amounts to subtracting an arbitrary function of $u$ with the only
restriction that it does not have a pole at $u=0$. In the
$\overline{\rm MS}$ scheme this function is entire \cite{BEN94} and
does not introduce new singularities at any $u$. In general,
it is quite non-trivial to determine this function in a
given renormalization scheme (for a single chain see \cite{BEN94},
appendix~A). We do not discuss
this point further, since our main interest concerns physical
quantities. In this case all subtractions necessary for UV finiteness
are implicitly contained in the definition of the renormalized
complete photon propagator.

Let us now turn to the diagrams c and d in Fig.~\ref{self}. Each
maximal forest contains two non-trivial elements, the entire graph
$\Gamma$ and a one-loop subgraph $\gamma$. The contributions from
most forests vanish, because the reduced graph $\Gamma/\gamma$ is
a tadpole graph, for instance in case of $\gamma=\{1,2,4,5\}$ for
diagram c. For diagram c only one non-vanishing forest
remains, ${\cal F}=\{\{2,4\},\Gamma_c\}$, for diagram d we have
$\{\{1,3,4\},\Gamma_d\}$ and the symmetric forest $\{\{2,4,5\},
\Gamma_d\}$. The leading singularity has two singular factors:
$1/(1-u)$ from $\Gamma/\gamma$ and $1/(1-u_j)$ ($j=1,2$) or $1/u_j$
from $\gamma$. Thus the leading singularity at $u=1$ arises from
the points $\uu_0=(1,0)$ and $\uu_0=(0,1)$ in regularization parameter
space.

We consider first diagram c and the vicinity of $(0,1)$. If we denote the
momentum of lines 2 and 5 by $Q-k_1$, the pole part of $\gamma=\{2,4\}$
is given by the second line of
eq.~(\ref{firstorderself}) with $u$ replaced by $u_2$ and
$Q$ by $Q-k_1$. The insertion of the polynomial in $Q-k_1$ associated
with the contraction of $\gamma$ into the reduced
graph is shown in Fig.~\ref{self}f.
The corresponding Feynman integral (up to constants
that can be read off eq.~(\ref{firstorderself})) is given by

\be\label{selfreduced}
\frac{1}{1-u_2}\int\frac{\dd^4 k_1}
{(2\pi)^4} \frac{\gamma_\mu\,(\!\not\!Q-\!\not\!k_1)
(Q-k_1)^2(\!\not\!Q-\!\not\!k_1)(\!\not\!Q-\!\not\!k_1)\gamma_\nu}
{(Q-k_1)^4}
\frac{1}{(k_1^2)^{1-u}}
\left(g_{\mu\nu}-\frac{k_\mu k_\nu}{k^2}\right)\,.
\ee

\n The denominator in $Q-k_1$ is cancelled and the integral vanishes.
In the vicinity of the other point, $u_1$=1, $u_2=0$, the potential
pole $1/u_2$ is absent due to finiteness of the one-loop self-energy
in Landau gauge as noted above. Thus diagram c does not contribute at
all to the leading singularity.
 Note that according to eq.~(\ref{lsing}) and the appendix, the
regularization parameter associated with the photon line of the reduced
diagram is $u=u_1+u_2$, rather than $u_1$. In other words, the reduced
diagram inherits the $u$-parameters from the contracted subgraph,
here $u_2$. The
reason for this is that the singularities in $\uu$ from an element
$\gamma_+\in {\cal F}$ (in the present example $\gamma_+=\Gamma$) are
determined by $u(\gamma_+)$, rather than $u(\gamma_+/\gamma)$, see Sect.~2.2.
A less formal argument comes from doing the subgraph $\gamma$
exactly. Simply by dimensional reasons the Borel transform is
proportional to $(\mu^2/(Q-k_1)^2)^{-u_2}$, so that the parameter $u_2$
is attached to the fermion line of the reduced graph. The
singularities of the reduced graph arise in turn from the integration
region $k_1\gg Q$, so that we may neglect $Q$ in this expression
and attach the parameter $u_2$ to the photon line, where it combines
with $u_1$ to $u$. This statement is true in general (see the appendix):
In calculating the {\em leading} singularity, the $u$-parameters
inherited from the contracted subgraph can be attached to an arbitrary
line of the reduced graph.

Turning to diagram d, consider the forest $\{\{1,3,4\},\Gamma_d\}$.
Again we have to analyze the points $\uu_0=(1,0)$ and $\uu_0=(0,1)$.
The second one does not contribute two singular factors, since the
one-loop vertex subgraph $\{1,3,4\}$ shown in Fig.~\ref{self}b is
UV finite in Landau gauge, so that no singularity $1/u_1$ occurs. In
the vicinity of the first point we require the singularity of the
vertex subgraph at $u_1=1$. With the momentum assignments of
Fig.~\ref{self}b we find

\bea\label{firstordervertex}
&&\int\frac{\dd^4 k}{(2\pi)^4} \frac{\gamma_\rho (\!\not\!p-\!\not\!k)
\gamma_\nu (\!\not\!p^\prime-\!\not\!k)\gamma_\sigma}
{(p-k)^2 (p^\prime-k)^2 (k^2)^{1-u_1}}
\left(g_{\mu\nu}-\frac{k_\mu k_\nu}{k^2}\right)
\nonumber\\
&&\stackrel{u_1\rightarrow 1}{=}\,
\frac{i}{16\pi^2}\frac{1}{1-u_1}\left[-\frac{2}{3} (Q^2\gamma_\mu-
\!\not\! Q Q_\mu) + \frac{1}{2}\left(\left(p^2+{p^\prime}^2\right)
\gamma_\mu + \!\not\! p\gamma_\mu \!\not\! p^\prime\right)\right]\,,
\eea

\n where $Q=p^\prime-p$ is the momentum of the external photon. Note
that the terms not containing $Q$ in the second line vanish on-shell
and in the Lorentz structure of the terms containing $Q$ one recognizes
the Feynman rule for insertion of the operator $\bar{\psi}\gamma_\nu
\psi \partial_\mu F^{\mu\nu}$.

In order to calculate the pole parts of one-loop integrals like the
vertex above, one should use the fact that the poles of interest are
of ultraviolet origin, that is from the region of integration, where
$k$ is much larger than all external momenta. Therefore we can expand
the denominator in $p/k$ and $p^\prime/k$. The IR divergences that
arise in this way can be regularized by a cut-off $|k|>\mu$ or
in any other convenient way. Lorentz
invariance then allows us to drop all terms with an odd number of
$k$'s in the numerator and to simplify the numerator by relations
such as $k_\rho k_\sigma \rightarrow g_{\rho\sigma}/4$ and its
generalizations to more factors of $k$. If the original integral was
logarithmically UV divergent, the result is expressed as the sum

\be
\sum_{n=0} {\cal P}_n(\uq) \,(-16 i\pi^2)
\intl_{|k|>\mu}\frac{\dd^4 k}{(2\pi)^4}\frac{1}{(k^2)^{2+n-u(\gamma)}}\,,
\ee

\n where ${\cal P}_n(\uq)$ is a polynomial of degree $n$ in the external
momenta. The $k$-integral gives a pure pole, so that ${\cal P}_n(\uq)$
can be identified with the residue in eq.~(\ref{lsing}) that is to be
inserted into the reduced graph. For the fermion-photon vertex
${\cal P}_0(\uq)$ vanishes and for ${\cal P}_1(\uq)$ we obtain
eq.~(\ref{firstordervertex}). Notice that the pole at $u=n$ is related
to the term $\dd^4 k/(k^2)^{2+n}$ in the expansion of the integrand.
The relation to Taylor operators is clarified further in the appendix.

The remainder of the calculation is straightforward.
Eq.~(\ref{firstordervertex}) is inserted into the reduced graph shown
in Fig.~\ref{self}g and the singular part of the reduced graph is extracted
in the same way as before. Adding the contribution from the symmetric
forest, we arrive at

\be G_{\Gamma_d}(Q;u_1,u_2) = \frac{i}{(4\pi)^2}\,Q^2\not\! Q
\,\left(-\frac{1}{3}\right)\frac{1}{1-u_1-u_2} \left[
\frac{1}{1-u_1}+\frac{1}{1-u_2}\right]\,.
\ee

\n After integration over $u_1$ and $u_2$ according to
eq.~(\ref{analytic1}), the final result for the leading
singularity at $u=1$ from the two-chain diagrams c and d is

\be \label{finalleadingself}
B_{\bar{\Gamma}_c}[\Sigma](u)+B_{\bar{\Gamma}_d}[\Sigma](u)
\stackrel{u\rightarrow 1}{=}
\frac{i}{4\pi}\left(-\frac{Q^2}{\mu^2} e^C\right) \not\!Q
\,\frac{1}{2}\frac{1}{1-u}\,\frac{1}{N_f}\left[
\ln(1-u) + {\cal O}(1)\right]\,,
\ee

\n which has to be compared with the leading order contribution,
eq.~(\ref{firstorderself}). (The factor $1/N_f$ originates from
$1/\beta_0$ in eq.~(\ref{analytic1}).) In this case we find
$\beta_0^n n!\,\ln n$ for large $n$ with a logarithmic enhancement
as compared to the leading order. The ${\cal O}(1)$ contribution
in brackets will be found in Sect.~3.4. This concludes the
illustration of the general method of extracting
the leading singularity. The algebraically more extensive case of
the vertex and vacuum polarization is treated in Sect.~4.

\subsection{Next-to-leading singularity and IR rearrangement}

At this point it is helpful to refer again to the large-order behaviour
of perturbative expansions, eq.~(\ref{renormalon}). Within the
expansion in the number of chains, we write

\be
K=K^{[1]}\,(1+\delta K^{[2]}+\ldots),\qquad b=b^{[1]}+b^{[2]}+\ldots\,,
\ee

\n where $\delta K^{[N]}$, $b^{[N]}$ denote contributions from $N$
chains.\footnote{To be precise, different $K$ and $b$ should be
introduced for each operator that appears in eq.~(\ref{dimsix}).} Since
we expand in chains before taking the large-$n$ limit,
eq.~(\ref{renormalon}) must be written as

\be\label{chainrenormalon}
r_n = K^{[1]}\,\beta_0^n\,n!\,n^{b^{[1]}}\left(1+\left\{b^{[2]}\ln n
+\delta K^{[2]}\right\} +\ldots\right) \left[1+\ON\right]\,,
\ee

\n where the ellipses represent terms like ${b^{[2]}}^2\ln^2 n$ etc. from
diagrams with three and more chains. Close to $u=1$, the corresponding
Borel transform eq.~(\ref{seriesdef}) is

\bea\label{chainborel}
B[G](\uq;u) &=& \frac{K^{[1]}\Gamma(1+b^{[1]})}{(1-u)^{1+b^{[1]}}}\left(
1+\left\{-b^{[2]}\left(\ln(1-u)-\psi(1+b^{[1]})\right)
+\delta K^{[2]}\right\}\ldots\right)
\nonumber\\
&&\times\left[
1+{\cal O}\left((1-u)\ln^k(1-u)\right)\right]\,,
\eea

\n where $\psi(x)$ is Euler's $\psi$-function. From this expression
we deduce that $b^{[2]}$ can be read off
from the leading singularities of diagrams with two chains, such as
those considered in the previous subsection. On the other hand to obtain
the first correction $\delta K^{[2]}$ to the normalization, we require
also some terms with only one singular factor. This can be either
$1/(1-u_1-u_2)$ or $1/(1-u_j)$ ($u_j=1,2$). After integration over $u_1$,
$u_2$, the second type produces a singularity $\ln(1-u)$ and contributes only
to $1/n$-corrections in eq.~(\ref{chainrenormalon}). Thus, we consider
only singular factors of type $1/(1-u)$.

This distinction is important both as far as the physical origin of
the singularity is concerned as well as its extraction. Recall that,
for example for the contribution from Fig.~\ref{self}d to the self-energy,
the leading singularity (two singular factors) comes from the loop
momentum regions

\be
Q\ll k_1\ll k_2\qquad Q\ll k_2\ll k_1\,.
\ee

\n To obtain at least one singular factor, we must consider the
regions

\bea
&& Q\ll k_1\sim k_2\,,
\nonumber\\
&& Q\sim k_1\ll k_2\qquad Q\sim k_2\ll k_1\,.
\eea

\n Only in the first case we obtain $1/(1-u)$, because both virtual
photons have momentum larger than the external momentum $Q$, so that the
large-momentum part of the diagram contains all available
$u$-parameters. Moreover,
because of this, the residue of the pole is polynomial in the external
momentum. On the other hand, the regions listed in the second line
lead to singularities $1/(1-u_j)$. The residue is no longer `local'
(polynomial in external momenta), but contains $\ln(Q^2/\mu^2)$. A simple
way to see this is to note that by dimensional reasons the Borel
transform for diagram d is proportional to $\!\not\!Q\,(Q^2/\mu^2)^u$.
This implies that the leading singularity at $u=1$
in fact arises in the combination

\be
\!\not\!Q\,Q^2\,\frac{1}{1-u_j}\left[\frac{1}{1-u}-\ln\frac{Q^2}{\mu^2}
\right]\,,
\ee

\n so that the coefficient of the logarithm is related to the
residue for the leading singularity. One may compare this with the
$1/\eps^2$ and $(1/\eps)\ln(Q^2/\mu^2)$ poles for a dimensionally
regularized two-loop integral before the subtraction of subdivergences
is taken into account.

We also see that if explicit renormalization (in the usual sense)
is needed as in case of the self-energy, the arbitrariness in
choosing finite subtractions affects only the residues of
$1/(1-u_j)$. Indeed, the effect of such a subtraction is described by
(taking $j=2$, so that the logarithmic UV divergence occurs in
the subgraph with photon line regularized by $u_1$)

\be
\frac{1}{1-u}\frac{a}{u_1}-\frac{1}{1-u_2}\left[\frac{a}{u_1}+G(u_1)
\right] = \frac{a}{(1-u) (1-u_2)}-\frac{G(u_1)}{1-u_2}\,,
\ee

\n where $G(u_1)$ has no pole at $u_1=0$, but is arbitrary otherwise.
Recall that $a$ is in fact zero for the self-energy (in Landau gauge).
We therefore conclude that the terms that contribute to
$\delta K^{[2]}$ are local and do not depend on the renormalization
scheme.

The fact that the residues of the singular factors that contribute
to the overall normalization of renormalon divergence are polynomial
in the external momenta allows us to use the method of {\em IR rearrangement}
\cite{Vlad,CS} (see also \cite{VS} for a review) which was
originally developed for and successfully
applied to renormalization group calculations.
In its original form, the method is based on polynomial dependence
of UV coun\-ter\-terms on momenta and masses. It consists of the following
two steps:
(a) Differentiation in external momenta and internal masses. The problem
thereby reduces
to the calculation of zero-degree polynomials. (b)~Putting to
zero all the internal masses and external momenta except one.
Usually, at step b one puts to zero all the momenta and masses and
chooses a new momentum that flows through the diagram in some
appropriate way. The goal is to make this choice in such a way that
the diagram becomes calculable.
In simple cases the problem reduces to calculation of the pole part of
primitively divergent propagator-type
integrals or vacuum integrals with one non-zero mass. The application
in the present context is illustrated in Sect.~3.4.

The arguments presented above can easily be generalized to the contribution
$\delta K^{[N]}$ to the normalization $K$ from $N$ chains. For such a
contribution we need at least one\footnote{If $b^{[1]}$ is different from
zero (as for the vacuum polarization, see Sect.~4), $b^{[1]}+1$ such factors
are needed.} singular factor of type $1/(1-u)$, (with
$u=u_1+\ldots+u_N$). Although no ordering of loop momenta is required,
all loop momenta $k_i$ have to be much larger than the external momenta $\uq$
in order to get such a singular factor.
Therefore the residue of these terms remains local for any $N$. Similarly,
if UV renormalization is necessary, the arbitrariness in choosing finite
subtractions affects the singularity of the Borel transform only at the
level of $\ln^k(1-u)$ (where typically $k<N$) and therefore modifies only
the $1/n$-corrections to the asymptotic behaviour in
eq.~(\ref{chainrenormalon}). A scheme-dependence enters the overall
normalization only through the subtraction constant $C$ for the
fermion loop \cite{BZ92}, see eq.~(\ref{fermionloop}). With the help
of IR rearrangement, the calculation of $K$ to an arbitrary order in
the expansion in chains reduces to extracting pole parts of
primitively divergent propagator-type
integrals or vacuum integrals with one non-zero mass.

Note that IR rearrangement can be applied also to the calculation of
$1/n$ corrections to the asymptotic behaviour. Because of the logarithms
in external momenta that enter in this place,
one must apply explicit subtractions in subgraphs,
which remove these logarithms. Since the combinatorial structure of
eq.~(\ref{singofg}) is completely analogous to the one that arises in
the context of usual renormalization, the combinatorial structure of
the subtraction operator is equally simple. We will return to this
point in Sect.~5.

\subsection{Self-energy: The next-to-leading singularity}

In this subsection we illustrate the calculation of $\delta K^{[2]}$
and the use of IR rearrangement for this purpose by continuing with
the fermion self-energy. When $\Gamma=\Gamma_c,\Gamma_d$ as shown in
Fig.~\ref{self}c and d, we have
for the integral in eq.~(\ref{analytic1}) the
following asymptotic behaviour at $u \to 1$:

\be
\int_0^u \dd u' \,G_{\Gm}(Q;u',u-u')
\stackrel{u\rightarrow 1}{=} \frac{1}{1-u} \left[ a_1 (Q) \ln (1-u) +
a_2 (Q) + {\cal O}\left((1-u)\ln(1-u)\right)\right]\,,
\label{AB}
\ee

\n where $a_i$ are proportional to $\Qslash Q^2$ with no logarithm
in $Q^2/\mu^2$ as explained in Sect.~3.3. ($a_1(Q)$ is given
in eq.~(\ref{finalleadingself}).) According to eq.~(\ref{singofg}),
the function  $G_{\Gm}(u_1,u_2)$ can be represented as

\be
G_{\Gm}(u_1,u_2) = \frac{1}{1-u_1-u_2} \left[
\frac{g_1 (u_1,u_2)}{1-u_1} + \frac{g_2 (u_1,u_2)}{1-u_2}
\right]
\equiv \frac{h(u_1,u_2)}{(1-u_1-u_2)(1-u_1)(1-u_2)}\,.
\ee

\n Now we apply the following proposition: If $f(u')$ is analytic
at $0 \leq u' \leq 1$ then

\be
\int_0^u \dd u'
\frac{f(u')}{(1-u')(1-u+u')} \stackrel{u\rightarrow 1}{=}
- [f(0) + f(1)] \ln (1-u)
\ee

\n when $u \to 1$. As a result we have

\bea
&&\int_0^u \dd u_1 \dd u_2 \,\delta(u_1+u_2-u)
\frac{h(u_1,u_2)}{(1-u_1-u_2)(1-u_1)(1-u_2)}
\nonumber\\
&&\hspace*{1cm} =\,- [h(1,0) + h(0,1)]\,\frac{\ln (1-u)}{1-u} + \frac{1}{1-u}
\int_0^1 \dd u' \overline{h} (u',1-u'),
\eea

\n when $u \to 1$. Here

\bea
\overline{h} (u',1-u') &=&
\frac{h(u',1-u')}{u' (1-u')} -
\frac{h(1,0)}{1-u'} - \frac{h(0,1)}{u'}
\nonumber\\
&=& \frac{g_1(u',1-u')-g_1(1,0)}{1-u'}+\frac{g_2(u',1-u')-g_2(0,1)}{u'}\,.
\eea

\n Define $b_1=g_1(1,0)$ and $b_2=g_2(0,1)$, so that the leading
singularity is determined by

\bea
G_{\Gm}(Q; u_1 ,u_2) =
\frac{1}{1-u_1-u_2} \frac{b_1}{1-u_1}\quad {\rm near}\quad
u_1=1,u_2=0\,,
\nonumber\\
G_{\Gm}(Q; u_1 ,u_2) =
\frac{1}{1-u_1-u_2} \frac{b_2}{1-u_2}\quad {\rm near}\quad
u_1=0,u_2=1\,.
\eea

\n Then the coefficients that enter eq.~(\ref{AB}) are expressed as

\bea\label{aandb}
a_1 &=& -b_1-b_2 ,
\nonumber\\
a_2 &=& \int_0^{1} \dd u' \left\{
\mbox{res}_{u-1} G_{\Gm}(Q;u',u-u')
- \frac{b_1}{1-u'} - \frac{b_2}{u'} \right\} .
\eea

Generally speaking,
one can directly use eq.~(\ref{aandb}) for calculation.
This is indeed possible
for the rainbow diagram of Fig.~\ref{self}c, which can be computed exactly
for arbitrary $u_1$ and $u_2$. After a straightforward calculation, we
get

\bea
a_1 &=& 0 \quad\mbox{(see Sect.~3.2)}
\nonumber\\
\overline{h} (u',1-u') &=& \frac{i}{(4\pi)^2}\,Q^2\!\not\! Q\,
\left(-\frac{3}{2}\right)\frac{1-u'}{2+u'}\,.
\eea

\n Integration according to eq.~(\ref{aandb}) and reinstating the factors
from eq.~(\ref{analytic1}) yields

\be\label{nlsdiagc}
B_{\bar{\Gamma}_c}[\Sigma](u)
\stackrel{u\rightarrow 1}{=}
\frac{i}{4\pi}\left(-\frac{Q^2}{\mu^2} e^C\right) \not\!Q
\,\frac{1}{2}\frac{1}{1-u}\,\frac{1}{N_f}\left[\frac{27}{4}\ln\frac{2}{3}
+\frac{9}{4}\right]\,.
\ee

In most cases, however, the analytically regularized Feynman integral is
hardly calculable for arbitrary values of $u_1$ and $u_2$. This is
true in particular already for diagram d in Fig.~\ref{self}. In this
situation one first calculates coefficients $b_1$ and $b_2$ (and if
necessary also the coefficients of terms $1/((1-u)u_j)$)
with the help of general statements regarding the leading singularity,
and then the function that enters the
integrand of eq.~(\ref{aandb}) using IR rearrangement.
To apply IR rearrangement to diagram d we
use the fact that the coefficients $a_i$ are polynomials in $Q$, in this
case $\Qslash Q^2$. Thus we differentiate three times. Since
$\pa_Q^2 \!\paslash_Q \Qslash Q^2=48$, the substitution

\be
G_{\Gm}(Q;u',u-u') \longrightarrow \,\Qslash Q^2\,\frac{1}{48}
\pa_Q^2 \!\paslash_Q G_{\Gm}(Q;u',u-u')
\ee

\n leaves the coefficient $a_2(Q)$ unchanged. To reduce the number of terms
that arise in the process of differentiation, we route the external
momentum $Q$ through the lines 1 and 5 in Fig.~\ref{self}d, so that
the corresponding momenta are $Q-k_1$ and $Q-k_2$, respectively. The
threefold differentiation of the product of photon propagator and fermion
propagator gives rise to six new diagrams $\Gamma_i$ with the same
topology as the original diagram
but with propagators differentiated in a special way. The residue

\be
\mbox{res}_{u-1} G_{\Gm_i}(Q;u',u-u')
\ee

\n for each diagram is now a number independent of the external momentum
$Q$. The second step in the IR rearrangement is to apply the equation

\be
\mbox{res}_{u-1} G_{\Gm_i} (Q;u',u-u') =
\mbox{res}_{u-1} G'_{\Gm_i} (p;u',u-u')\,.
\label{IRR}
\ee

\n The meaning of this equation is as follows: Instead of the initial
external momentum $Q$, one introduces a new external momentum $p$
which flows through the diagram in some appropriate way. In our
example diagram d (and its descendents $\Gamma_i$) we choose $p$
to flow through line 5 (so that the vertex to which lines 1, 4, 5 are
attached becomes external). Due to this choice of external
momentum the `new' diagram
$G'_{\Gm_i} (p;u',u-u')$ becomes explicitly calculable by
consecutive application of one-loop integrations.

The extensive algebra connected with numerators of integrals after
differentiation can be done with the help of
{\sf REDUCE} \cite{reduce}. A generic
integral has the form

\be
\int\frac{\dd^4 k_1}{(2\pi)^4}\frac{\dd^4 k_2}{(2\pi)^2} \frac{N(k_1,k_2,p)}
{(k_1^2)^a ((k_1-k_2)^2)^b (k_2^2)^c ((p-k_2)^2)^d}\,.
\ee

\n We can take the $k_1$-integral. Since the remaining $k_2$ satisfies
$k_2\gg p$ to
obtain a singularity $1/(1-u)$, we can then extract the pole factor from the
$k_2$-integral by expansion in the external momentum as in Sect.~3.2.
Summing over all six contributions from differentiation, we obtain

\bea
a_1 &=& \frac{i}{(4\pi)^2}\,Q^2\!\not\!Q\left(-\frac{2}{3}\right)
 \quad\mbox{(see Sect.~3.2)}
\nonumber\\
\overline{h} (u',1-u') &=& \frac{i}{(4\pi)^2}\,Q^2\!\not\! Q\,
\frac{1}{3}
\left[\frac{3}{2}-\frac{1}{2+u'}+\frac{1}{1+u'}+\frac{1}{2-u'}-
\frac{1}{3-u'}\right]\,.
\eea

\n Integration according to eq.~(\ref{aandb}) and reinstating the factors
from eq.~(\ref{analytic1}) yields

\be\label{nlsdiagd}
B_{\bar{\Gamma}_d}[\Sigma](u)
\stackrel{u\rightarrow 1}{=}
\frac{i}{4\pi}\left(-\frac{Q^2}{\mu^2} e^C\right) \not\!Q
\,\frac{1}{2}\frac{1}{1-u}\,\frac{1}{N_f}\left[\ln(1-u)+
2\ln2-\ln 3+\frac{3}{4}\right]\,.
\ee

\n When combined with eq.~(\ref{nlsdiagc}), this results in

\be\label{deltakcd}
\delta K^{[2]}_{c+d} = \frac{1}{4}\left(35\ln 2 - 31\ln 3+12\right)\,.
\ee

\subsection{Self-energy: Propagator corrections and summary}

To complete the calculation of all contributions at subleading order
in $1/N_f$ (for the $N_f$-counting one rescales $a=\alpha N_f$) we
have to analyze diagram e in Fig.~\ref{self} (plus the self-energy type
contributions to the vacuum polarization insertion, not shown in the
figure). This diagram is treated differently from c and d, because
it appears as the first correction to approximating the complete
photon propagator in eq.~(\ref{borelcomplete}) by a chain, the first
term on the right hand side of eq.~(\ref{borelcomplete}). Corrections
of this type have been considered previously in \cite{BRO93,BEN95a}.
To include the second term in eq.~(\ref{borelcomplete}), we write,
as in \cite{BEN95a},

\be
F_{\rm reg}(u) = F(u)+\frac{1}{u}\,, \qquad
G_{\rm reg}(u) = G(u)+\frac{1}{u}\,,
\ee

\n where $F(u)$ is given in eq.~(\ref{f}) and $F_{\rm reg}(u)$ and
$G_{\rm reg}(u)$ are finite at $u=0$. When the second term on the
right hand side of eq.~(\ref{borelcomplete}) is inserted for the
single complete photon propagator in eq.~(\ref{someintermediateformula}),
we obtain

\bea\label{threeterms}
B_{\bar{\Gamma}_e}[\Sigma](u) &=& \frac{\beta_1}{\beta_0^2}
\intl_0^u\frac{d v\,v}{u-v}\left(B_{\bar{\Gamma}_a}[\Sigma](u)-
B_{\bar{\Gamma}_a}[\Sigma](v)\right)
\nonumber\\
&&-\,\frac{\beta_1}{\beta_0^2}\intl_0^u
d v\,v\left(B_{\bar{\Gamma}_a}[\Sigma](u)\,
F_{\rm reg}(u-v)-B_{\bar{\Gamma}_a}[\Sigma](v)\,
G_{\rm reg}(u-v)\right)\,,
\eea

\n where $B_{\bar{\Gamma}_a}[\Sigma](u)$ is the leading order contribution
given in eq.~(\ref{firstorderself}). The vacuum polarization insertion
requires renormalization and the function $G(u)$ takes into account the
arbitrariness of finite subtractions, that is, the arbitrariness in
defining the renormalized coupling $\alpha=\alpha(\mu)$. We evaluate
eq.~(\ref{threeterms}) in two schemes: The $\overline{\rm MS}$ scheme,
in which case $G_{\rm reg}(u)$ is an entire function \cite{BBB1,BEN95a},
and the so-called MOM scheme, where the renormalized coupling is defined
by

\be
\alpha_{\rm MOM}(\mu) = \frac{\alpha_R(\mu')}{1+\Pi_R(\mu/\mu',\mu')}\,,
\ee

\n where $R$ denotes any renormalization scheme or bare quantities. In
this scheme we have $F(u)=G(u)$ \cite{BBB1}
and eq.~(\ref{threeterms}) takes the
particularly simple form

\be\label{threetermsMOM}
B_{\bar{\Gamma}_e}^{\rm MOM}[\Sigma](u) = \frac{\beta_1}{\beta_0^2}
\intl_0^u d v\,v\left[\frac{1}{u-v}-F_{\rm reg}(u-v)\right]
\left(B_{\bar{\Gamma}_a}[\Sigma](u)-
B_{\bar{\Gamma}_a}[\Sigma](v)\right)\,.
\ee

\n Contrary to the $\overline{\rm MS}$ scheme, the finite subtractions
are themselves factorially divergent in this scheme.

The right hand side of the first line of eq.~(\ref{threeterms}) is
easily evaluated as $u\to 1$, given that $B_{\bar{\Gamma}_a}[\Sigma](v)
\sim 1/(1-v)$ close to $v=1$. We get

\be\label{firstterm}
\frac{\beta_1}{\beta_0^2}
\intl_0^u\frac{d v\,v}{u-v}\left(B_{\bar{\Gamma}_a}[\Sigma](u)-
B_{\bar{\Gamma}_a}[\Sigma](v)\right)
\stackrel{u\to 1}{=}-\frac{\beta_1}{\beta_0^2}\,
B_{\bar{\Gamma}_a}[\Sigma](u)\left(\ln(1-u)+1
\right)
\ee

\n up to ${\cal O} ((1-u)\ln (1-u))$ terms.
To evaluate the second line in eq.~(\ref{threeterms}), we note first
that we can drop the contribution from $G_{\rm reg}$ as long as we
are not interested in singularities weaker than $1/(1-u)$. In the
$\overline{\rm MS}$ scheme, this follows from $G_{\rm reg}(u)$ being
entire, so that one gets only a $\ln(1-u)$-singularity from $v$ close
to one. In the MOM scheme, we also observe singular behaviour when
$v$ is close to zero, but since $B_{\bar{\Gamma}_a}[\Sigma](v)$ is finite
as $v\to 0$, one gets again only $\ln(1-u)$. Therefore, to the present
accuracy, the results in the $\overline{\rm MS}$ and MOM scheme are
the same. Next we separate the double pole of $F_{\rm reg}(u)$
at $u=1$, since it gives rise to a logarithmic
singularity of the $v$-integral at $u=1$. In the remaining terms we can set
$u=1$. Adding the result from the previous equation we obtain

\be\label{selfprop}
B_{\bar{\Gamma}_e}[\Sigma](u) \stackrel{u\to 1}{=}
B_{\bar{\Gamma}_a}[\Sigma](u)\left[\left(-\frac{\beta_1}{\beta_0^2}-
\frac{1}{N_f}
\right)\ln(1-u) + \frac{1}{N_f}\,\delta K_{\rm prop}\right]\,,
\ee

\n where

\bea
\delta K_{\rm prop} &=& -\frac{2}{3}+\frac{233}{12}\ln 2-\frac{157}{12}\ln 3
\nonumber\\
&&\,-\,24\sum_{k=3}^\infty\frac{(-1)^k k}{(k^2-1)^2}\left\{\frac{1}{2 k^2}
+\frac{1}{2}\ln\frac{k^2-1}{k^2-4}+\frac{3 k^2-1}{k^3}\left[
\mbox{arctan}\frac{1}{k}-\mbox{arctan}\frac{2}{k}\right]\right\}
\nonumber\\[0.1cm]
&=& -1.7266\ldots\,.
\eea

\n Recall that $\beta_1/\beta_0^2=9/(4 N_f)={\cal O}(1/N_f)$.
We can now combine eqs.~(\ref{nlsdiagc}), (\ref{nlsdiagd}) and
(\ref{selfprop}) to obtain the final result for the self-energy to
subleading order in $1/N_f$:

\bea\label{nlsself}
B[\Sigma](u)
&\stackrel{u\rightarrow 1}{=}&
\frac{i}{4\pi}\left(-\frac{Q^2}{\mu^2} e^C\right) \not\!Q
\,\frac{1}{2}\frac{1}{1-u}\,
\Bigg[1-\frac{\beta_1}{\beta_0^2}\ln(1-u)+\frac{1}{N_f}\left(
\delta K^{[2]}_{c+d}
+\delta K_{\rm prop}\right)
\nonumber\\
&&\vspace*{1cm} +\,{\cal O}\!\left((1-u)\ln(1-u),\frac{1}{N_f^2}
\right)\Bigg]\,,
\eea

\n with $\delta K^{[2]}_{c+d}$ given in eq.~(\ref{deltakcd}). By comparison
of eq.~(\ref{chainborel}) with eq.~(\ref{chainrenormalon}), this result
can be translated into the large-order behaviour of the self-energy. Notice
that there is a cancellation of $\ln(1-u)/(1-u)$ terms between diagrams
d and e of Fig.~\ref{self} and the only remaining logarithmic
enhancement is proportional to $\beta_1/\beta_0^2$. This remaining term
can be traced to eq.~(\ref{firstterm}) and then further to the coefficient
of the pole of $F(u)$ at $u=0$. It is therefore unambiguously identified
as the expected correction to $b$ in eq.~(\ref{renormalon}) due to
two-loop evolution of the coupling (see Sect.~5).

\begin{figure}[t]
   \vspace{0cm}
   \centerline{\epsffile{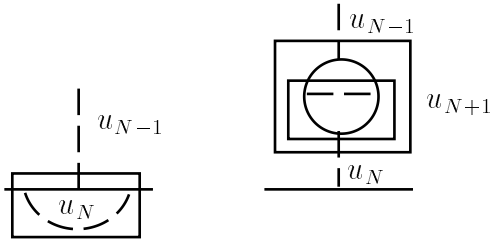}}
   \vspace*{0.0cm}
\caption{\label{cancel1} Cancellation of leading singularities between
vertex and propagator corrections. Since these graphs should be inserted
as subgraphs, the external photon lines are drawn as chains.}
\end{figure}

The cancellation between vertex and propagator corrections
is of general nature and recurs in further examples in
Sect.~4. It can be phrased as a cancellation between the two subgraphs
shown in Fig.~\ref{cancel1}. Note that these subgraphs are of the same
order in $1/N_f$ and for any diagram that contains the left subgraph
there is a diagram that contains the right one and vice versa. Assume
that these diagrams have $N$ and $N+1$ chains respectively. The
contraction of the left vertex subgraph gives $c/(1-u_N)$
with some constant $c$. Now consider the other subgraph. Contraction
of the vertex subgraph in the inner box gives $c/(1-u_{N+1})$ with
the same constant $c$ when $u_{N+1}$ is close to one.
The subsequent contraction of the second box
turns the simple pole into a double pole. If one takes into account
all factors, including the different power of $\beta_0$ in
eq.~(\ref{analytic1}), one finds that the coefficient of the double
pole is $(-i) c$. After this contraction, the lower chain carries
regularization parameter $u_N+u_{N+1}$. A factor $(-i)$ from
one photon propagator is obtained
from joining the two chains into one (this can be done,
since both have the same
momentum). One can then make use of the $\delta$-function constraint in
eq.~(\ref{analytic1}). From the left diagram we get simply

\be
\frac{c}{1-(u-u_1\ldots-u_{N-1})}
\ee

\n and from the right

\be
\int\limits_0^{u-u_1-\ldots-u_{N-1}} \!\!\! d u_N\frac{(-c)}{
(1-(u-u_1\ldots-u_{N-1}-u_N))^2}\,.
\ee

\n Both contributions cancel each other up to non-singular terms. As a
result the leading singularity drops out in the sum of the two diagrams that
contain the graphs of Fig.~\ref{cancel1} as subgraphs. This cancellation
does not apply to the next-to-leading singularity of the diagrams.

Finally we note that eq.~(\ref{threeterms}) is universal and can be used
for any Green function with the
obvious replacement for the corresponding leading order Borel transform.

%%%%%%%%%%%%%%%%%%% SECTION 4 %%%%%%%%%%%%%%%%%%%%%%%%%%%%

\section{Vertex function and vacuum polarization}
\setcounter{equation}{0}

In this section we turn to the classical example of the photon vacuum
polarization. For reasons that will become clear it is useful to consider
first the fermion-photon vertex function. We restrict ourselves to the
leading singularity at $u=1$.

The photon vacuum polarization is gauge-independent. One can introduce a
gauge parameter $\xi$, so that the Lorentz structure in
eq.~(\ref{rulephoton}) takes the familiar form for covariant gauges.
The result must be independent of $\xi$. In particular, one can take
Feynman gauge and considerably simplify the algebra connected with
numerators. The price to pay for this simplification is the presence
of singular factors like $1/((1-u)u_j)$ in individual diagrams,
which are absent in Landau gauge, because the one-loop self-energy
and vertex function are separately UV finite
in this gauge. To render each diagram
separately finite, one would have to add the usual counterterms.
In practice we do not have to bother about these subtractions, since
we know that all such singular factors must cancel in the final
result. We have done the calculation for the vacuum polarization in
an arbitrary covariant gauge and verified independence of $\xi$
explicitly. Gauge-dependent intermediate expressions that follow
below are given in Landau gauge, as before.

\subsection{Vertex function}

The result for the leading order diagram with one chain,
Fig.~\ref{self}b, is obtained from eq.~(\ref{firstordervertex}) after
restoration of overall factors:

\be\label{leadvertex}
B_{\rm LO}[\Gamma](u)\stackrel{u\rightarrow 1}{=}\,
-\frac{1}{4\pi\mu^2}\,e^C\,
\frac{1}{1-u}\left[-\frac{2}{3} (Q^2\gamma_\mu-
\!\not\! Q Q_\mu) + \frac{1}{2}\left(\left(p^2+{p^\prime}^2\right)
\gamma_\mu + \!\not\! p\gamma_\mu \!\not\! p^\prime\right)\right]\,.
\ee

\n The corresponding large-order behaviour is proportional to
$\beta_0^n n!$. Recall that the coup\-ling $(-ig)$ to the external photon
is not included in the Borel transform.

\subsubsection{Vertex diagrams with two chains}

\begin{figure}[t]
   \vspace{0cm}
   \centerline{\epsffile{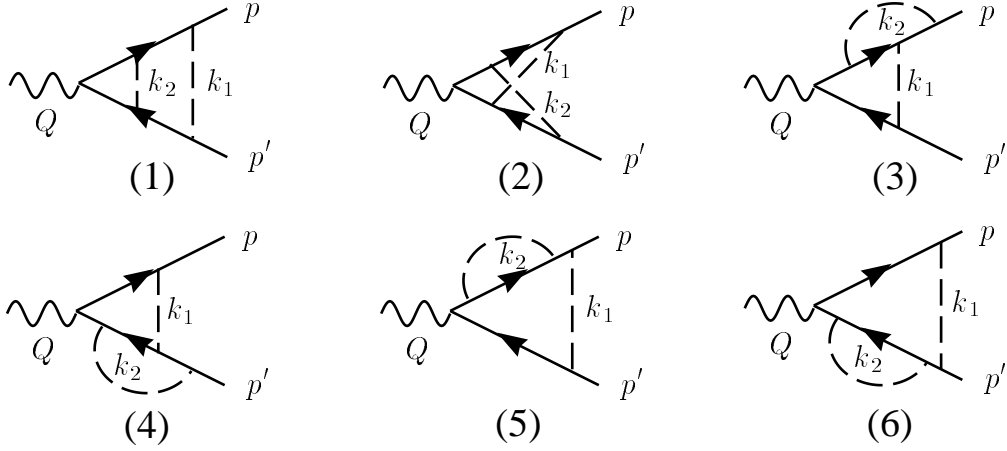}}
   \vspace*{0.4cm}
\caption{\label{vertex} Vertex diagrams with two chains.}
\end{figure}

The diagrams with two chains are shown in Fig.~\ref{vertex}.
Any maximal forest of any of these diagrams has two 1PI
elements, one of which is the diagram itself. According to
eq.~(\ref{lsing}), the leading singularity arises from the
combinations

\be
\frac{1}{(1-u)^2}\,,\,\,\frac{1}{1-u}\frac{1}{1-u_j}\,,\,\,
\frac{1}{1-u}\frac{1}{u_j}\,,
\ee

\n where the last combination is in fact absent. The first combination
arises from `box' subgraphs as discussed in Sect.~2.2.
For each diagram, in each forest,
there are three choices for the second 1PI element $\gm\not=\Gm$,
which is a one-loop subgraph. For each of the
18 distinct combinations we have to consider the region $u(\gamma)=1$,
where $u(\gamma)$ is the sum of $u$-parameters
of the lines of the corresponding one-loop subgraph. The result is
denoted by (Nx), where N=$1,\ldots,6$ for the six diagrams and
x=a,b,c for the three possible one-loop subgraphs.
Seven of the 18 combinations lead to insertions
into reduced graphs that vanish. The remaining eleven ones are described
as follows, according to their subgraph $\gamma$:\\

(1a): The `small' vertex consisting of lines with momenta
$k_2,p^\prime-k_1-k_2,p-k_1-k_2$.

(1b): The `box' subgraph.

(2a): The subgraph consisting of lines with momenta
$k_2,p-k_1-k_2,p^\prime-k_1-k_2,p^\prime-k_2$.

(2b): The subgraph symmetric to (2a) with lines
$k_2,p-k_1,p-k_1-k_2,p^\prime-k_1-k_2$.

(2c): The `crossed box' subgraph.

(3a): The subgraph with lines $k_1,p-k_1-k_2,p-k_1,
p^\prime-k_1$.

(3b): The vertex subgraph on the upper fermion leg.

(4a): The subgraph with lines $k_1,p-k_1,p^\prime-k_1,
p^\prime-k_1-k_2$, symmetric to (3a).

(4b): The vertex subgraph on the lower fermion leg.

(5a): The self-energy subgraph on the upper leg.

(6a): The self-energy subgraph on the lower leg.\\

\n In addition to subgraphs of self-energy [(5a),(6a)] and vertex
type [(1a),(3b),(4b)], which have superficial degree of divergence
$\om(\gm)=0$, we also encounter subgraphs with four external fermion
legs [(1b),(2c)] and two fermion and two photon legs
[(2a),(2b),(3a),(4a)] with $\om(\gm)=-2$.

Specifically, for the box-type subgraphs one obtains

\be \label{boxes}
\frac{i}{1-u}\frac{3\pi}{N_f}
\left[\pm \frac{3}{4}\gamma_\mu\otimes\gamma^\mu - \frac{3}{2}
\gamma_\mu\gamma_5\otimes\gamma^\mu\gamma_5 \right]\,,
\ee

\n with the upper sign for (1b) and the lower sign for (2c). In the
sum of both contributions only the structure
$\gamma_\mu\gamma_5\otimes\gamma^\mu\gamma_5$ survives. Similar
cancellations occur between other subgraphs and one should combine
them before insertion in the reduced diagram. It is then
straightforward to calculate the combinations

\bea\label{prettylong}
\mbox{(I)} &=& \mbox{(1a)}
\nonumber\\
&=&-\frac{1}{12}\,(p^2+{p^\prime}^2)\,\gamma_\mu+\frac{1}{3}\,(\!\not\!p
p_\mu+\!\not\!p^\prime p^\prime_\mu)-\frac{3}{2}\,(\!\not\!p p^\prime_\mu
+\!\not\! p^\prime p_\mu)+2p\cdot p^\prime \gm_{\mu}
+\!\not\!p\gamma_\mu\!\not\!
p^\prime
\nonumber\\
\mbox{(II)} &=& \mbox{[(2a)+(3a)]+[(2b)+(4a)]}
\nonumber\\
&=&\frac{1}{12}\,(p^2+{p^\prime}^2)\,\gamma_\mu-\frac{1}{3}\,(\!\not\!p
p_\mu+\!\not\!p^\prime p^\prime_\mu)+\frac{1}{2}\,(\!\not\!p p^\prime_\mu
+\!\not\! p^\prime p_\mu)-2p\cdot p^\prime \gm_{\mu}
-\frac{1}{2}\,
\!\not\!p\gamma_\mu\!\not\! p^\prime
\nonumber\\
\mbox{(III)} &=& \mbox{(3b)+(4b)} = \frac{8}{9}\,(Q^2\gamma_\mu-
\!\not\! Q Q_\mu)
\\
&&-\,\frac{3}{4}\,(p^2+{p^\prime}^2)\,\gamma_\mu+\frac{1}{3}\,(\!\not\!p
p_\mu+\!\not\!p^\prime p^\prime_\mu)-\frac{1}{2}\,(\!\not\!p p^\prime_\mu
+\!\not\! p^\prime p_\mu)+2 p\cdot p^\prime \gm_{\mu}
-\frac{1}{6}\!\not\!p\gamma_\mu\!\not\!p^\prime
\nonumber\\
\mbox{(IV)} &=& \mbox{(5a)+(6a)}
\nonumber\\
&=&\frac{1}{12}\,(p^2+{p^\prime}^2)\,\gamma_\mu-\frac{1}{3}\,(\!\not\!p
p_\mu+\!\not\!p^\prime p^\prime_\mu)+\frac{3}{2}\,(\!\not\!p p^\prime_\mu
+\!\not\! p^\prime p_\mu)-2p\cdot p^\prime \gm_{\mu}
-\!\not\!p\gamma_\mu\!\not\!
p^\prime
\nonumber\\
\mbox{(V)} &=& \mbox{(1b)+(2c)} = 2\,(Q^2\gamma_\mu-
\!\not\! Q Q_\mu)
\nonumber
\eea

\n It is understood that each line is multiplied by $1/(16\pi^2)$ and
$1/(1-u)(1-u_1)$ except for (V), where the latter factor is
replaced by $1/(1-u)^2$. Note that (I) and (IV) cancel against
each other. The other terms add to

\bea\label{intermedvertex}
G_{\bar{\Gm}_{\rm (1)-(6)}}[\Gm](u) &\stackrel{\rm LS}{=}&
-\frac{1}{12\pi^2}\Bigg\{\frac{1}{1-u}\frac{1}{1-u_1}
\left[-\frac{2}{3} (Q^2\gamma_\mu-
\!\not\! Q Q_\mu) + \frac{1}{2}\left(\left(p^2+{p^\prime}^2\right)
\gamma_\mu + \!\not\! p\gamma_\mu \!\not\! p^\prime\right)\right]
\nonumber\\
&&\hspace*{1cm}-\,\frac{3}{2}\,\frac{1}{(1-u)^2}\,(Q^2\gamma_\mu-
\!\not\! Q Q_\mu)\Bigg\}\,.
\eea

\n The first line has exactly the same structure as the leading
order result, eq.~(\ref{leadvertex}). After integration over the
$u$-parameters, it leads to a singularity of the form
$\ln(1-u)/(1-u)$,

\be
B_{\rm LO}[\Gamma](u)\,\frac{1}{N_f}\ln(1-u)\,.
\ee

\n This is not the only contribution of this form,
however. As in Sect.~3.5, eq.~(\ref{selfprop}),
the universal propagator corrections add

\be
B_{\rm LO}[\Gamma](u)\,\left(-\frac{\beta_1}{\beta_0^2}-\frac{1}{N_f}
\right) \ln(1-u)\,,
\ee

\n and by the same cancellation as for the self-energy only the term
with $\beta_1/\beta_0^2$ is left.

Adding everything together, we
find that, including diagrams with two chains and propagator corrections,
eq.~(\ref{leadvertex}) is extended to

\bea\label{nextleadvertex}
B_{\rm NLO}[\Gamma](u) \!\!&\stackrel{u\rightarrow 1}{=}&\!\!
-\frac{1}{4\pi\mu^2}\,e^C\Bigg\{
\frac{1}{1-u}\left(1-\frac{\beta_1}{\beta_0^2}\ln(1-u)\right)
\bigg[-\frac{2}{3} (Q^2\gamma_\mu-
\!\not\! Q Q_\mu) + \frac{1}{2}\left(\Big(p^2+{p^\prime}^2\right)
\gamma_\mu
\nonumber\\
&&\hspace*{2cm}+\,\!\not\! p\gamma_\mu \!\not\! p^\prime\Big)\bigg]
+\,\frac{3}{2 N_f}\,\frac{1}{(1-u)^2}\,(Q^2\gamma_\mu-
\!\not\! Q Q_\mu)\Bigg\}\,.
\eea

\n The first line has a similar structure as in case of the self-energy.
The two-loop evolution of the coupling leads to a logarithmic
enhancement in the large-order
behaviour, $\beta_0^n n!\,\ln n$, while being formally of order $1/N_f$.
The second line incorporates the contributions from box type subgraphs,
which were absent for the self-energy (at the level of two chains).
While this contribution starts at order $1/N_f$, it leads to an enhanced
divergence in large orders, $\beta_0 n!\,n$ for the reasons explained
in Sect.~2.2. In the language of operator insertions, introduced in
Sect.~1 and pursued in Sect.~5, we can identify this contribution with
an insertion of the operator $\bar{\psi}\gamma_\mu\gamma_5\psi
\bar{\psi}\gamma^\mu\gamma_5
\psi$ in place of the box subgraph, see eq.~(\ref{boxes}). The
pattern of large-order behaviour for the vertex function with two chains
is the same as for the vacuum polarization in \cite{VAI94}. With the
help of IR rearrangement one can also compute the NLS, i.e. all terms
that yield $1/(1-u)$. In particular, the contribution from propagator
corrections $\delta K_{\rm prop}$ is the same as for the self-energy
in Sect.~3.5.

\subsubsection{Vertex diagrams with one fermion loop and three chains}

\begin{figure}[t]
   \vspace{0cm}
   \epsfysize=12cm
  \epsfxsize=12cm
   \centerline{\epsffile{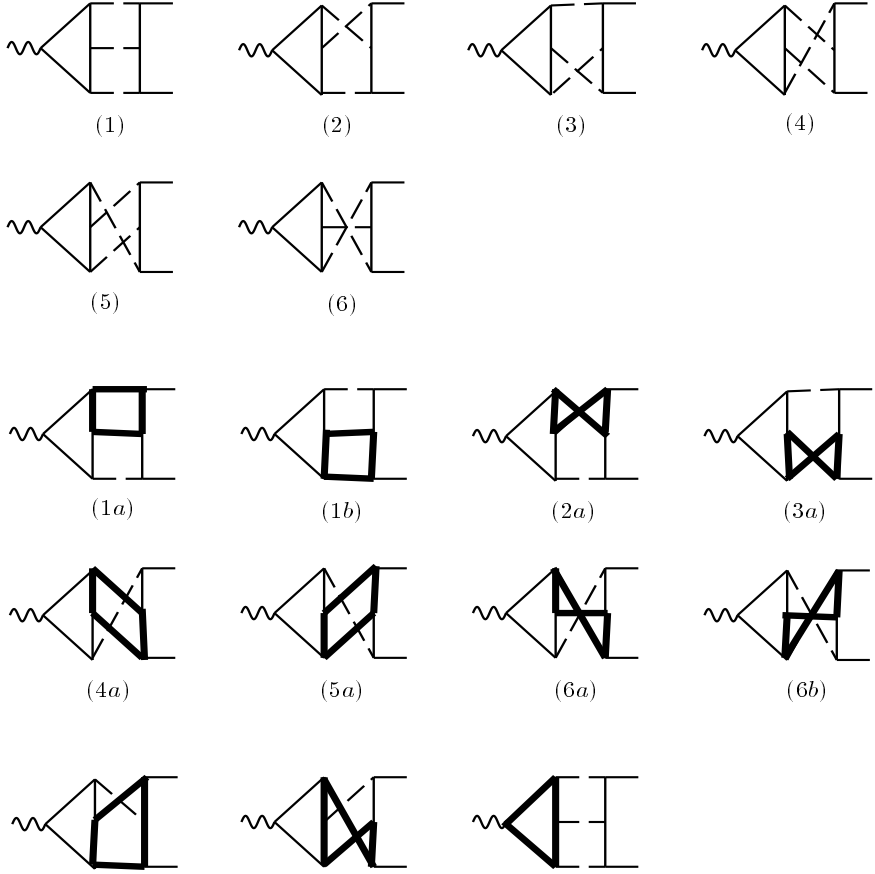}}
   \vspace*{0.8cm}
\caption{\label{vertex31} (1)--(6): Diagrams with three chains that
contribute to the vertex at order $1/N_f^2$. (1a)--(6b): The eight
subgraphs which contribute to the leading singularity at $u=1$. The
last line shows subgraphs with non-vanishing reduced graphs that do
not contribute at $u=1$.}
\end{figure}

In this subsection we analyze the set of graphs, labelled (1) to (6)
in Fig.~\ref{vertex31}. These diagrams have three chains, but
contribute at order $1/N_f^2$, because the fermion loop brings a
factor $N_f$. This set of diagrams is separately gauge-independent
and we use Feynman gauge to simplify the algebra. The leading
singularity is determined by the contributions with the maximal number
of singular factors in eq.~(\ref{singofg}) which is three for the
diagrams in Fig.~\ref{vertex31} rather than two as for those in
Fig.~\ref{vertex}. Because of fermion loops, the counting in $1/N_f$
and chains is no longer equivalent beyond a certain order. In
general it is the number of chains and not the power in $1/N_f$,
which determines the strength of the renormalon singularity and the
divergence in large orders of perturbation theory, so that the
diagrams in Fig.~\ref{vertex31} can be expected to dominate the
large-order behaviour at order $1/N_f^2$. We determine the singularity
at $u=1$ up to terms $\ln(1-u)/(1-u)$. This includes those parts
of the next-to-leading singularity that produce $1/(1-u)^2$.

As usual, the leading singularity is determined by subsequent
contractions of one-loop subgraphs. The eight subgraphs that contribute
to the singularity at $u=1$ are enumerated (1a) to (6b) in
Fig.~\ref{vertex31}. There are many more subgraphs, which after
contraction lead to vanishing (tadpole-type) reduced graphs. Other
subgraphs like those in the last row of Fig.~\ref{vertex31} have
non-vanishing reduced graphs, but do not contribute at $u=1$. For
the first two graphs in that row, the contraction can be seen
to correspond to operators of dimension eight, which contribute
to $u=2$ and higher poles only. In other words, the degree of divergence
is $\om(\gm) \leq -3$. The last graph need not be considered,
because the subgraph contains only fermion lines and no
$u$-parameter.

\begin{figure}[t]
   \vspace{0cm}
   \centerline{\epsffile{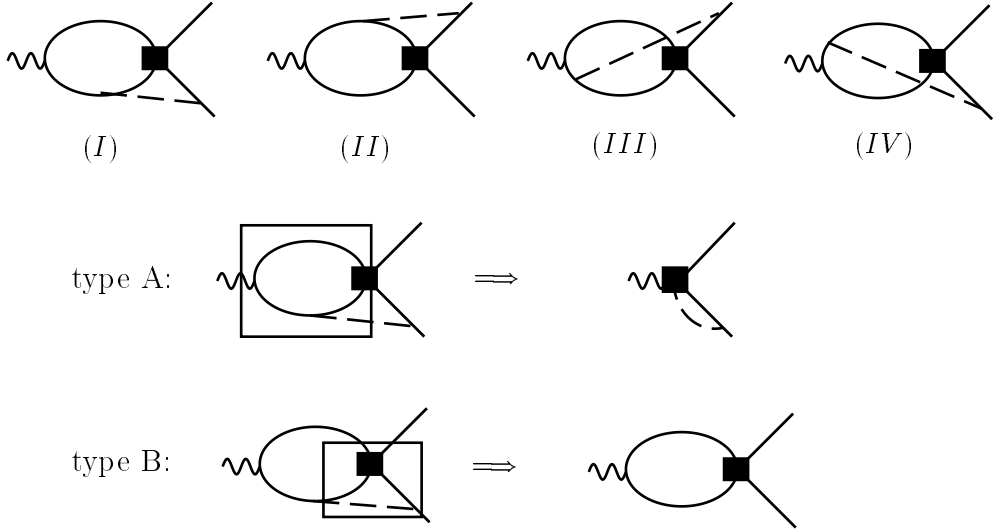}}
   \vspace*{0.6cm}
\caption{\label{vertex32} $(I)$--$(IV)$: The reduced graphs corresponding
to contraction of the subgraphs in (1a)--(6b) in the previous figure. Each
reduced graph has a subgraph of type A and B, whose contraction gives
rise to the reduced graph shown on the right side.}
\end{figure}

The subgraphs of all eight non-vanishing contributions have the form
of boxes or `crossed boxes'. When each box is combined with its
corresponding crossed box, contraction of these subgraphs reduces
to the insertion of $\bar{\psi}\gamma_\mu\gamma_5\psi
\bar{\psi}\gamma^\mu\gamma_5\psi$, see eq.~(\ref{boxes}).
The four reduced
graphs are shown in Fig.~\ref{vertex32}, where the black box
is accompanied by a factor $1/(1-u_i-u_j)$ ($i,j$ refer to the labels
of the two contracted chains). Each of the graphs $(I)$ to $(IV)$
has two subgraphs with non-vanishing reduced graphs, called `type A'
and `type B' in Fig.~\ref{vertex32}. The contraction of type A
yields another factor $1/(1-u_i-u_j)$ ($i,j$ being the same as in the
previous contraction), while the contraction of type B produces
$1/(1-u)$. The final contraction of the one-loop graphs to the
right of the arrows in Fig.~\ref{vertex32} gives $1/(1-u)$ in both
cases. After integrating over the $u_i$ with

\be \intl_0^u\,du_1 du_2 du_3\,\delta(u-u_1-u_2-u_3)
\ee

\n the leading singularity at $u=1$ is of form

\bea
\mbox{type A}: && \frac{1}{(1-u)^2} \nonumber\\
\mbox{type B}: && \frac{\ln(1-u)}{(1-u)^2}\,.
\eea

\n In fact, we find that for each of the four diagrams
of type A the traces vanish and there is no double pole contribution
from this set. The contraction of type B results only in the tensor
structure $\gamma_\mu \otimes \gamma^\mu$ but no
$\gamma_\mu \gamma_5\otimes \gamma^\mu\gamma_5$. As evident from
Fig.~\ref{vertex32} the pole term obtained in this contraction is
related to operator mixing between $\bar{\psi}\gamma_\mu\gamma_5\psi
\bar{\psi}\gamma^\mu\gamma_5\psi$ and $\bar{\psi}\gamma_\mu\psi
\bar{\psi}\gamma^\mu\psi$. The final result for the leading
singularity is then

\be\label{leadvertex3}
B_{\bar{\Gm}_{\rm Fig.\ref{vertex31}}}[\Gm](u)
\stackrel{u\to 1}{=}-\frac{1}{4\pi\mu^2}
\,e^C\,\frac{27}{2 N_f}\,
\frac{\ln(1-u)}{(1-u)^2}\,
(Q^2\gamma_\mu-\not\! Q Q_\mu)\,,
\ee

\n where $Q$ is the momentum of the external photon.

To obtain all terms $1/(1-u)^2$ in the final result, we
have to collect in addition all contributions with {\em two} singular
denominator factors $1/(1-u)$, since the contributions from terms
with three singular factors (type A above) vanish. Since one
such factor comes from the last contraction, all such terms arise
from two-loop four-fermion graphs, obtained by cutting the fermion loop
in the diagrams (1) to (6) in Fig.~\ref{vertex31}. The problem
therefore reduces to computing first the leading and next-to-leading
singularity for the four-fermion Green function with three chains,
extending eq.~(\ref{boxes}) to the next order in the chain expansion.
The resulting expression is then inserted into a reduced graph
of type B,
shown to the right of the arrow in Fig.~\ref{vertex32}.

As in the case of the self-energy, we use the method of infrared rearrangement
to obtain the coefficient of the singular factor $1/(1-u)$ of the
two-loop (three chain) four-fermion graphs. Since the singularity is
local and proportional to $Q^0$, no differentiation is required.
Therefore we can set all external momenta to zero from the start and
introduce a new external momentum $p$ that flows through a single line.
We then encounter integrals of the form

\be
\int\frac{\dd^4k_1}{(2\pi)^4}\frac{\dd^4k_2}{(2\pi)^4}
\frac{\gamma_\alpha\!\not\! k_2\gamma_\beta \!\not\! k_1
\gamma_\gamma\otimes \gamma^\gamma\!\not\! k_1\gamma^\beta \!\not\! k_2
\gamma^\alpha}{((p+k_1)^2)^{3-u_1} (k_2^2)^{3-u_2}
((k_1+k_2)^2)^{1-u_3}}
\ee

\n which are easily evaluated. Then, as a straightforward generalization
of the formulae of Sect.~3.4, one subtracts the contributions from the
leading singularity before integration over $u$-parameters.
We can set $u=u_1+u_2+u_3=1$ after this step and perform the integration.
We then obtain for the Borel transform of the two-loop
four-fermion graphs near $u=1$

\be\label{doubleboxes}
-\frac{i}{1-u}\frac{3\pi}{N_f}
\frac{27}{2 N_f}\left[\ln(1-u)+C_{\rm box}\right] \,
\gamma_\mu\otimes \gamma^\mu\,,
\ee

\n where $C_{\rm box}=1.016\ldots$ results from numerical integration
over a product of $\Gamma$-functions.

The corresponding contribution to the vertex follows from
inserting the previous line in place of the black box in the
last diagram of Fig.~\ref{vertex32}. The result is

\be\label{nextleadvertex3}
B_{\bar{\Gm}_{\rm Fig.\ref{vertex31}}}[\Gm](u)
\stackrel{u\to 1}{=}-\frac{1}{4\pi\mu^2}
\,e^C\,\frac{27}{2 N_f}\,
\frac{1}{(1-u)^2}\left[\ln(1-u)+C_{\rm box}\right]\,
(Q^2\gamma_\mu-\not\! Q Q_\mu)\,,
\ee

\n extending eq.~(\ref{leadvertex3}) including all contributions
of order $1/(1-u)^2$. Note that these diagrams produce the dominant
pole in the vertex function at order $1/N_f^2$.

\subsubsection{Summary}

Collecting all contributions at order $1/N_f^2$,
eq.~(\ref{nextleadvertex}) and eq.~(\ref{nextleadvertex3}), the final
result for the large-order behaviour of the vertex function reads

\be
\Gamma(p,p^\prime;\alpha) = (-i g)\sum_{n=0}^\infty
r_n\alpha^{n+1}
\ee

\n with

\bea\label{vertexlargeorder}
r_n \!\!&\stackrel{n\to\infty}{=}&\!\!
-\frac{1}{4\pi\mu^2}\,e^C\,\beta_0^n\,n!\Bigg\{
\left(1+\frac{\beta_1}{\beta_0^2}(\ln n-\psi(1))\right)
\bigg[-\frac{2}{3} (Q^2\gamma_\mu-
\!\not\! Q Q_\mu) + \frac{1}{2}\left(\Big(p^2+{p^\prime}^2\right)
\gamma_\mu
\nonumber\\
&&\hspace*{-2cm}+\!\not\! p\gamma_\mu \!\not\! p^\prime\Big)\bigg]
+\,\left(\frac{3}{2 N_f} n+\frac{27}{2 N_f} n\,\left(-\ln n+\psi(2)+
C_{\rm box}+{\cal O}\!\left(\frac{\ln n}{n}\right)
\right)\right)\,(Q^2\gamma_\mu-
\!\not\! Q Q_\mu)\Bigg\}\,.
\eea

\n The missing ${\cal O}(1/n \ln n)$ terms come from the uncalculated
$\ln(1-u)/(1-u)$ terms of the diagrams with three chains and one
fermion loop.

\subsection{Vacuum polarization}

With the results for the vertex function, the results for the
vacuum polarization are immediate. A specialty of the vacuum
polarization is that the skeleton diagrams corresponding to
diagrams with one chain have two loops. This gives an additional
factor $n$ in large-orders as compared to the vertex already in
lowest order.

\subsubsection{One chain}

The diagrams with a single chain are part of Fig.~\ref{chain}. The
diagram where the chain forms a self-energy subgraph does not
contribute to the leading singularity. The reason is that the only
forest that does not lead to tadpole reduced graphs is the one
with the self-energy subgraph. If $p$ is the external momentum of
this subgraph, the self-energy close to $u=1$ is proportional to
$p^2\!\not\! p$, see eq.~(\ref{firstorderself}), and the insertion
of this polynomial into the reduced graph eliminates any $1/p^2$
from loop lines of the reduced graph. Thus the reduced graph is
effectively a tadpole and vanishes. We are left with the first graph in
the second line of Fig.~\ref{chain}. The contraction of the vertex
subgraph leads to the insertion of eq.~(\ref{leadvertex}), so that

\bea\label{leadingvacuum}
B_{\rm LO}[\Pi](u) &\stackrel{u\to 1}{=}& 2\cdot \left(-\frac{N_f}{4\pi\mu^2}
\right)\,e^C\left(-\frac{2}{3}\right)
\frac{1}{1-u}\int\frac{\dd^4 p}{(2\pi)^4}
\frac{{\rm tr}\,(\gamma_\mu\!\not\! p (Q^2\gamma_\nu-\!\not\! Q
Q_\nu) (\!\not\!p+\!\not\! Q))}{(p^2)^{2-u} (p+Q)^2}
\nonumber\\
&\stackrel{u\to 1}{=}& (-i)\,(g_{\mu\nu} Q^2-Q_\mu Q_\nu) \left(
-\frac{Q^2}{\mu^2} e^C\right)\frac{N_f}{36\pi^3}\frac{1}{(1-u)^2}
\,.
\eea

\n Notice that the terms involving $p$ and $p^\prime$ in
eq.~(\ref{leadvertex}) do not contribute, when inserted into the
reduced diagram, because they eliminate one of the two denominators.
The overall factor two accounts for the two possible insertions
into the left or right vertex.
The second line above reproduces\footnote{
There is a factor $-N_f/(4\pi)$ difference due to different
conventions. Here the two couplings $(-ig)$ to the external photon
are not included by definition, while in \cite{BEN93} the Borel
transform of $\Pi/a$ is given with $a=\alpha N_f$.} the asymptotic
behaviour of the exact result of \cite{BEN93}. The corresponding
large-$n$ behaviour is proportional to $\beta_0^n n!\,n$. To
compute the $1/n$ corrections one would use IR rearrangement with
additional subtractions of non-local terms in external momenta.

\subsubsection{Two chains}

\begin{figure}[t]
   \vspace{-1cm}
   \epsfysize=8cm
   \epsfxsize=12cm
   \centerline{\epsffile{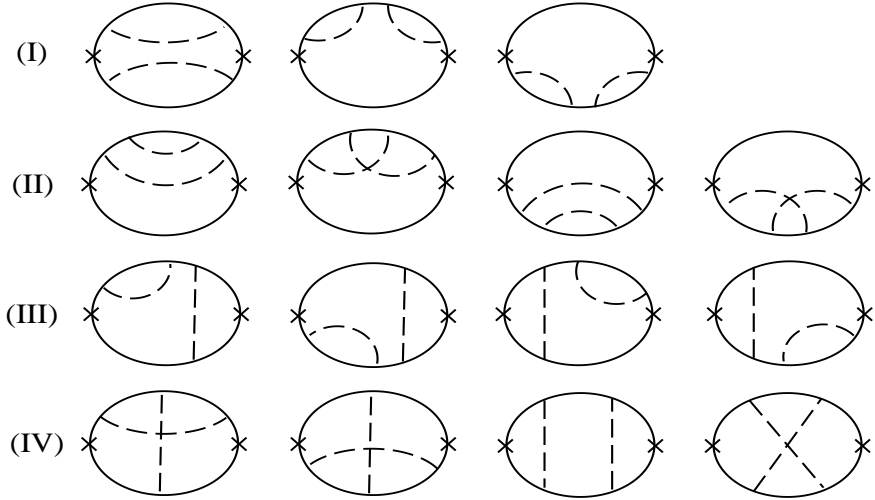}}
   \vspace*{0cm}
\caption{\label{vacpol} Vacuum polarization diagrams with two chains.
The crosses denote insertions of external photon legs.}
\end{figure}

We calculate the leading singularity for the diagrams with two
chains, see Fig.~\ref{vacpol}. A maximal forest contains three
1PI elements. Now we encounter for the first time forests, where
non-trivial elements are disjoint and do not form a nested sequence,
for instance for the third diagram in class (IV). However, close
to $u=1$, such forests do not contribute to the leading singularity,
since they produce at most $\ln(1-u)/(1-u)$. When the non-trivial
elements form a nested sequence, the reduced diagram is a tadpole,
whenever the smallest subgraph in this sequence contains one or both
of the external vertices.
\phantom{\ref{vacpol}}

With these restrictions in mind, let us turn to the diagrams (I)
in Fig.~\ref{vacpol}. The smallest graph in each non-vanishing
maximal forest is a self-energy subgraph. Close to $u_i=1$,
where $u_i$ is the regularization parameter of this subgraph, it reduces
to an insertion proportional to $p^2\!\not\! p$, with $p$ the
external momentum of the subgraph. Again this insertion eliminates
one denominator, so that the reduced graph in fact vanishes.
One can convince oneself that for class (II) a non-tadpole
reduced graph always requires contraction of the two-loop self-energy
subgraph. We then get the same cancellation as for (I). Thus
there is no contribution to the leading singularity
from the first seven diagrams.

Considering class (III) and (IV), we can divide all maximal forests
into three groups: (i) A forest of type $\{{\cal F}_{\rm vertex},\Gamma
\}$, where ${\cal F}_{\rm vertex}$ is a forest that has already been
evaluated for the two-chain vertex (Fig.~\ref{vertex}) and $\Gm$ is
the entire diagram. (ii) Forests with nested 1PI elements, not
belonging to group (i). (iii) Forests with disjoint one-loop elements.
It turns out that (ii) always leads to reduced graphs of tadpole type
and (iii) does not contribute to the leading singularity
as noted above. Thus we arrive at the same conclusion as in the case of
a single chain that the leading singularity of the vacuum polarization
is given by insertion of the result for the vertex function, in the
present situation eq.~(\ref{intermedvertex}) or
eq.~(\ref{nextleadvertex}), if one includes already the propagator
corrections. Again, only the terms involving $Q$ contribute for the
same reason as above. Since such terms do not arise from class (III)
diagrams, only the final four diagrams in Fig.~\ref{vacpol}
contribute to the leading singularity. (This could be deduced without
knowing the result for the vertex function, applying the same
argument as to class (I) and (II).)
The insertion leads to exactly the same integral for the
reduced graph as in eq.~(\ref{leadingvacuum}). Therefore we
get (including propagator corrections)

\bea
B_{\rm NLO}[\Pi](u) &\stackrel{u\to 1}{=}&
(-i)\,(g_{\mu\nu} Q^2-Q_\mu Q_\nu) \left(
-\frac{Q^2}{\mu^2} e^C\right)\frac{N_f}{36\pi^3}\frac{1}{(1-u)^2}
\Bigg[1-\frac{\beta_1}{\beta_0^2}\ln(1-u)
\nonumber\\
&&-\frac{9}{4 N_f}\,
\frac{1}{1-u}\Bigg]\,.
\eea

\n and the corresponding large-order behaviour of the perturbative
expansion (including now the `light-by-light' scattering
contributions)

\bea\label{finalvacuum}
r_n &\stackrel{n\to\infty}{=}& (-i)\,(g_{\mu\nu} Q^2-Q_\mu Q_\nu)
\left(-\frac{Q^2}{\mu^2} e^C\right)\frac{N_f}{36\pi^3}\,\beta_0^n\,
n!\,n\Bigg(1+\frac{\beta_1}{\beta_0^2}\,(\ln n-\psi(2))
\nonumber\\
&&\,- \frac{9}{8 N_f} n - \frac{81}{8 N_f} n\left\{-\ln n+\psi(3)
+C_{\rm box}\right\}
\Bigg)\,.
\eea

\n The corrections to this result are ${\cal O}(1/N_f,\ln^2 n/N_f^2)$
relative to the unity in brackets. The pattern of divergence is
the same as for the vertex function in that contractions that lead
to four-fermion operator insertions are enhanced by a factor of $n$
as discussed in Sect.~2.2. In analogy with the vertex function,
Sect.~4.1.2, the `light-by-light' diagrams with three chains and
two fermion loops, not shown in Fig.~\ref{vacpol},
contribute to the vacuum polarization at subleading
order in $1/N_f$.

Setting the expression in curly brackets to zero,
eq.~(\ref{finalvacuum}) agrees with the corresponding result in the
erratum to \cite{VAI94}, where the light-by-light
contributions are eliminated by a certain charge assignment to the fermions.
Compared to the approach taken there,
based on the background field technique, the
diagrammatic formalism presented here lacks the elegance of
maintaining explicit gauge invariance. This leads to a proliferation
of terms such as in eq.~(\ref{prettylong}), most of which cancel
in the final result. On the other hand, the conceptual framework
developed in Sect.~3 allows us to systematically compute corrections,
either from more then two chains or contributions subleading in $n$
for large $n$, such as the first correction to the normalization
$K$ of the large-order behaviour (see the example of the
self-energy in Sect.~3) and thus elucidates the origins and organization
of divergence for arbitrarily complicated diagrams. The understanding
of the systematics of the chain expansion to all orders is the subject
of the following section.

%%%%%%%%%%%%%%%%%%%% SECTION 5 %%%%%%%%%%%%%%%%%%%%%%%%%%%%%%%

\section{UV renormalon counterterms and \newline renormalization group}
\setcounter{equation}{0}

The previous examples illustrate that diagrams with a larger number
of chains (suppressed by powers of $1/N_f$) lead to stronger divergence
in large orders of perturbation theory. Except for an enhancement of
$n$ due to the new class of box-type subgraphs, the enhancement is
logarithmic in $n$. For each additional chain one can obtain at most
one $\ln n$, as there is one additional singular factor in the
$u$-parameters. The expansion parameter of the chain expansion is in
fact $\ln n/N_f$ and any finite order approximation in $1/N_f$ does
not provide the correct asymptotic behaviour in $n$.

The situation is reminiscent of standard applications of the
renormalization group, which allow us to deduce the large-momentum
behaviour of Green functions, which can not be determined from
finite-order perturbative expansions in $\alpha$. The key idea is
to prove factorization of the dependence on the variable under
consideration and relate this dependence to one on an artificial
factorization scale. In the present case, factorial divergence
in $n$ is equivalent to poles in the Borel parameter $u$, which in
turn are related to the regions of large loop momenta. Specifically,
for the contribution from a given forest, the leading singularity
could be determined from the $\dd^4k/k^6$-piece in the expansion of
the smallest non-trivial element of the forest in its external
momenta and thereby associated with insertion of a dimension six
operator. Generalizing this observation, the desired factorization
would be summarized by the statement \cite{PAR78}
that the leading UV renormalon
could be compensated in all Green functions by adding

\be \label{dimsix2}
{\cal L}_{\rm ct}^{(6)} = \frac{1}{\mu^2}\sum_i E_i(\alpha(\mu))
\,{\cal O}_i\,,
\ee

\n to the renormalized Lagrangian, where ${\cal O}_i$ are dimension
six operators and $E_i(\alpha(\mu))$ factorially divergent
series in $\alpha$ with finite coefficients that depend on the
renormalization conditions implied for the (usual) renormalized
Lagrangian as well as the operators ${\cal O}_i$.

Such a statement implies that UV renormalons are local. Indeed, all
explicit results for large-$n$ behaviour given so far were
straightforwardly obtained as  local quantities (polynomial in external
momentum). But the expectation to get each time a local result is
too naive and logarithms of the external momenta arise at the
level of $1/n$-corrections.\footnote{See the discussion
in Sect.~3.3 after eq. (3.13).} The situation is rather similar
to the calculation of UV counterterms. If one does not remove
subdivergences and calculates the pole part
of a diagram in $\eps=(4-d)/2$ (where $d$ is the space-time dimension)
the result is inevitably non-local.
To get a proper answer one should first take care of all the
subdivergences of the diagram by inserting the corresponding
counterterms which subtract the subdivergences. In the case of
UV renormalons, one should also subtract, for subgraphs of a
given graph, the factorial divergence associated with subgraphs.
This procedure is of course implicit in writing eq.~(\ref{dimsix2})
and the non-trivial assertion is that the remaining `overall'
UV renormalon divergence for the given graph is indeed local.
The UV renormalon subtractions will be organized in the same way
as usual subtractions of UV divergences that enter the
standard UV $R$-operation (i.e. renormalization at the diagrammatical
level).

In Sect.~5.1 we illustrate the procedure on the simplest example,
then defining the general ${\cal R}$-operation in Sect.~5.2. The
analogy with the usual $R$-operation suggests eq.~(\ref{dimsix2}),
but we do not actually prove this result. The resulting combinatorial
structure of the ${\cal R}$-operation enables us to
straightforwardly translate our result
into the Lagrangian language. The best way to establish this property
manifestly is to apply formulae of the so-called
counterterm technique \cite{AZ,CS1,VS} which are essentially based
on combinatorial properties of the $R$-operation and locality
of counterterms. We shall list the corresponding formulae in
Sect.~5.3. These formulae do not explicitly give the operators
which are added to the Lagrangian in every concrete situation.
In Sect.~5.4 we characterize these operators, the corresponding
renormalon counterterms and derive renormalization group equations for
the coefficient functions $E_i$ that enter eq.~(\ref{dimsix2}).
The solutions to these equations sum all contributions of order
$(\ln n/N_f)^k$ and provide the correct asymptotic behaviour in $n$,
up to an overall constant, that can be determined only order
by order in $1/N_f$. For the vacuum polarization, this solution is
found in Sect. 5.5.

\subsection{Counterterms: An example}

The simplest case of subtraction of subdivergences, considered
already in \cite{DIC95}, is the photon vacuum polarization with
a single chain.

\begin{figure}[t]
   \vspace{0cm}
   \centerline{\epsffile{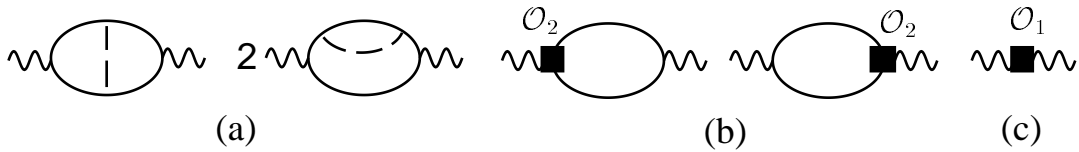}}
   \vspace*{0.0cm}
\caption{\label{counter1} Leading order vacuum polarization: Diagrams
(a) and counterterm insertions (b), (c).}
\end{figure}

Let $G_{\bar{\Gamma}}(\uq;\alpha)$ represent the perturbation series in
$\alpha$, generated by the set of diagrams $\bar{\Gamma}$,
such as shown in Fig.~\ref{counter1}a for the vacuum polarization.
Assuming that we know how to extract the factorial divergence for
a given set of diagrams including $1/n^k$-corrections to the leading
asymptotic behaviour up to some specified value $k_0$ of $k$, we define
an operation ${\cal K}$ that does this. If the factorial divergence
comes from a branch point in the Borel plane as is generally the
case, $k_0$ cuts off the
infinite series of corrections to the leading asymptotic behaviour.
If it comes from a pole, as for a single chain, this series
terminates
by itself. Since there is only one regularization parameter $u$,
the singularities at $u\to 1$ are of the form $1/(1-u)^2$ and
$1/(1-u)$.
The residue of the simple pole of
the corresponding skeleton diagram with analytically regularized
photon propagator involves a logarithm of the external momentum. In
terms of large-$n$ behaviour, we find

\bea \label{K1}
{\cal K} \Pi_{\bar{\Gamma}}(Q;\alpha) &=&\nonumber
(-i)\,(g_{\mu\nu} Q^2-Q_\mu Q_\nu) \left(
-\frac{Q^2}{\mu^2} e^C\right) \sum_n
\frac{N_f}{36\pi^3}\,\beta_0^n\,n!
 \alpha(\mu)^{n+1}
\\
&&\,\times\left\{n-\ln\left(-\frac{Q^2}{\mu^2}e^C\right)+\frac{17}{6}
\right\}\,,
\eea

\n see eq.~(\ref{leadingvacuum}). The $1/n$-corrections are taken from
the exact result \cite{BEN93}. Because of the logarithm, the previous
equation can not be obtained from the insertion of

\be
\frac{1}{\mu^2} E_4^\prime(\alpha(\mu))\,{\cal O}_4^\prime \; ,
\qquad {\cal O}_4^\prime = \partial_\nu F^{\nu\rho}\partial^\mu F_{\mu\rho}
\ee

\n into the two-point function.\footnote{$F_{\mu\nu}$ is the field
strength tensor of the photon.
We use `primes' on coefficient functions and operators here,
because later we introduce another basis of operators.}

To arrive at a local result for the overall counterterm associated with
the vacuum polarization diagrams one should insert the (renormalon)
counterterms $\ODl(\gm_i)$
for two 1-loop vertex subgraphs $\gm_1$ and $\gm_2$ of the first
diagram in Fig.~\ref{counter1}a
and deal with the quantity

\be
{\cal R}'\Pi_{\bar{\Gamma}}(q;u) \equiv (1 + \ODl(\gm_1) + \ODl(\gm_2) )
\Pi_{\bar{\Gamma}}(q;u)\,.
\label{2chpl0}
\ee

\n Here  ${\cal R}'$ is the incomplete (`renormalon')
${\cal R}$-operation, i.e. without the last (overall)
counterterm. The various symbols that
appear in the previous equation will be defined more precisely in the
following subsection.

The counterterms connected with the vertex subgraphs can be deduced from
eq.~(\ref{leadvertex}). Adding the term

\be\label{e2}
\frac{1}{\mu^2} E_3^\prime(\alpha(\mu))\,{\cal O}_3^\prime \equiv
-\sum_n\frac{e^C}{6\pi\mu^2}\,\beta_0^n n!\,\bar{\psi}
\gamma_\mu\psi\partial_\nu F^{\nu\mu}
\ee

\n to the Lagrangian, the vertex function $\Gamma$ is free from the
first UV
renormalon to the leading order in $1/N_f$.\footnote{To be precise,
additional operator structures must be added to compensate the terms
involving $p$ and $p^\prime$ in eq.~(\ref{leadvertex}). We may ignore
these, because their insertion leads to vanishing reduced graphs in the
present example, see also the discussion in Sect.~4.2.2.} This new term
in the Lagrangian contributes to the vacuum polarization at leading
order through the diagrams shown in Fig.~\ref{counter1}b, which
correspond to the two additional terms in eq.~(\ref{2chpl0}). Their
evaluation gives

\bea\label{K2}
(\ODl(\gm_1) + \ODl(\gm_2))\,
\Pi_{\bar{\Gamma}}(q;\alpha) &=&
i\,(g_{\mu\nu} Q^2-Q_\mu Q_\nu) \left(
-\frac{Q^2}{\mu^2} e^C\right) \nonumber\\
&&\hspace*{-3cm}
\times\sum_n \frac{N_f}{36\pi^3}\,\beta_0^n\,n!
 \alpha(\mu)^{n+1}\left[\left(\frac{1}{\eps}-\gamma_E+\ln 4\pi\right)
- \ln\left(
-\frac{Q^2}{\mu^2}\right)+\frac{5}{3}\right]\,.
\eea

\n The insertion of counterterms for subdiagrams
generates new UV divergences.
Their subtraction corresponds to a renormalization
prescription for the set of dimension six operators. Since the
$u$-parameters are not sufficient to regularize these UV divergences, we
have applied dimensional regularization. Choosing the $\overline{\rm
MS}$ subtraction scheme, operator renormalization on the diagrammatic
level is accomplished, when the insertion of the (finite) renormalon
counterterm is followed
by the action of the $\overline{\rm MS}$ $R$-operation.  Thus we apply
the product

\be
R {\cal R}' = 1 + (1+ \Dl(\Gm/\gm_1))\ODl(\gm_1)
+ (1+ \Dl(\Gm/\gm_2))\ODl(\gm_2) ,
\label{2chpl}
\ee

\n instead of eq.~(\ref{2chpl0}). Here $1+ \Dl(\Gm/\gm_1) \equiv
R(\Gm/\gm_i),\; i=1,2$
are  $R$-operations for the reduced diagrams. Combining eqs.~(\ref{K1})
and (\ref{K2}), we obtain

\bea\label{K3}
{\cal K} R {\cal R}'\Pi_{\bar{\Gamma}}(Q;\alpha) &=&
(-i)\,(g_{\mu\nu} Q^2-Q_\mu Q_\nu) \left(
-\frac{Q^2}{\mu^2} e^C\right)\sum_n
\frac{N_f}{36\pi^3}\,\beta_0^n\,n!
 \alpha(\mu)^{n+1}
\nonumber\\
&&\,\times\left[n+\frac{7}{6}-C
%-\frac{5}{3}
\right]\,.
\eea

\n The logarithm of $Q^2$ has dropped out and the overall
counterterm, represented by diagram c of Fig.~\ref{counter1}, is
local as required. It is straightforward to read off the coefficient
$E_4^\prime$ from the previous equation, so that the vacuum polarization is
free from the first UV renormalon, if the operators ${\cal O}_3^\prime$ and
${\cal O}_4^\prime$ are added to the Lagrangian with appropriately chosen
coefficients.

\begin{figure}[t]
   \vspace{0cm}
   \centerline{\epsffile{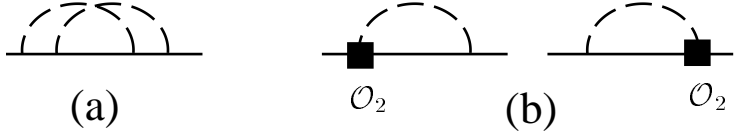}}
   \vspace*{0.0cm}
\caption{\label{counter2} A self-energy diagram and its counterterms.}
\end{figure}

For diagrams with two chains, the diagrams with counterterm insertions
also have to be calculated in the chain expansion. Consider the diagram
shown in Fig.~\ref{counter2}a that contributes to the self-energy.
Combining results from Sect.~3, the Borel transform of this (set of)
diagram(s) in the vicinity of $u=1$ can be represented as

\bea\label{example21}
B[\Sigma_{\bar{\Gamma}_c}](u)&\stackrel{u\rightarrow 1}{=}&
\frac{i}{4\pi}\left(-\frac{Q^2}{\mu^2} e^C\right) \not\!Q
\,\frac{1}{2}\frac{1}{N_f}\Bigg[
\frac{\ln(1-u)}{1-u} + \frac{\delta K^{[c]}}{1-u}
\nonumber\\
&&\,+\ln(1-u)\left\{-\ln\left(-\frac{Q^2}{\mu^2}\right) + {\,\rm const}
\right\} + \ldots \Bigg]\,,
\eea

\n where $\delta K^{[c]}$ has been given in Sect.~3.4 and
the unspecified constant is scheme-dependent. The two diagrams with
insertions of ${\cal O}_3^\prime$, Fig.~\ref{counter2}b, contain one chain
and its contribution is given by the product of the series for
$E_2(\alpha(\mu))$ with the series for the renormalized diagram with
operator insertion. The Borel transform is given by the convolution

\be
2\cdot \frac{1}{\beta_0}\int\limits_0^u d u_1 d u_2\,\delta(u-u_1-u_2)\,
\frac{1}{\mu^2} B[E_2](u_1)\,\frac{(-i)}{4\pi}\,Q^2\!\not\!Q
\left[\left(-\frac{Q^2}{\mu^2} e^C\right)^{-u_2} \!\! P(u_2)-R(u_2)
\right]\,,
\ee

\n where both, $P(u_2)$ and $R(u_2)$, behave as $-1/(2 u_2)$ as
$u_2\to 0$ and the factor of two accounts for the two identical
diagrams. The function $P(u_2)$ arises from straightforward
calculation of these diagrams with analytically regularized photon
propagator. Since the diagrams are quadratically divergent, the
series of ultraviolet poles of $P(u_2)$ starts from $u_2=-1,0,1,\ldots$.
The function $R(u_2)$ incorporates the ultraviolet subtractions
in the adopted scheme for the operator renormalization (in this
case mixing of ${\cal O}_3^\prime$ into
$\bar{\psi}\partial^2\!\!\not\!\partial
\psi$). In the $\overline{\rm MS}$ scheme only the logarithmic
divergences are subtracted, so that $R(u_2)$ is analytic except at
$u_2=0$. The function $R(u_2)$
can be calculated with the methods of \cite{BEN94}.
Using the series for $E_3^\prime$ given above, the previous integral can be
evaluated and yields

\be\label{example22}
B[\Sigma_{\bar{\Gamma}_{c,\rm ct}}](u)\stackrel{u\rightarrow 1}{=}
\frac{i}{4\pi}\left(-\frac{Q^2}{\mu^2} e^C\right) \not\!Q
\,\frac{1}{2}\frac{1}{N_f}\,\ln(1-u)\left\{\ln\left(-\frac{Q^2}
{\mu^2}\right) + {\,\rm const}'
\right\}\,.
\ee

\n The sum of eqs.~(\ref{example21}) and (\ref{example22}) can again
be absorbed into a local counterterm proportional to
$\bar{\psi}\partial^2\!\not\!\partial \psi$. The logarithm
of $Q^2$ arises only from $u_1\to 1$, and the unspecified constant
depends on the subtraction scheme for the dimension six
operators.

It appears that there is also a logarithmic singularity in $u$
from the other integration boundary, $u_2\to 1$, because $P(u_2)$
is singular at this point. This contribution would be proportional
to $\not\!Q (Q^2/\mu^2)^2$. However, we can always choose the series
for $E_3^\prime$ to start at some sufficiently large $n_0$, so that the
region $u_2\to 1$ is suppressed by a zero of $B[E_3^\prime](u_1)$ and
is non-singular.

It is worth emphasizing that the non-local terms are related to
{\em logarithmic} divergences of the reduced diagrams (diagrams with
insertion of dimension six operators), while their coefficient
is also unambiguously related to the leading singularity of the
original diagram. (As in dimensional regularization a pole in
$\eps$ is accompanied by a logarithm in momenta.) This property
allows us to sum the leading singularities to all orders in $1/N_f$
with a variant of standard renormalization group techniques.

We also note that as long as we consider only the first UV renormalon
at $u=1$, we do not have to consider multiple insertions of dimension
six operators. For instance, for two insertions, to be
sensitive at once, to the singularity at $u(\gm_1)=1$ and at
$u(\gm_2)=1$, we need at least $u=2$, since the $\gm_1$ and
$\gm_2$ can not have common $u$-parameters. The same conclusion
follows from

\be
(\beta_0^n n!)\,(\beta_0^n n!) \sim 2 \left(\frac{\beta_0}{2}\right)^{2n}
(2n)!\,\sqrt{\pi n}
\ee

\n in the language of large-$n$ behaviour.

\subsection{${\cal R}$-operation}

The illustrated counterterm procedure can be generalized to
arbitrary diagrams. Moreover, since according to eq.~(\ref{singofg})
the combinatorial structure of poles in the $u$-parameters that
eventually give rise to singularities in $u$ of the Borel transform
is identical to the combinatorial structure of logarithmic
UV divergences, the `renormalon ${\cal R}$-operation'
can be constructed in analogy with the usual $R$-operation.

Let us recall that the standard $R$-operation applied to
a Feynman integral $F_{\Gm}$ acts as \cite{BP,H,Z,BM}

\be
R\,F_{\Gm} = \sum_{\gm_1, \ldots, \gm_j}
\Dl(\gm_1) \ldots \Dl(\gm_j) F_{\Gm}
\equiv R'\,F_{\Gm} + \Dl(\Gm) \,F_{\Gm} ,
\label{R}
\ee

\n where $\Dl(\gm)$ is the corresponding counterterm operation, and
the sum is over all sets $\{\gm_1, \ldots, \gm_j\}$
of disjoint divergent 1PI subgraphs,
with $\Dl(\emptyset)=1$. The `incomplete' $R$-operation $R'$
by definition includes all the counterterms except the overall
one $\Dl(\Gm)$. Within dimensional renormalization, the action of
the counterterm
operation on the given Feynman integral is described as

\be
\Dl(\gm) \,F_{\Gm} = F_{\Gm / \gm} \circ P_{\gm} ,
\label{CT}
\ee

\n where $F_{\Gm / \gm}$ is the Feynman integral corresponding to the
reduced graph $\Gm / \gm$ and the right-hand side of eq.~(\ref{CT})
denotes the Feynman integral
that differs from  $F_{\Gm / \gm}$ by insertion of the
polynomial $P_{\gm} (\uq,\um)$ into the
vertex $v_{\gm}$ to which the subgraph $\gm$ was
collapsed. In the MS-scheme the
coefficients of these polynomials are represented
as linear combinations of poles in $\eps=(4-d)/2$.
It is implied that dimensional regularization is still
present in eqs.~(\ref{R}) and (\ref{CT}).

In the framework of the renormalization schemes based on dimensional
regularization the $P_{\gm} (\uq,\um)$ are polynomials
with respect to masses $\um$ and external momenta $\uq$  of $\gm$
\cite{Cl} (remember that it is sufficient
to consider the pure massless case in our problem).
The degree of each $P_{\gm}$ equals the degree of divergence
$\om(\gm)$. In the MS scheme these polynomials are defined
recursively by equations

\be
P_{\gm} \equiv  \Dl(\gm) \,F_{\gm} = - \HK_{\eps} R'\,F_{\gm}
\label{Pgm}
\ee

\n for the graphs $\gm$ of a given theory.
Here $\HK_{\eps}$ is the operator
that picks up the pole part of the Laurent series in $\eps$.
Note that the essential part of the basic
theorem on the $R$-operation in the framework of dimensional
renormalization
\cite{BM} is just the above polynomial dependence of diagrammatic
counterterms $P_{\gm} (\uq,\um)$ on masses and external momenta.

Let us now define an operation ${\cal R}$, with the same
combinatorial structure as $R$ in eq.~(\ref{R}), which
removes a (finite) number of terms in the large order behavior of a
given set of diagrams (that is, the leading large-$n$ behaviour,
including $1/n^k$-corrections up to some $k_0$, is removed).
This operation is applied to {\em sets} of
diagrams $\bar{\Gamma}$, defined
as in Sect.~2.1 except that the photon propagators
in $\bar{\Gamma}$ are chains rather than complete photon propagators.
We assume that this set of diagrams includes the usual UV counterterms,
if necessary, so that $G_{\bar{\Gamma}}(\uq;\alpha)$
is a series in $\alpha$ with
finite coefficients, when the regularization is removed. We define
the ${\cal R}$ recursively by

\be
{\cal R}\,G_{\bar{\Gm}} = \sum_{\gm_1, \ldots, \gm_j}
\ODl(\bar{\gm}_1) \ldots \ODl(\bar{\gm}_j) G_{\bar{\Gm}}
\equiv {\cal R}'\,G_{\bar{\Gm}} + \ODl(G) \,G_{\bar{\Gm}}\,,
\label{R1}
\ee
\be
\ODl(\bar{\gm}) \,G_{\bar{\Gm}} = \frac{1}{\mu^2} G_{\overline{\Gm/
\gm}} \circ {\cal P}_{\bar{\gm}} \,,
\label{CT1}
\ee

\n where the sum is over all sets $\{\gm_1, \ldots, \gm_j\}$
of disjoint 1PI skeleton subgraphs with degree of divergence
$\omega(\gm_i)\geq -2$. Every polynomial ${\cal P}_{\bar{\gm}}$ is
represented as a finite sum of
terms of the form

\be
\sum_n e^C\beta_0^n\,n! n^j\ln^l n\,p_{jl} (\uq^{\gm})
\ee

\n where $p_{jl}$ are polynomials of degree $\om (\gm) +2$
in external momenta of $\gm$.
These counterterms ${\cal P}_{\gm}$ are determined by

\be
{\cal P}_{\gm}\equiv \ODl(\bar{\gm}) \,G_{\bar{\gm}}
= - {\cal K} R {\cal R}' G_{\gm}(\uq,\alpha)\,.
\label{fc}
\ee

\n The operator ${\cal K}$ extracts the factorial divergence
(including $1/n^k$-corrections up to the chosen $k_0$) of the
series in $\alpha$ to its right.
The incomplete ${\cal R}$-operation ${\cal R}'$ is described by
eq.~(\ref{R1}).
It is implied that the initial $R$-operation in the MS-scheme
that subtracts the (usual) UV divergences in $\bar{\gm}$ is
implicit in $G_{\bar{\gm}}(\uq,\alpha)$. (Alternatively it could be
included in the definition of ${\cal R}$. In terms of Borel transforms
these subtractions correspond to the subtraction of poles at
$u(\gm)=0$ plus a series in $u(\gm)$.) The insertion of ${\cal P}$
for subgraphs of a given graph generated new UV divergences,
which are subtracted by subsequent application of
$R \equiv R^{\rm MS}$, the usual renormalization in the MS-scheme.
Note that these extra MS-counterterms (associated with the
renormalization of dimension six operators) are not
inserted in the usual way: We have counterterms ${\cal P}_{\bar{\gm}}$
accompanied by such counterterms but do not
have these counterterms alone.

When calculating the counterterms ${\cal P}_{\bar{\gm}}$
with the help of eq.~(\ref{fc}) we introduce
dimensional regularization which will be switched off only
after the action of the operation $R$ but before extracting the
large-order behaviour with ${\cal K}$. As in the usual calculation of
counterterms IR rearrangement  can be used (as in Sect.~3.3)
to simplify the calculation of ${\cal P}_{\bar{\gm}}$,
i.e. we differentiate $\om (\gm) +2$ times in the external momenta,
then put them to zero for all resulting diagrams and
introduce, in the simplest way for the calculation,
a new external momentum.
If there is no way to avoid IR divergences when
following this  prescription,
we can apply the so-called $R^*$-operation \cite{CS} which
removes these
spurious divergences. In this case, we have (after differentiation)
the product $R^* {\cal R}'$  in eq.~(\ref{fc}) instead of $R {\cal R}'$.
Note that it is natural to apply the dimensional $R^*$ because we then
get zero not only for massless vacuum diagrams and tadpoles
themselves but as well for the corresponding
`$R^*$-normalized' values.

We state without detailed proof that the ${\cal P}_{\bar{\gm}}$
will indeed be local (polynomial), so that the corresponding
subtractions can be implemented in terms of dimension six operators
in the Lagrangian. A complete proof would follow the same strategy
as for the usual $R$-operation.

\subsection{Counterterms in the Lagrangian}

The operation ${\cal R}$ was defined at the diagrammatic level. However,
the combinatorial structure of the $R$-operation, eq.~(\ref{R})
together with the locality of the counterterms
enable one to express the
renormalization procedure by inserting counterterms into the Lagrangian.
An explicit way to do this is to apply formulae\footnote{For brevity,
we list the formulae for normally ordered Lagrangians. In the case
without normal ordering which is really implied, the corresponding
generalizations differ by inserting some additional operator --
see details in \cite{CS1,VS}.}
of the
counterterm technique \cite{AZ,CS1,VS}. For the generating functional
of Green functions the basic formula looks like

\be
R \exp\{iS(\phi)\} = \exp\{iS_r (\phi)\},
\label{CTT}
\ee

\n where $\phi$ stands for the collection of all the fields of a given
theory, $S$ is the interaction part of the action and
$S_r$ is the renormalized action (interaction plus counterterms)
which is explicitly expressed through the counterterm operation as

\be
S_r (\phi) = \Dl ( \exp\{iS(\phi)\}-1 ).
\label{sr}
\ee

\n By definition, the $R$-operation and counterterm operations, as applied
to functionals, act on each diagram involved in their diagrammatic
expansions.

Now we observe that the same combinatorial arguments
that resulted in
eqs.~(\ref{CTT}) and (\ref{sr}) can be used to describe the
${\cal R}$-operation in eq.~(\ref{R1}) in the Lagrangian language. We have

\be
{\cal R} \exp\{iS(\phi)\} = \exp\{ i \overline{S}_r (\phi) \} ,
%\label{CTT1}
\;\;\;\;
\overline{S}_r (\phi) = \ODl ( \exp\{ i S(\phi) \} -1 ) ,
\label{sr1}
\ee

\n where application of the counterterm operation $\ODl$ to the functional
in the right-hand side of eq.~(\ref{sr1})
reduces to the action of the diagrammatic counterterm operation
(given by eqs.~(\ref{CT1}) and (\ref{fc})) on
the whole classes of diagrams with vacuum polarization insertions
into the photon propagator.

Eqs.~(\ref{sr1}) provide an explicit realization of the fact that
the renormalons can be compensated by inserting counterterms
into the Lagrangian. The calculation of these counterterms is performed
by use of eq.~(\ref{fc}) -- see examples in the previous section.
Remember that to absorb the large order
behavior into  the Lagrangian we restrict ourselves to the
dimension six finite
extra counterterms while the infinite extra counterterms serve to
renormalize diagrams which are generated by insertion of these
dimension six operators.

\subsection{Renormalization group equations}

Since we do not necessarily want to modify the QED Lagrangian by
higher-dimen\-sional operators, the main relevance
of the statement that
UV renormalons could be compensated in this way is that it
provides information
about the large-$n$ behaviour of perturbative expansions
in the unmodified theory beyond the $1/N_f$-expansion.

To develop the argument it is convenient to adopt a different language
for the counterterms (and we stress that the physical content stays
the same -- in particular it is entirely perturbative, although some
expressions may not appear so). Given the Borel transform $B[G](u)$,
we can formally recover $G(\alpha)$ by

\be
G(\alpha) = \frac{1}{\beta_0}\int\limits_0^\infty d u\,
e^{-u/(\beta_0\alpha)}\,B[G](u)\,.
\ee

\n Let us assume that the perturbative coefficients of $G(\alpha)$
are made real by multiplication of appropriate factors of $i$ so
that $B[G](u)$ is real.
Because of the singularities on the integration contour, starting
at $u=1$, we define the integral with a contour in the complex plane
slightly above the real axis. $G(\alpha)$ acquires an imaginary
part, whose dependence on $\alpha$ is in one-to-one correspondence
with the nature of the singularity. If $B[G](u)$ has a simple pole
at $u=1$,

\be
B[G](u)\,\sim\frac{c}{1-u} \Rightarrow {\rm Im}\,G(\alpha) =
\frac{\pi c}{\beta_0}\,e^{-1/(\beta_0\alpha)}\,.
\ee

\n A double pole gives an additional factor $1/(\beta_0\alpha)$.
Instead of choosing $E_3^\prime$ as series in
$\alpha$ to compensate factorial
divergence, let us now choose $E_3^\prime$ to compensate the imaginary
parts of the Borel integral
generated by factorial divergence. In this language $E_3^\prime$ in
eq.~(\ref{e2}) is replaced by

\be
-\frac{1}{\mu^2}\frac{\pi}{\beta_0}\frac{e^C}{6\pi}\,
e^{-1/(\beta_0\alpha)}\,.
\ee

\n By construction the Green functions computed from the
modified Lagrangian
(i.e. including dimension six counterterms) satisfy

\be
{\rm Im}\,G^{\rm mod}(\alpha) = {\cal O}\!\left(e^{-2/(\beta_0
\alpha)}\right)\,,
\ee

\n where the remaining imaginary part is related to singularities
at $u=2$ and higher. Thus,

\be\label{compensate}
{\rm Im}\,G(\alpha) = -\frac{1}{\mu^2}\sum_i
E_i(\alpha)\,G_{{\cal O}_i}(\alpha) + {\cal O}\!\left(e^{-2/(\beta_0
\alpha)}\right)
\ee

\n for the Green functions in the unmodified theory. Here
$G_{{\cal O}_i}(\alpha)$ denotes the Green function $G$ with one
zero-momentum
insertion of the dimension six operator ${\cal O}_i$. As discussed
in Sect.~5.1 multiple insertions contribute only to singularities
at $u=2$ or higher. A simple equation like the previous one can
not be expected to hold for these singularities.

A Green function $G(\alpha)$, and therefore
${\rm Im}\,G(\alpha)$, and
$G_{{\cal O}_i}(\alpha)$ satisfies a standard renormalization group
equation. Comparing the renormalization group equations for
${\rm Im}\,G(\alpha)$ and $G_{{\cal O}_i}(\alpha)$,
we obtain the evolution equation for
the coefficient functions,

\be\label{oprge}
\left[\left(\mu^2\frac{\partial}{\partial\mu^2} + \beta(\alpha)
\frac{\partial}{\partial\alpha}\right)
\delta_{ij} + \gamma_{ji}(\alpha)\right]
\left(\frac{1}{\mu^2}\,E_j(\alpha(\mu))\right) = 0\,,
\ee

\n where

\be
\gamma_{ij}(\alpha) = \mu^2 \frac{d Z_{ik}}{d\mu^2}\,(Z^{-1})_{kj}\, ,
\qquad
{\cal O}_i^{\rm ren} = Z_{ij} {\cal O}_j^0
\ee

\n are the anomalous dimension matrix and renormalization constants
for the set of dimension six operators. (We assume dimensional
renormalization so that operators with different dimension do not
mix.) Since $E_i(\alpha(\mu))$ does not depend on $\mu$ explicitly,
we arrive at

\be\label{ordrge}
\left[\left(\beta(\alpha)\frac{d}{d\alpha}-1\right) \delta_{ij}
+\gamma^T_{ij}(\alpha)\right]
E_j(\alpha) = 0\,,
\ee

\n where $\gm^T$ denotes the transpose of $\gm$. The solution is
given by

\be\label{sol}
E_i(\alpha(\mu)) = \exp\left(\int\limits_{\alpha_0}^{\alpha(\mu)}
d\alpha'\frac{\delta_{ij}-\gamma^T_{ij}(\alpha')}{\beta(\alpha')}
\right)\hat{E}_j
\ee

\n and $\alpha$-independent initial conditions $\hat{E}_j$.
Consequently the $\alpha$-dependence of ${\rm Im}\,G(\alpha)$
and therefore the $n$-dependence of the large-$n$ behaviour of
the perturbative expansion of $G(\alpha)$ is completely determined by
renormalization group considerations. (To obtain the
complete $1/n^k$-corrections to the leading large-$n$ behaviour one
has to compute $G_{{\cal O}_i}(\alpha)$ to order $\alpha^k$ in
addition.) The initial conditions
$\hat{E}_j$ provide the overall normalization, which has to
be determined, for each operator, from a suitably chosen
Green function. This
calculation can be done systematically only through expansion
in $1/N_f$. A solution of form (\ref{sol}) has been obtained
previously by Parisi \cite{PAR78} and employed in \cite{BEN95a} in a
heavy quark effective theory context. Similar ideas, although in a
technically somewhat different set-up, were used by Vainshtein
and Zakharov \cite{VAI94} to find the asymptotic behaviour of a
current two-point function in a QED-like model.

Let us now discuss the set of operators ${\cal O}_i$. Since we
consider operator insertions into Green functions, which can have
off-shell external momenta, we keep
operators which vanish by equations of motion as well as
gauge-variant operators in the general case. However, not all
operators are of importance for any given Green function $G$,
since the contribution to ${\rm Im}\,G$ from an operator ${\cal O}_i$
with coefficient $E_i$ (which is independent of $G$) can have
additional factors of $\alpha$ from $G_{{\cal O}_i}(\alpha)$, which
lead to $1/n$-suppression in the contribution to the large-order
behaviour of $G$.

Specifically, in the example to follow in Sect.~5.5, we consider the
operators

\bea\label{opset}
{\cal O}_1 &=& \bar{\psi} \gamma_\mu\gamma_5\psi \bar{\psi} \gamma^\mu
\gamma_5\psi
\\
{\cal O}_2 &=& \bar{\psi} \gamma_\mu\psi \bar{\psi} \gamma^\mu\psi\,,
\nonumber\\
{\cal O}_3 &=& g \bar{\psi}\gamma_\mu\psi \left(D_\nu F^{\nu\mu} +
g \bar{\psi}\gamma^\mu\psi\right)
\nonumber\\
{\cal O}_4 &=& g^2 D_\nu F^{\nu\rho} \left(D^\mu F_{\mu\rho} +
g \bar{\psi}\gamma_\rho\psi\right)
\nonumber
\eea

\n Note that ${\cal O}_3$ and ${\cal O}_4$ vanish by the equation of
motion for the photon field, but should be kept, because for the
vertex function and vacuum polarization we consider an off-shell
external photon. We omit the operators that vanish by the equation of
motion for the fermion field from the above list, because in the
examples to follow we assume the external fermion lines to be on-shell
for simplification. We have also omitted the gauge-variant operators.

\subsection{Asymptotic behaviour}

\subsubsection{Solution to the renormalization group equation}

We now solve the renormalization group
equations, eq.~(\ref{oprge}), which allows us to derive a
`renormalization group improved' form for the large-order behaviour
of the vertex function and vacuum polarization which replaces
the next-to-leading $1/N_f$-result given in Sect.~4. This
improved form sums all $(\ln n/N_f)^k$-contributions to all orders
in $1/N_f$ and provides the correct asymptotic behaviour in $n$ with
corrections being of order $1/n$ or $1/N_f$.  In the case of the
vertex function, we consider only the case when the external
fermions are on-shell, so that the Dirac structures involving
$p$ and $p^\prime$ in eq.~(\ref{leadvertex}) can be neglected. As a
consequence, eq.~(\ref{opset}) exhausts the list of all possible
dimension six operators, since we have to consider zero-momentum
insertions only and restrict ourselves to gauge-invariant
quantities.

In the operator basis chosen above, the anomalous dimension matrix
is block triangular, since the operators that vanish by the
equation of motion do not mix into those that don't. The result for
the anomalous dimension matrix is

\be\label{andimmatrix}
-\gamma^T_{ij}(\alpha) =
\left(
\begin{array}{cccc}
0\cdot\alpha & \frac{3\alpha}{2\pi} & 0 & 0\\[0.1cm]
\frac{11\alpha}{6\pi} & \frac{2 N_f+1}{3}\frac{\alpha}{\pi}
& 0 & 0\\[0.1cm]
-\frac{1}{12\pi^2} & -\frac{2 N_f+1}{12\pi^2}
& \frac{N_f\alpha}{3\pi} & 0\cdot \alpha^2\\[0.1cm]
0 & 0 & \frac{N_f}{12\pi^2} & \frac{N_f\alpha}{3\pi}
\end{array}
\right)
\,.
\ee

\n All entries are given to lowest order in $\alpha$ for each
entry. Corrections of order $\alpha$ will generate only
$1/n$-corrections to the asymptotic behaviour in $n$ and we do
not consider such corrections. Eq.~(\ref{ordrge}) is
easily solved with this anomalous
dimension matrix. It is possible and advantageous to
choose the solution such that $\alpha$ appears only in the
combination $a\equiv\beta_0\alpha$. We first solve
for $E_1$ and $E_2$ by
diagonalizing the $2\times 2$-matrix pertaining to operators
${\cal O}_1$ and ${\cal O}_2$.
Then we integrate to obtain first $E_3$ and then $E_4$. Introducing

\be\label{eigenvalues}
\lambda_{1,2}=\frac{2 N_f+1}{6}\pm\sqrt{\left(\frac{2 N_f+1}{6}
\right)^2+\frac{11}{4}}\qquad\quad
\gm_i=-\frac{\lambda_i}{\pi\beta_0}\,,
\ee

\n we obtain

\bea\label{ee1}
E_1(\alpha) &=& e^{-1/a} a^{-\beta_1/\beta_0^2}\,(1+{\cal O}(a))
\left(\left(\frac{2 N_f+1}{3}\right)^2+11\right)^{-1/2}\nonumber\\
&&\times\left(-\lambda_2 a^{-\gm_1} \left[\hat{E}_1+\frac{6\lambda_1}
{11}\hat{E}_2\right] + \lambda_1 a^{-\gm_2} \left[\hat{E}_1
+\frac{6\lambda_2}{11}\hat{E}_2\right]\right) ,
\eea
\bea\label{ee2}
E_2(\alpha) &=& e^{-1/a} a^{-\beta_1/\beta_0^2}\,(1+{\cal O}(a))
\left(\left(\frac{2 N_f+1}{3}\right)^2+11\right)^{-1/2}\nonumber\\
&&\times\left(\frac{11}{6} a^{-\gm_1} \left[\hat{E}_1+\frac{6\lambda_1}
{11}\hat{E}_2\right] - \frac{11}{6} a^{-\gm_2} \left[\hat{E}_1
+\frac{6\lambda_2}{11}\hat{E}_2\right]\right) ,
\eea
\bea\label{ee3}
E_3(\alpha) &=& e^{-1/a} a^{-\beta_1/\beta_0^2}\,(1+{\cal O}(a))
\Bigg[\frac{1}{a}\,\hat{E}_3 +
\left(\left(\frac{2 N_f+1}{3}\right)^2+11\right)^{-1/2}\nonumber\\
&&\times\Bigg\{\frac{\gm^T_{31}}{a}
\left(-\frac{\lambda_2}{\gm_1} a^{-\gm_1}
\left[\hat{E}_1+\frac{6\lambda_1}
{11}\hat{E}_2\right] + \frac{\lambda_1}{\gm_2} a^{-\gm_2}
\left[\hat{E}_1 +\frac{6\lambda_2}{11}\hat{E}_2\right]\right)
\nonumber\\
&&\,+\frac{\gm^T_{32}}{a}
\left(\frac{11}{6\gm_1} a^{-\gm_1}
\left[\hat{E}_1+\frac{6\lambda_1}
{11}\hat{E}_2\right] - \frac{11}{6\gm_2} a^{-\gm_2}
\left[\hat{E}_1 +\frac{6\lambda_2}{11}\hat{E}_2\right]\right)
\Bigg\}\Bigg] ,
\eea
\bea\label{ee4}
E_4(\alpha) &=& e^{-1/a} a^{-\beta_1/\beta_0^2}\,(1+{\cal O}(a))
\Bigg[\frac{1}{a}\,\hat{E}_4 + \frac{\gm^T_{43}}{a^2}\,\hat{E}_3 +
\left(\left(\frac{2 N_f+1}{3}\right)^2+11\right)^{-1/2}\nonumber\\
&&\hspace*{-1.8cm}\times\Bigg\{\frac{\gm^T_{43}}{a}\frac{\gm^T_{31}}{a}
\left(-\frac{\lambda_2}{\gm_1 (1+\gm_1)} a^{-\gm_1}
\left[\hat{E}_1+\frac{6\lambda_1}
{11}\hat{E}_2\right] + \frac{\lambda_1}{\gm_2 (1+\gm_2)} a^{-\gm_2}
\left[\hat{E}_1 +\frac{6\lambda_2}{11}\hat{E}_2\right]\right)
\\
&&\hspace*{-1.8cm}\,+\frac{\gm^T_{43}}{a}\frac{\gm^T_{32}}{a}
\left(\frac{11}{6\gm_1 (1+\gm_1)} a^{-\gm_1}
\left[\hat{E}_1+\frac{6\lambda_1}
{11}\hat{E}_2\right] - \frac{11}{6\gm_2 (1+\gm_2)} a^{-\gm_2}
\left[\hat{E}_1 +\frac{6\lambda_2}{11}\hat{E}_2\right]\right)
\Bigg\}\Bigg] .\nonumber
\eea

\n These results can be translated into the original language, where
the $E_i(\alpha)$ were factorially divergent series, by the rule

\be\label{rule}
e^{-1/a}\,a^b \longrightarrow
\frac{1}{\pi}\sum_n \beta_0^n\,
n! n^{-b}\,.
\ee

\n One can then determine the large-order behaviour of any given
Green function directly from

\be
G(\alpha) \stackrel{n\to\infty}{=} -\frac{1}{\mu^2}\sum_i
E_i(\alpha)\,G_{{\cal O}_i}(\alpha)\,.
\ee

\subsubsection{Matching}

It remains to determine the integration constants $\hat{E}_i$
by matching the solution to the renormalization group equation above
with the results of explicit calculations in the chain expansion.
We do this interpreting $E_i(\alpha)$ as
factorially divergent series in $\alpha$.
Comparison with explicit calculation provides a non-trivial
consistency check as it must be reproducible with four
{\em $n$-independent} constants $\hat{E}_i$.\\

{\bf Four-fermion scattering.} Since the insertion of
${\cal O}_3$ and ${\cal O}_4$ into the four-point function vanishes,
the four-point function allows us to determine $\hat{E}_1$ and
$\hat{E}_2$. As mentioned before, we consider on-shell external
fermion legs. Most of the ingredients for the calculation of
four-fermion scattering to next-to-next-to-leading (NNLO) order
in $1/N_f$ are scattered among previous results and we quote
results only after putting them together.

At leading order, we have a tree diagram with a single chain. It
does not yield divergences in large orders. At next-to-leading order,
the set of 1PI diagrams consists of the box diagrams. From
eq.~(\ref{boxes}) we obtain

\be
r_n^{\rm 1PI,NLO} \stackrel{n\to\infty}{=} -\frac{1}{\mu^2} e^C
\,2\left(-\frac{9\pi}{N_f}\right)\beta_0^n n!\,\gm_\mu\gm_5
\otimes\gm^\mu\gm_5\,,
\ee

\n where $r_n$ is the coefficient of $i\alpha^{n+1}$ in the perturbative
expansion and the factor of two comes from an equal contribution
of scattering and annihilation type diagrams. There are two contributions
from one-particle reducible (1PR) diagrams. One from a vertex-correction
to the LO diagram and the other from inserting a chain into one of
the fermion loops of the chain in the LO diagram. Each contribution
individually gives $(n-1)!$ for large $n$, but this leading terms
cancels in the sum of both by the cancellation discussed at the
end of Sect.~3. As a result we find

\be
r_n^{\rm 1PR,NLO} \stackrel{n\to\infty}{=} -\frac{1}{\mu^2} e^C
\,{\rm const}\cdot\beta_0^n (n-2)!\,\gm_\mu\otimes\gm^\mu\,,
\ee

\n where the constant is determined by the $1/n$-correction
to the asymptotic behaviour of the leading order vacuum polarization,
see e.g. Sect.~5.1. Compared to the 1PI diagrams, the 1PR ones are
suppressed by $1/n^2$ for large $n$.

 To compare this with the solution to the renormalization group equation,
we expand eqs.~(\ref{ee1}) and (\ref{ee2}) in $1/N_f$ and note that
up to $1/n$-corrections, the large-order behaviour of the four-point
function is given by

\be
-\frac{4}{\mu^2}\left(E_1(\alpha)\,\gm_\mu\gm_5
\otimes\gm^\mu\gm_5 + E_2(\alpha)\,\gm_\mu\otimes\gm^\mu\right)\,.
\ee

\n With

\be \gm_1=-2-\frac{1}{N_f}-\frac{99}{8 N_f^2}+\cdots
\qquad\quad
\gm_2=\frac{99}{8 N_f^2}+\cdots
\ee

\n we get

\be
E_1(\alpha) = \sum_n\beta_0^n n! \alpha^{n+1}\,\hat{E}_1
\qquad\quad
E_2(\alpha) = \sum_n\beta_0^n (n-2)! \alpha^{n+1}\,\hat{E}_2\,,
\ee

\n where the $1/n^2$-suppression of $E_2$ follows from
$\gm_1=-2$ at leading order. By comparison with the NLO
calculation one determines

\be
\hat{E}_1=-\frac{9\pi}{2 N_f} e^C\,,
\ee

\n while $\hat{E}_2$ is related to the $1/n$-correction to the
leading order vacuum polarization.

At NNLO we separate again the 1PI from the 1PR diagrams. All
terms that involve $\gm_\mu\otimes\gm^\mu$ have been evaluated
in Sect.~4.1.2. All other 1PI diagrams that did not contribute
to the vertex function in Sect.~4.1.2 are proportional
to $\gm_\mu\gm_5\otimes\gm^\mu\gm_5$ in large orders and do not
yield logarithmic enhancement. We can summarize their contribution
by a constant $D_{\rm box}$ in order to obtain

\bea\label{NNLO1PI}
r_n^{\rm 1PI,NNLO} &\stackrel{n\to\infty}{=}& -\frac{1}{\mu^2}\,e^C
\,2\left(-\frac{9\pi}{N_f}\right)\beta_0^n n!
\Bigg[\left\{\frac{\beta_1}{\beta_0^2}\,(\ln n-\psi(1)) +
\frac{D_{\rm box}}{N_f}\right\} \gm_\mu\gm_5
\otimes\gm^\mu\gm_5
\nonumber\\
&& \,+\left\{-\frac{9}{2 N_f}\,(\ln n-\psi(1)) + \frac{9 C_{\rm box}}
{2 N_f}\right\} \gm_\mu\otimes\gm^\mu\Bigg]
\,,
\eea

\n with $C_{\rm box}$ as in Sect.~4.1.2. The 1PR diagrams involve
insertions of the next-to-leading order results for the vertex
function and vacuum polarization collected in Sect.~4. There is
only a partial cancellation between vertex and vacuum polarization
insertions and the 1PR diagrams are not suppressed by $1/n^2$
in comparison to the 1PI ones any more:

\be\label{NNLO1PR}
r_n^{\rm 1PR,NNLO} \stackrel{n\to\infty}{=} -\frac{1}{\mu^2}\,e^C
\,2\left(-\frac{9\pi}{N_f}\right)\beta_0^n n!
\left\{\frac{9}{2 N_f}\,(\ln n-\psi(1)) - \frac{9 C_{\rm box}}
{2 N_f} - \frac{11}{4 N_f} \right\}\gm_\mu\otimes\gm^\mu \, .
\ee

\n In the sum of 1PR and 1PI contributions only the last term
survives in the $\gm_\mu\otimes\gm^\mu$-structure. Expanding
eqs.~(\ref{ee1}) and (\ref{ee2}) to NNLO, we have

\bea
E_1(\alpha) &=& \sum_n\beta_0^n n! \alpha^{n+1}
\left[\hat{E}_1 \left(1+\frac{\beta_1}{\beta_0^2}\ln n
\right)-\frac{9}{4 N_f}
\hat{E}_2\right] \, ,
\nonumber\\
E_2(\alpha) &=& \sum_n\beta_0^n n! \alpha^{n+1}
\left[-\frac{11}{4 N_f}\hat{E}_1\right]\,,
\eea

\n neglecting $1/n$-corrections. Given $\hat{E}_1$ above,
$E_2(\alpha)$ is completely determined and coincides with the
$\gm_\mu\otimes\gm^\mu$-term in the explicit calculation, the
sum of eqs.~(\ref{NNLO1PI}) and (\ref{NNLO1PR}) -- a highly
non-trivial consistency check. The comparison
of the $\gm_\mu\gm_5\otimes\gm^\mu\gm_5$-term determines the
$1/N_f^2$-corrections to $\hat{E}_1$.

Although the initial conditions can only be determined order by
order in $1/N_f$, the $n$-dependence in large-orders is
determined already by the anomalous dimension matrix. Since
$\gm_2 > \gm_1$, the final result for four-fermion scattering
can be written as

\bea
r_n &\stackrel{n\to\infty}{=}& -\frac{1}{\mu^2}\,e^C
\,\beta_0^n n!\,n^{\beta_1/\beta_0^2+\gm_2}\left(1+{\cal O}\!
\left(\frac{1}{n}\right)\right)
\nonumber\\
&&\hspace*{-1cm}\times
\,2\left(-\frac{9\pi}{N_f}\right) \left(1+{\cal O}\!
\left(\frac{1}{N_f}\right)\right)
\left\{\gm_\mu\gm_5\otimes\gm^\mu\gm_5-\frac{11}{4 N_f}\,
\gm_\mu\otimes\gm^\mu\right\}\,.
\eea

\n For $N_f=1$, $\beta_1/\beta_0^2+\gm_2=4.48\ldots$. In the above
equations all terms $(\ln n/N_f)^k$ are summed to all orders in
$1/N_f$ and corrections are of order $1/n$ and $1/N_f$ to the
normalization as indicated.\\

{\bf Vacuum polarization.} The asymptotic behaviour up to $1/n$-corrections
follows from insertion of ${\cal O}_4$ into the photon two-point function,
which results in

\be \label{vacins}
2 g^2 \left(-\frac{Q^2}{\mu^2}\right) (Q^2 g_{\mu\nu}-Q_\mu Q_\nu)\,
E_4(\alpha)\,.
\ee

\n Again we check that the renormalization group equation sums
correctly the $\ln n$-terms in the finite-order $1/N_f$-result. Since
we did not compute $1/n$-corrections to the leading asymptotic behaviour,
we can drop the terms containing
$n^{\gm_1}\sim1/n^2$ ($a^{-\gm_1}$ in eq.~(\ref{ee4}))
in comparison to those with $n^{\gm_2}$. Furthermore,
since $\lambda_2\sim 1/N_f^2$ and $\hat{E}_1$ and $\hat{E}_2$ are of the
same order in $1/N_f$, we may also neglect all terms that contain
$\lambda_2 \hat{E}_2$ in the remaining terms. With these simplifications

\bea
E_4(\alpha) &=& \sum_n\beta_0^n n! n^{\beta_1/\beta_0^2}
\left(1+{\cal O}\left(\frac{1}{n}\right)\right)
\Bigg[n\,\hat{E}_4 + \gm^T_{43}\,n^2\,\hat{E}_3
\nonumber\\
&&\,+\left(\left(\frac{2 N_f+1}{3}\right)^2+11\right)^{-1/2}
\gm^T_{43} \left\{\gm^T_{31}\lambda_1-\gamma^T_{32}\frac{11}{6}
\right\}\,\frac{n^2 n^{\gm_2}}{\gm_2 (1+\gm_2)}\,\hat{E}_1
\Bigg]\,.
\eea

\n Notice the term $n^{\gm_2}/\gm_2$, which, upon expansion in $1/N_f$,
produces

\be
n\,{\rm -\,independent} + \ln n +{\cal O}\!\left(\frac{1}{N_f}\right)\,
\ee

\n with $n$-independent terms proportional to $N_f^2$. These terms would be
absent, had we expanded the renormalization group equation first in
$1/N_f$ and then integrated it. In this case we would have obtained
$\ln n$ for $n^{\gm_2}/\gm_2$, when it occurs in $E_3(\alpha)$ and
$\ln n+1$, when it occurs in $E_4(\alpha)$. (The additional `$+1$' comes
from expansion of $1/(1+\gm_2)$ in $E_4(\alpha)$, when the expansion
in $1/N_f$ is performed after integration of the renormalization
group equation.) The terms
proportional to positive powers of $N_f$ are indeed irrelevant and can be
absorbed into a redefinition of $\hat{E}_3$. In the following this
redefinition will be understood implicitly, so that we may substitute
$n^{\gm_2}/\gm_2$ by $\ln n$ (in $E_3(\alpha)$) and $\ln n+1$
(in $E_4(\alpha)$) to the order in $1/N_f$, where we can
compare with explicit calculation. Finally, we obtain

\be
E_4(\alpha) = \sum_n\beta_0^n n!\,n \left(1+\frac{\beta_1}{\beta_0^2}
\ln n\right) \Bigg[\hat{E}_4 + \frac{N_f}{12\pi^2}\,n\,\hat{E}_3
+\frac{1}{2}\,
e^C\,\frac{N_f}{36\pi^3} \frac{81}{8 N_f}\,n\,(\ln n+1)\Bigg]\,,
\ee

\n where the leading order result for $\hat{E}_1$ has been used. Explicit
calculation, including the light-by-light scattering diagrams, yielded

\bea
r_n &\stackrel{n\to\infty}{=}&
g^2 \left(-\frac{Q^2}{\mu^2}\right) (Q^2 g_{\mu\nu}-Q_\mu Q_\nu)\,
\beta_0^n n!\, n \left(1+\frac{\beta_1}{\beta_0^2}\ln n
\right)
\nonumber\\
&&\times\frac{N_f}{36\pi^3} \left[1-\frac{9 n}{8 N_f}
-\frac{81 n}{8 N_f}\left\{-\ln n+\psi(3)+C_{\rm box}\right\}
\right]
\eea

\n as coefficient of $i\alpha^{n+1}$. Comparison of the
previous two equations shows that the $\ln n$-term is indeed
correctly reproduced (recall the factor of two in eq.~(\ref{vacins}))
and the others can be accounted for, if

\be
\hat{E}_4=\frac{N_f}{72\pi^3}\, e^C
\ee

\n and the terms proportional to $n$ in square brackets, including
$C_{\rm box}$ from light-by-light scattering, are absorbed into
$\hat{E}_3$. The correct asymptotic behaviour of the vacuum
polarization is

\be\label{asympvacpol}
r_n \stackrel{n\to\infty}{=} {\rm const}\times \beta_0^n n!\,
n^{2+\beta_1/\beta_0^2+\gm_2}\,,
\ee

\n the constant again being determined only as expansion in $1/N_f$.

Our solution for $E_4(\alpha)$ that determines
the asymptotic behaviour of the vacuum polarization differs
from \cite{VAI94}, because we consider a different theory, QED with $N_f$
fermions of identical charge. In \cite{VAI94}, the condition
$\sum_i Q_i=0$ on the charges of the fermions has been imposed, so
that the light-by-light contributions to the vertex and vacuum
polarization vanish. This condition modifies the anomalous dimension
matrix and consequently the solution for $E_i(\alpha)$ for $i=3,4$.
As far as we can tell, our result would coincide with \cite{VAI94}, if
the same condition were imposed, although our derivation and formalism
is technically different (but physically equivalent).\\

{\bf Vertex function.} Since all $\hat{E}_i$ are now fixed (to leading order
in $1/N_f$), we obtain a prediction for the vertex function without free
parameters. For the three-point function, the 1PR diagrams with an
insertion of ${\cal O}_4$ into the `external' photon line are not
suppressed for large $n$, since $E_4(\alpha)/E_3(\alpha)\sim n$. The
large-order behaviour is given in terms of $E_3(\alpha)$ and
$E_4(\alpha)$ by

\be
-g \frac{1}{\mu^2}\,(Q^2\gamma_\mu-\!\not\!Q Q_\mu)\,[E_3(\alpha)-g^2
E_4(\alpha)]\,.
\ee

\n After expansion in $1/N_f$, the terms involving $\hat{E}_i$ for
$i=1,2,3$ almost cancel in the combination $E_3(\alpha)-g^2
E_4(\alpha)$, the difference arising entirely from the different
interpretation of $n^{\gm_2}/\gm_2$ for $E_3(\alpha)$ and
$E_4(\alpha)$, when expanded in $1/N_f$, as mentioned before. We are left
with

\be
r_n \stackrel{n\to\infty}{=} -g \frac{1}{\mu^2}\,e^C\,\frac{1}{6\pi}\,
(Q^2\gamma_\mu-\!\not\!Q Q_\mu)\,\beta_0^n n!\,\alpha^{n+1}
\left(1+\frac{\beta_1}{\beta_0^2}\ln n
\right)\left[1-\frac{81 n}{8 N_f}\right]\,.
\ee

\n An analogous cancellation between 1PI and 1PR diagrams occurs in
the explicit calculation to NLO in $1/N_f$, for which we obtain

\bea
r_n &\stackrel{n\to\infty}{=}& -g \frac{1}{\mu^2}\,e^C\,
(Q^2\gamma_\mu-\!\not\!Q Q_\mu)\,\beta_0^n n!
\left(1+\frac{\beta_1}{\beta_0^2}\ln n
\right)\alpha^{n+1}
\nonumber\\
&&\times\,\Bigg\{\left[-\frac{1}{6\pi}+\frac{3 n}{8\pi N_f} +
\frac{27 n}{8\pi N_f}\, (-\ln n+\psi(2)+C_{\rm box})\right]
\nonumber\\
&&+\,\left[\frac{1}{3\pi}-\frac{3 n}{8\pi N_f} -
\frac{27 n}{8\pi N_f}\, (-\ln n+\psi(3)+C_{\rm box})\right]
\Bigg\}
\\
&=&\,-g \frac{1}{\mu^2}\,e^C\,\frac{1}{6\pi}\,
(Q^2\gamma_\mu-\!\not\!Q Q_\mu)\,\beta_0^n n!\,\alpha^{n+1}
\left(1+\frac{\beta_1}{\beta_0^2}\ln n
\right)\left[1-\frac{81 n}{8 N_f}\right]\,.
\nonumber
\eea

\n The second line comes from the vertex function in Sect.~4.1.3 and the third
from insertion of the vacuum polarization (Sect.~4.2.2) into the external
line. The sum is in agreement with the previous result. To obtain the
1PR vertex function, one would have to amputate the external photon line
and the corresponding contribution to the large-$n$ behaviour that comes
from the photon vacuum polarization.

%%%%%%%%%%%%%%%%%%%% SECTION 6 %%%%%%%%%%%%%%%%%%%%%%%%%%%%%%%

\section{Conclusion and outlook}
\setcounter{equation}{0}

We have described a systematic approach to factorial
divergence in large orders of perturbation theory
generated by large-momentum regions of loop integrations.
In this approach we
reorganize the full perturbation series in terms of diagrams
with a fixed number of chains, equivalent to an expansion in $1/N_f$.
We can then exploit a standard technique applied for analysis
of UV/IR divergences and singularity structure of
analytically regularized Feynman amplitudes to classify
the UV renormalon singularities of the Borel transform for any given
class of diagrams. The problem of finding the overall normalization
$K$ of the large-order behaviour reduces to calculating the residues
of these singularities. The coefficient of the strongest singularity
can be found algebraically by repeated contraction of
one-loop subgraphs, but provides information
on anomalous dimensions only. A non-trivial
contribution to $K$ requires calculation of part of the next-to-leading
singularity. However this part is local so that it can be obtained with
the help of infrared rearrangement. Still, diagrams with an arbitrary
number of chains contribute to the overall normalization.

Using the similarity of our UV renormalon calculus with the analysis
of usual logarithmic UV divergences, we have shown that the
contributions from UV renormalons to the large-order behaviour are
naturally characterized as insertions of higher-dimensional operators
into Green functions, a hypothesis originally forwarded by
Parisi \cite{PAR78}. These insertions assume the form
$\sum_i E_i(\alpha) G_{{\cal O}_i}(\uq;\alpha)$, where the $E_i$
are independent of external momenta
and account for factorial divergence from loop momentum
regions $\underline{k}\gg\uq$, where all loop momenta are much
larger than the external momenta. On the other hand higher-order corrections
in $\alpha$ to the Green function with operator insertion
$G_{{\cal O}_i}(\uq;\alpha)$ take into account subleading contributions
in $n$ for large $n$ from loop momentum regions, where at least one
of the loop momenta is of order of the external momenta.

This systematic organization comes at a price, the introduction of an
artificial expansion parameter $1/N_f$. In particular, diagrams with
a larger number of chains (and thus suppressed in $1/N_f$) are not
suppressed for large $n$, the order of perturbation theory. Quite
to contrary, they are typically enhanced by a factor of $\ln n$ per chain,
so that any finite order in the $1/N_f$-expansion does not provide
the correct asymptotic behaviour in $n$ in the full theory for any
$N_f$. However, just as renormalizability of the theory allows
us to go beyond a finite-order expansion in $\alpha$ using
renormalization group equations, the fact that UV renormalons can be
compensated by counterterms of higher-dimensional operators to any
order in $1/N_f$ allows us to transcend a fixed-order expansion
in $1/N_f$, using low-order results only. The application of
renormalization group ideas then leads to resummation of
$(\ln n/N_f)^k$-terms and the correct large-$n$ behaviour is found
in terms of anomalous dimensions of dimension-six operators (see
eq.~(\ref{asympvacpol}) for the photon vacuum polarization). The
overall normalization is determined by matching order by order
in $1/N_f$. The behaviour of the $1/N_f$-expansion for this constant
remains a problem that can not be addressed within this framework.
Because the overall normalization remains undetermined for all
practical purposes in the realistic situation ($N_f\sim 1$), a
modified QED without UV renormalons (quite similar to the
suggestion of \cite{GRU94}) -- and therefore potentially Borel
summable -- resembles in effect a non-renormalizable theory. For each
higher-dimensional operator an unknown constant $\hat{E}_i$ has to be
introduced. Of course, this is just the statement that for all
practical purposes QED can be considered as an effective theory
below the Landau pole, not withstanding its hypothesized `triviality'.

On the technical side, we have somewhat glossed over the details of two
issues. The first is the calculation of $1/n$-corrections to the
leading asymptotic behaviour. Here one could also use IR rearrangement,
provided one properly subtracts the non-local terms that arise from
subdivergences. This procedure can lead to a non-trivial interplay
of analytic and dimensional regularizations. A similar interplay affects
the second issue of how to implement the correct $\overline{\rm MS}$
subtractions (or any other) implied to obtain the renormalized
perturbation series in the first place. In QED this issue concerns only
the overall subtractions for the photon vacuum polarization. While
these subtractions do not involve factorial divergence at the level
of a single chain, their insertion as counterterm into more complicated
diagrams contributes to factorial divergence at the level of
$1/n$-corrections to the leading asymptotic behaviour of diagrams with
two or more chains. Thus both issues must be considered when computing
$1/n$-corrections, but do not affect the status of the overall
normalization $K$, which is independent of subtractions
up to the trivial factor $e^C$.

There exist several obvious extensions of this work. The techniques
developed here are applicable without modification to the next
UV renormalons ($u=2,3,\ldots$), which are related to operators
with dimension larger than six. However, one has to cope with a
significant increase in complexity, because multiple insertions of
dimension six (and in general, lower-dimensional) operators contribute
and the Borel transform is a multi-valued function. When applying the
formalism the Heavy Quark Effective Theory (to be precise, an abelian
version of it), owing to the different power-counting for the
static quark propagator, one can obtain singularities at half-integer
values of $u$. A special feature of this theory is that the self-energy
is linearly UV divergent. If one repeats the analysis of the heavy
quark self-energy \cite{BEN95a} along the lines of Sect.~3, one finds
that the terms $N_1^c$ and $N_1^d$, not computed in \cite{BEN95a},
add to zero.

A natural extension would include infrared renormalons. The
corresponding poles at negative $u$ can again be classified in
terms of singularities of analytically regularized integrals, now from
regions of small loop momentum flow through chains. The modifications
of our formalism arise only from the different notions of irreducibility
of subgraphs in the infrared and ultraviolet. Thus 1PI (UV-irreducible)
subgraphs in Sect.~3.1 should be replaced by IR-irreducible subgraphs
\cite{VS}, (UV) forest by IR forests etc. For the general
structure of large-$n$ behaviour associated with diagrams with
multiple chains, we expect little difference from the ultraviolet
case. In the language of counterterms in the Lagrangian, IR renormalons
would be interpreted as non-local operators \cite{PAR79}, since
IR renormalons are local in momentum space (while UV renormalons
are local in coordinate space).

The ultimate goal is still a diagrammatic understanding of renormalons
in non-abelian gauge theories, that is QCD. Given that even in QED
with $N_f\sim 1$ there is little to say about the overall normalization
of large-order behaviour, there is little to hope that one could obtain
more than the exponent $b$ in the expected behaviour

\be
K\,{\beta_0^{\rm NA}}^n \,n! n^b\,.
\ee

\n Already this form poses the challenge to understand, on a diagrammatic
level, how the contributions from many diagrams with many loops combine
to produce the first coefficient $\beta_0^{\rm NA}$ of the non-abelian
$\beta$-function, which, contrary to QED, is not related to
loop-insertions into the gluon propagator. Although the physics described
by non-abelian gauge theories changes discontinuously with $N_f$ for
sure, the structure of Feynman diagrams does not, so that we may still
try to approach the question from large $N_f$, provided we consider
the expansion in $N_f$ as formal. Writing $\beta_0^{\rm NA} =
\beta_0+\delta \beta_0^{\rm NA}$, where $\beta_0$ denotes the abelian
part, the previous equation is expanded as

\be
K^{[1]} \,\beta_0^n \,n! n^{b^{[1]}} \left\{1+
\left[\delta \beta_0^{\rm NA} n + \delta b^{[2]} \ln n + \delta
K^{[2]}+\ldots\right]\right\}\,,
\ee

\n so that for each chain (defined as in the abelian theory as gluon
propagator with fermion loops) we need an additional enhancement of
$n$ (rather than $\ln n$) that combines exactly to ${\beta_0^{\rm NA}}^n$
to all orders in $1/N_f$. In QCD, contrary to QED (where these
contributions cancel between various diagrams due to the Ward identity),
one also has to take into account the possibility of first inserting
a (usual) UV counterterm and then an UV renormalon
counterterm, in addition to the situation discussed at length where
one first inserts a UV renormalon counterterm and then a (usual)
UV counterterm, related to renormalization of a dimension-six
operator. A more detailed investigation reveals that
the first contributions combine to produce the term
$\delta \beta_0^{\rm NA} n$ above, while the others are interpreted
in terms of dimension-six operators just as in the abelian theory.

\vspace*{0.5cm}
{\bf Acknowledgments.} We would like to thank Gerhard Buchalla,
Konstantin Chetyrkin,
Arkady Vainshtein and Valentine Zakharov for numerous helpful discussions
during the course of this work. We are especially
thankful to Arkady Vainshtein for
his insistent criticism of an earlier version of Sect.~5.5.
We are grateful to Wolfgang Hollik,
Thomas Mannel and the particle theory group at the University of
Karlsruhe for their hospitality, when part of this work was done.
We thank the organizers of the Ringberg Workshop `Advances in perturbative
and non-perturbative techniques', where this work was started, and
the organizers of the Aspen Workshop on `Higher order corrections
in the Standard Model', where this work was
(almost) completed, for
inviting us both. Thanks to the Aspen Center for Physics for
creating an enjoyable atmosphere during the
Workshop. The work of M.~B. is supported by
the Alexander von Humboldt-foundation.

\appendix

\section{Singularity structure of analytically regularized Feynman integrals}
\setcounter{equation}{0}

\subsection{$\al$-representation}

A general theorem \cite{Speer} regarding the
analytical structure of an analytically
regularized Feynman integral $F_{\Gm}(\uq,\um;\ulm)$
states that it is a meromorphic function
of the regularization parameters.

A standard way to prove this property and to get
more concrete information about singularities with respect to $\ulm$
is to apply the well-known $\al$-representation of the Feynman
integrals\footnote{For simplicity, we consider Euclidean Feynman integrals.
The analytical properties in $\ulm$ are the same in Minkowski
and Euclidean spaces.} which is obtained by rewriting propagators as

\be
\frac{Z_l(p)}{(p^2+m_l^2)^{1+\lm_l}} = \frac{1}{\Gm(\lm_l+1)}
Z_l(\pa/\pa \xi_l)
\int_0^{\infty} \dd \al_l \; \al_l^{\lm_l}
\left. \exp\{-(p^2+m_l^2)\al + \xi_l \} \right|_{\xi_l=0}\,.
\label{pral}
\ee

\n and performing Gau{\ss} integration in the loop momenta in the integrand
of the integral in $\al_1,\ldots\al_L$.
Since the parameters $r_l$ can be taken into account by a redefinition of
$\lm_l$ we suppose now that $r_l=0$ for all $l$.

After that the analytically regularized
Feynman integral takes the form

\bea
F_{\Gm}(\uq,\um;\ulm) &=& C(\ulm)
\int_{R_+^L} \dd \ual \prod_l \al_l^{\lm_l}
D(\ual)^{-2} Z(\uq, \ual)
\exp\left\{ -A(\uq,\ual)/D(\ual) - \sum_{l=1}^L m^2_l \al_l\right\}
\nn \\
&\equiv&
\int \dd \ual \prod_l \al_l^{\lm_l}
I(\uq,\um,\ual;\ulm)\,,
\label{A}
\eea

\n where

\be
C(\ulm)
= \frac{1}{(16\pi^2)^h} \prod_l \frac{1}{\Gm(\lm_l+1)}\,,
\label{C} \nn
\ee
\be
Z(\uq, \ual) =
\prod_l Z_l(\pa/\pa u_l) \left.
\exp\left\{ (B(\uq,\uxi, \ual) +
K(\uq,\uxi, \ual))/D(\ual)\right\}
\right|_{\uxi=0}\,.
\label{Z}
\ee

\n Here $\uxi= (\xi_1, \ldots, \xi_L)$,
$\ual= (\al_1, \ldots, \al_L)$,
$\dd \ual= \dd\al_1 \ldots\dd \al_L$. The integration is over the
domain of non-negative parameters $\ual$, and $D,A,B,K$
are homogeneous functions that are constructed for the graph
$\Gm$ according to well-known rules (see, e.g., \cite{BM,VS,AZ}).
The homogeneity degrees
of the functions $D(\ual)$ and $A(\uq,\ual)$
with respect to the set of variables $\ual$ are respectively $h$
and $h+1$.
The function $A(\uq,\ual)$ is quadratic with respect to $\uq$.

\subsection{Sectors}

The problem of characterizing the singularities in $\ulm$ can be
reduced to resolving singularities of integrands in the
$\al$-representation which is a problem of algebraic geometry.
A natural approach is to locally introduce new
variables in which the original
complicated singularities are factorized.
In renormalization theory however one applies a simple change of
variables to
achieve this goal. First, the integration domain in (\ref{A}) is
decomposed into subdomains that are called sectors. Second, in each sector
new (sector) variables are introduced. In these variables
the singularities of all the functions
involved are factorized so that the problem reduces to power counting in
the sector variables.

We will not deal with  IR problems so that it is sufficient to
introduce
sectors and sector variables associated with 1PI subgraphs and
corresponding to maximal generalized forests. Let us call a subgraph
{\em UV-irreducible} if it is either 1PI or consists of a single line
which is not a loop line.
A set $F$ of UV-irreducible subgraphs is called a
{\em generalized forest} if
the following conditions hold: (a) for every pair of 1PI subgraphs
$\gm,\gm'\in F$ one has
either $\gm \subset \gm'$, or $\gm' \subset \gm$, or $\gm$ and $\gm'$
are disjoint (with respect to the set of vertices);
(b)  if $\gm_1,\ldots,\gm_j \in F$ are
pairwise disjoint with respect to the set of lines then the subgraph
$\cup_i \gm_i$ is one-particle-reducible (1PR).
Generalized forests which do not involve
single lines are nothing but forests in the common sense.
Thus every generalized forest $F$ is represented as the union
$F_h \cup F_r$ of a forest $F_h$ consisting of 1PI elements and
a family $F_r$ of 1PR single lines.

Let $\cal F$ be a {\em maximal generalized forest}
so that for any $\gm$ which does not belong to ${\cal F}$
the set
${\cal F} \cup \{\gm\}$ is no longer a generalized forest
Let $\sg$ be the mapping defined by
$\sg(\gm)\in\gm$ and $\sg(\gm)\not\in\gm'$ for any
$\gm'\subset \gm, \; \gm' \in {\cal F}$ so that $\sg(\gm)$ is the line
that belongs to $\gm$ and does not belong to subgraphs smaller than
$\gm$.

Let us define the following family of domains (sectors)
\cite{Pohl,Speer,BM} associated with
the maximal generalized forests of $\Gm$:

\be
{\cal D}_{\cal F} = \{ \ual | \al_l \leq \al_{\sg (\gm)} \;\;
\forall l\in\gm \in {\cal F} \} \,.
\label{sector}
\ee

\n The intersection of any two distinct (corresponding to different
generalized forests) sectors has zero measure and the
union of all the sets $\bigcup_{\cal F} {\cal D}_{\cal F}$ is the whole
integration domain in the $\al$-representation.

Let $F_{\Gm}^{\cal F}(\uq,\um;\ulm)$ be the contribution
of a sector ${\cal D}_{\cal F}$ to the given Feynman integral.
Let us introduce the following (sector) variables:

\be
\al_l = \prod_{l\in\gm\in {\cal F}} t_{\gm}\,.
\label{sv}
\ee

\n The corresponding Jacobian equals
$\prod_{\gm\in {\cal F}} t_{\gm}^{L(\gm)-1}$.
The inverse formulae are

\be
t_{\gm} = \al_{\sg (\gm)} / \al_{\sg (\gm_+)}\,,
\label{svi}
\ee

\n where $\gm_+$ denotes the minimal element of $\cal F$ that contains $\gm$
(we put $\al_{\sg (\Gm_+)} \equiv 1$).

Due to factorization
properties of the homogeneous functions of the $\al$-representation,
$F_{\Gm}$ can be represented as

\be
F_{\Gm}^{\cal F}(\uq,\um;\ulm) =
\int_0^{\infty}
t_{\Gm}^{\lm(\Gm)-[\om(\Gm)/2]-1}  \dd t_{\Gm}
\prod_{\gm\in {\cal F}; \gm\neq \Gm} \int_0^1
t_{\gm}^{\lm(\gm)-[\om(\gm)/2]-1}  \dd t_{\gm} f(\uq,\um,\ut;\ulm)\,,
\label{fact}
\ee

\n where $f$ is an infinitely differentiable function, the square brackets
denote the integer part of a number, and

\be
\lm(\gm) = \sum_{\l\in\gm} \lm_l\,.
\label{LG}
\ee

Thus the UV convergence analysis and therefore the analysis
of the analytic properties in $\ulm$ is characterized with the help of
properties of one-dimensional distribution $x^{\lm}_+$ which is a
meromorphic function with respect to $\lm$, with simple poles at
the points $\lm=-1-n, \; n=0,1,2, \ldots$ where

\be
x^{\lm}_+ = \frac{(-1)^{n}}{n!} \frac{1}{\lm+n} \dl^{(n)}(x) +
{\cal O}(1)\,.
\label{xl}
\ee

\subsection{Taylor operators}

To analyze analytical structure of the Feynman integral with respect
to complex parameters $\ulm$ let us introduce standard
Taylor operators and present them in various forms.

Let ${\cal T}^{N}_{\ldots}$ be the operator that picks up the terms up
to the $N$-th order of the Taylor expansion in the corresponding set of
variables. Then the formal Taylor expansion in masses and (independent)
external momenta is
described in the $\al$-parametric language as

\bea
M^{N}_{\Gm} F_{\Gm}(\uq,\um;\ulm) &\equiv&
{\cal T}^{N}_{\uq,\um} F_{\Gm}(\uq,\um;\ulm)
\equiv
\left. {\cal T}^{N}_{\ka} F_{\Gm}(\ka\uq,\ka\um;\ulm) \right|_{\ka=1} \nn \\
&=& \int \dd \ual
\prod_l \al_l^{\lm_l}
\left. {\cal T}^{N}_{\ka}
\ka^{4h(\Gm) + a(\Gm)}
I(\uq,\um,\ual';\ulm)
\right|_{\ka=1} \,,
\label{ME}
\eea

\n where $\al'_l = \ka^2 \al_l$.

Note that this is just a formal expansion operator in the sense that it
would certainly generate IR divergences when naively applied to the
Feynman integral. But we shall usually apply such operators not to the whole
integrals but to its sector contributions. These
operators are by definition applied to the integrand of the
$\al$-integrals involved.

Let $\gm$ be a subgraph of $\Gm$. Let us perform its mass and momentum
expansion
by the operator $M_{\gm}$ and then insert the result into the reduced
graph $\Gm/\gm$. We denote the insertion of a polynomial $\cal P$ in
external momenta of the subgraph $\gm$ into the reduced diagram by
$F_{\Gm/\gm} \circ {\cal P}$. Thus we are dealing with
$F_{\Gm/\gm} \circ M_{\gm} F_{\gm}$ which is by definition the action of
the operator $M_{\gm}$ associated with $\gm$  on the whole Feynman
integral $F_{\Gm}$.

To describe this procedure in the $\al$-parametric language let us
remember that the $\al$-representation is obtained by performing integrals
over the loop momenta. If one first integrates over thee loop momenta of the
subgraph $\gm$, one obtains the $\al$-representation for the Feynman
integral associated with the subgraph. After using formula (\ref{ME})
and continuing integration over the rest loop momenta (in fact
associated with the reduced graph $\Gm/\gm$) one arrives at the
relation

\bea
M_{\gm}^{N} F_{\Gm}(\uq,\um;\ulm) &\equiv&
(F_{\Gm/\gm} \circ M_{\gm} F_{\gm})(\uq,\um;\ulm)
\nn \\
&=& \int \dd \ual
\prod_l \al_l^{\lm_l}
\left. {\cal T}^{N}_{\ka}
\ka^{4h(\gm) + a(\gm)}
I(\uq,\um,\ual';\ulm)
\right|_{\ka=1} \,,
\label{MEgm}
\eea

\n where $\al'_l = \ka^2 \al_l$ if $l\!\not\!\in\gm$ and
$\al'_l = \al_l$ if $l \in\gm$.

These simple arguments would be rigorous if one could control
convergence of the integrals in the loop momenta and in the $\al$-parameters
involved and easily prove that one may change the order of the
integration.
It happens that it is simpler to provide a direct proof of
(\ref{MEgm}) --- see \cite{BM} (Lemma~5).
More  details can be found in \cite{VS} (Lemma~11.1).

Let $\gm \in {\cal F}$. Note that by
(\ref{sv}) multiplication of all $\al_l$ with
$l\in\gm$ by $\ka^2$ is equivalent to the replacement
$t_{\gm}\to \ka^2 t_{\gm}$. Therefore the action of the operator
$M_{\gm}^{N}$ on the sector contribution takes the form

\be
M_{\gm}^{N} F_{\Gm}^{\cal F} (\uq,\um;\ulm)
= \int_0^{\infty}
t_{\Gm}^{\lm(\Gm)-[\om(\Gm)/2]-1}  \dd t_{\Gm}
\prod_{\gm'\in {\cal F}; \gm'\neq \Gm} \int_0^1
t_{\gm'}^{\lm(\gm')-[\om(\gm')/2]-1}  \dd t_{\gm'}
{\cal T}^{N}_{t_{\gm}} f(\uq,\um,\ut;\ulm) \,,
\label{Tfact}
\ee

\n where $f$ is the function that enters (\ref{fact}).

\subsection{Analytic structure in $\lm$}

Representation (\ref{fact}) shows that
the sector contribution $F_{\Gm}^{\cal F}(\uq,\um;\ulm)$ to the
analytically regularized Feynman integral $F_{\Gm}(\uq,\um;\ulm)$
can be represented as a

\be
\prod_{\gm\in {\cal F}:\om(\gm)\geq 0}
\frac{1}{\lm(\gm)} \; g_{\cal F} (\ulm) \,,
\label{F01}
\ee

\n where the function $g_{\cal F} $ is analytical in a
vicinity of the point $\ulm=\underline{0}$.
The whole Feynman integral is therefore looks like
the sum (\ref{as}) of the terms (\ref{F01}) over all
the maximal forests of $\Gm$.

The {\em leading singularity} of the given Feynman
integral is by definition
the sum of terms (\ref{F01}) with the maximal number of the factors
$\lm(\gm)$ in the denominator.
According to (\ref{F01}) this number is equal to or less than
the maximal number of divergent
subgraphs that can belong to the same forest.
For simplicity, let us now consider only the vicinity of the point
$\lm_l=0$.
%where $l$ are lines that are associated with the analytically regularized
%propagators.

To calculate the leading singularity let us observe
that the singular factor $1/\lm(\gm)$ originates from the
integral over the sector variable $t_{\gm}$ in (\ref{fact}).
Due to (\ref{xl}), evaluation of the residue at this pole amounts to
the action of the Taylor operator ${\cal T}^{\om(\gm)}_{t_{\gm}}$ which
in turn is equivalent to the action of the operator
$M_{\gm}^{N} $, with $N=\om(\gm)$,
given by the right-hand side of (\ref{MEgm}).
Using this relation we observe that the problem reduces to
evaluation of the residue of the Feynman integral $F_{\gm}(\ulm)$
with respect to $\lm(\gm)$ and
insertion of the result into the reduced graph.

Consider, for example, a situation when
a given maximal forest contains only two divergent subgraphs $\gm$ and
$\Gm$. Let $\HK_{\lm}$ be the operator
that picks up the pole part of Laurent series in $\lm$.
Then we see that the leading singularity $LS (F^{\cal F}_\Gm)$ is given by

\be
\HK_{\lm(\Gm)} \left(
F^{{\cal F}/\gm}_{\Gm/\gm}
(\ulm_{\Gm/\gm}, \{\lm(\gm)\})
\circ
\HK_{\lm(\gm)} \overline{F}^{{\cal F}(\gm)}_{\gm}(\ulm) \right)\,,
\label{LS1}
\ee

\n Remember that $\lm(\gm)$ is given by (\ref{LG}).
Here ${\cal F}/\gm$ is a generalized forest of the reduced graph
(it is obtained from ${\cal F}$ by `dividing' each element $\gm'$ with
$\gm \subset \gm'$ by $\gm$)
and ${\cal F}(\gm)$ is the generalized forest of $\gm$ which is
the `projection' of $\cal F$ to $\gm$.
(Note that
there is one-to-one correspondence between
generalized forests  $\cal F$ that satisfy $\gm\in {\cal F}$
and generalized forests of $\gm$ and $\Gm/\gm$.)
By $F^{\cal F/\gm}_{\Gm/\gm}(\ldots,\{\lm\})$ we denote the given sector
contribution regularized by inserting the factor
$\al_{\sg(\gm_+)}^{\lm} \equiv
t_{\gm_+}^{\lm} t_{(\gm_+)_+}^{\lm} \ldots t_{\Gm}^{\lm}$
into the integrand.
Finally,
$\overline{F}^{{\cal F}(\gm)}_{\gm}(\ulm)$
differs from
$F^{{\cal F}(\gm)}_{\gm}(\ulm)$ by the reduction of
the region of integration to $\al_l \leq \al_{\sg(\gm_+)}, \; l\in\gm$.

Let us now observe that

\be
\left.
\HK_{\lm(\gm)} \overline{F}^{{\cal F}(\gm)}_{\gm}(\ulm)
=\HK_{\lm(\gm)} F^{{\cal F}(\gm)}_{\gm}(\ulm)
=\HK_{\lm} F^{{\cal F}(\gm)}_{\gm}(\lm)
\right|_{\lm=\lm(\gm)}
\equiv \frac{1}{\lm(\gm)} \mbox{res}_{\lm} F^{{\cal F}(\gm)}_{\gm}(\lm)\,.
\label{LS2}
\ee

\n where it is implied that $F^{{\cal F}(\gm)}_{\gm}(\lm)$
is regularized by introducing the regularization parameter $\lm$
into any line of $\gm$. Similarly, we have

\be
\left.
\HK_{\lm(\Gm)} \left( F^{{\cal F}/\gm}_{\Gm/\gm}
(\ulm_{\Gm/\gm}, \{\lm(\gm)\}) \circ {\cal P} \right)
=\HK_{\lm} \left( F^{{\cal F}/\gm}_{\Gm/\gm} (\lm) \circ {\cal P} \right)
\right|_{\lm=\lm(\Gm)}
\equiv \frac{1}{\lm(\Gm)}
\mbox{res}_{\lm} \left( F^{{\cal F}/\gm}_{\Gm/\gm} (\lm)
\circ {\cal P} \right)\,,
\label{LS3}
\ee

\n Here ${\cal P}$ is the polynomial (\ref{LS2}) that is
inserted into the reduced graph.

Let us represent a maximal forest $\cal F$ as
${\cal F}_h \cup {\cal F}_r \equiv {\cal F}_{h,div}
\cup {\cal F}_{h,conv} \cup {\cal F}_r $ where the subscripts `div' and
`conv' denote 1PI elements respectively with $\om(\gm) \geq 0$ and
$\om(\gm) < 0$. Let $\max_{\cal F} |{\cal F}_{h,div} | = n_0$
be the maximal possible number of divergent subgraphs that can belong
to the same maximal forest.\footnote{Note that a single line with
the propagator $1/(p^2)^{\lm}$ is formally UV-divergent.
However the corresponding pole is cancelled by
a $\Gamma$-function in
(\ref{A}). Therefore it is sufficient to consider the usual forests
rather than generalized forests
when calculating the number of singular factors that enter the
leading singularity.}

Using straightforward
generalization of the arguments used in the above example with
two divergent subgraphs
and performing summation over all the maximal forests
of the given graph
we come to the following simple representation for the leading singularity:

\be
LS (F_\Gm) = \sum_{{\cal F}: |{\cal F}_{h,div} | = n_0}
\prod_{\gm\in {\cal F}_{h,div}}
\frac{1}{\lm(\gm)}
\left(
\mbox{res}_{\lm} F_{\gm/\gm_-} (\lm)
\right)\, .
\label{LS4}
\ee

\n Here $\gm_-$ is the set of maximal
elements  $\gm'\in {\cal F}$ with $\gm'\subset\gm$.
Each factor
$\mbox{res}_{\lm} F_{\gm/\gm_-} (\lm)$ is a
polynomial with respect to external momenta and internal masses of
$\gm/\gm_-$.
It is implied that these factors are partially ordered
and before calculation the residue
$\mbox{res}_{\lm} F_{\gm/\gm_-} (\lm)$
all polynomials associated with the set $\gm_-$
are inserted into this `next' reduced diagram.
Remember that when calculating these residues it does not
matter into which line the regularization parameter $\lm$
is introduced.

\end{document}